\documentclass[usenatbib]{mn2e}
\pdfoutput=1

\usepackage{newtxtext,newtxmath}
\usepackage[T1]{fontenc}
\usepackage{ae,aecompl}
\usepackage{docmute}
\usepackage{float}
\usepackage{comment}
\usepackage{color}
\usepackage{caption}
\usepackage{multirow}
\usepackage{docmute}
\usepackage{import}
\usepackage[dvipsnames]{xcolor}
\usepackage{graphicx}	
\usepackage{amsmath}	
\usepackage{amssymb}	
\usepackage{amsfonts}
\usepackage{booktabs}
\usepackage{siunitx}
\usepackage{pifont}
\usepackage{fixltx2e}
\usepackage{caption}
\usepackage{threeparttable}
\usepackage{tikz}
\usepackage{threeparttable}
\usepackage{float}
\usepackage{aas_macros}
\usepackage{hyperref}

\newcommand{\cmark}{\ding{51}}%
\newcommand{\xmark}{\ding{55}}%

\hypersetup{
  linkcolor  = blue, 
  citecolor  = blue, 
  urlcolor   = RubineRed,
  colorlinks = true,
}

\title[Feedback-induced DM core profile]{A dark matter profile to model diverse feedback-induced core sizes of \boldmath$\Lambda$CDM haloes}

\author[A. Lazar et al.]{Alexandres Lazar$^{1}$\thanks{\href{mailto:aalazar@uci.edu}{aalazar@uci.edu}},
James S. Bullock$^{1}$,
Michael Boylan-Kolchin$^{2}$,
T.K. Chan$^{3,4}$, \newauthor 
Philip F. Hopkins$^{5}$,
Andrew S. Graus$^{2}$, 
Andrew Wetzel$^{6}$,
Kareem El-Badry$^{7}$, \newauthor 
Coral Wheeler$^{4}$, 
Maria C. Straight$^{8,2}$,
Du\v{s}an Kere\v{s}$^{4}$,
Claude-Andr\'e Faucher-Gigu\`ere$^{9}$, \newauthor 
Alex Fitts$^{2}$
and Shea Garrison-Kimmel$^{5}$
\\
$^{1}$Department of Physics and Astronomy, University of California, Irvine, CA 92697 USA\\
$^{2}$Department of Astronomy, The University of Texas at Austin, 2515 Speedway, Stop C1400, Austin, Texas 78712-1205, USA\\
$^{3}$Institute for Computational Cosmology, Durham University, South Road, Durham, DH1 3LE, UK\\
$^{4}$Department of Physics, Center for Astrophysics and Space Science, University of California at San Diego, 9500 Gilman Dr.\\
$^{5}$California Institute of Technology, TAPIR, Mailcode 350-17, Pasadena, CA 91125, USA\\
$^{6}$Department of Physics, University of California, Davis, CA 95616, USA\\
$^{7}$Department of Astronomy and Theoretical Astrophysics Center, University of California Berkeley, Berkeley, CA 94720\\
$^{8}$Department of Physics, Whitworth University, 300 West Hawthorne Road, Spokane, Washington 99251, USA \\
$^{9}$Department of Physics and Astronomy and CIERA, Northwestern University, 2145 Sheridan Road, Evanston, IL 60208, USA 
\vspace{-0.4cm}
}

\date{Working Draft\vspace{-0.6cm}}
\pubyear{2020}

\begin{document}
\label{firstpage}
\pagerange{\pageref{firstpage}--\pageref{lastpage}}
\maketitle
\raggedbottom

\begin{abstract}
We analyze the cold dark matter density profiles of 54 galaxy halos simulated with FIRE-2 galaxy formation physics, each resolved within $0.5\%$ of the halo virial radius. These halos contain galaxies with masses that range from ultra-faint dwarfs ($M_\star \simeq 10^{4.5} M_{\odot}$) to the largest spirals ($M_\star \simeq 10^{11} M_{\odot}$) and have density profiles that are both cored and cuspy. We characterize our results using a new, analytic density profile that extends the standard two-parameter Einasto form to allow for a pronounced constant-density core in the resolved innermost radius. With one additional core-radius parameter, $r_{c}$, this three-parameter {\em core-Einasto} profile is able to characterize our feedback-impacted dark matter halos more accurately than other three-parameter profiles proposed in the literature. In order to enable comparisons with observations, we provide fitting functions for $r_{c}$ and other profile parameters as a function of both $M_\star$ and $M_{\star}/M_{\rm halo}$. In agreement with past studies, we find that dark matter core formation is most efficient at the characteristic stellar-mass to halo-mass ratio $M_\star/M_{\rm halo} \simeq 5 \times 10^{-3}$, or $M_{\star} \sim 10^9 \, M_{\odot}$, with cores that are roughly the size of the galaxy half-light radius, $r_{c} \simeq 1-5$ kpc. Furthermore, we find no evidence for core formation at radii $\gtrsim 100\ \rm pc$ in galaxies with $M_{\star}/M_{\rm halo} < 5\times 10^{-4}$ or $M_\star \lesssim 10^6 \, M_{\odot}$. For Milky Way-size galaxies, baryonic contraction often makes halos significantly more concentrated and dense at the stellar half-light radius than DMO runs. However, even at the Milky Way scale, FIRE-2 galaxy formation still produces small dark matter cores of $\simeq 0.5-2$ kpc in size. Recent evidence for a ${\sim} 2$ kpc core in the Milky Way's dark matter halo is consistent with this expectation.
\end{abstract}



\section{Introduction}
The theory of Cold Dark Matter with the inclusion of the cosmological constant ($\Lambda$CDM) has been the benchmark paradigm in cosmological studies, as its framework has been successful in modeling the distribution of large-scale structure of our universe. However, on small scales, there are potential inconsistencies between predictions made by the $\Lambda$CDM paradigm and what is observed in real galaxies. One of these inconsistencies concerns the distribution of dark matter in centers of galaxies. This known as the {\it cusp-core} problem: dark matter halos simulated without baryons in $\Lambda$CDM have {\it cusped} dark matter densities at small radii, i.e.  $\rho(r) \propto r^{\alpha}$ with $\alpha \sim -1$ \citep{dubinkski1991,navarro1997universal,navarro2004inner}, while observations of some dark matter dominated galaxies appear to suggest profiles are better described by constant-density {\it cores} at small radii, i.e. $\alpha \sim 0$ \citep{flores1994observational, moore1994evidence,salucci2000dark,swaters2003central,gentile2004cored,spekkens2005cusp,walter2008things,oh2011central,relatores2019inner}. Another potentially related discrepancy is called the {\it Too Big to Fail} problem \citep{boylan2011too}: Milky Way satellite galaxies are observed to have much smaller inner dark matter densities compared to the surplus of subhalos predicted from (dark matter only) cosmological $N$-body simulations. This problem also persists in other dwarf galaxies of the Local Group and local field \citep{garrison2014too,tollerud2014,papastergis2015too}.

Most of the above-mentioned problems were posed from dark matter only simulations, which lack the effects of baryons. One way galaxy formation can affect dark matter is by boosting central dark matter densities as a result of baryons clustering at the center of the halo \citep{blumenthal1986contraction}. This denoted as ``baryonic contraction'' in the literature and it is an effect that is particularly important for Milky Way-mass galaxies \citep[e.g.][]{gnedin2004response,chan2015impact}. Alternatively, the inner dark matter density can decrease in response to repetitive energetic outflows from stellar feedback, a process often referred to as ``feedback-induced core formation'', and one that is most effective in galaxies that are somewhat smaller than the Milky Way  \citep{navarro1996cores,read2005shallow,governato2010bulgeless,governato2012cuspy,pontzen2012supernova,teyssier2013cusp,di2013dependence,chan2015impact,brook2015local,tollet2016nihao}. Another possibility is that dynamical friction from small accretion events \citep{elzant2001cusp,tonini2006angular,romano2008erasing,goerdt2010core,cole2011weakening} can flatten the dark matter density profile.

The effects of feedback on core formation depend sensitively on the total amount and precise nature of star formation. For example, \citet{penarrubia2012coupling} showed that galaxies with too few stars (and therefore, too few supernovae) are unlikely to have feedback-induced cores owing to an insufficient amount energy from supernovae to substantially transform the dark matter profile. \cite{mashchenko2006removal} showed that concentrated star formation episodes that are {\em spatially} displaced from halo centers can drive bulk gas flows, alter dark matter particle orbits, and increase the likelihood for dark matter core formation. Time-repetitive ``bursty'' star formation also affects core formation, allowing for dark matter particle orbits to be affected significantly over time as gas is expelled and re-accreted in the baryon cycle \citep{pontzen2012supernova}. The timing of star formation relative to dark matter halo growth can also affect core formation; in cases where dark matter rich mergers occur after core-producing star formation, cusps can be reborn \citep{onorbe2015forged}. Dark matter core formation is seen in many fully self-consistent cosmological simulations that resolve star formation on small spatial scales \citep[e.g.][]{governato2010bulgeless,munshi2013stellar,brooks2014why,madau2014core,onorbe2015forged,el2016breathing,tollet2016nihao,fitts2017fire}. One common aspect of these simulations is that they have relatively high gas density thresholds for star formation. Cosmological simulations with lower density thresholds for star formation, e.g. {\footnotesize APOSTLE} and {\footnotesize Auriga} \protect\citep{bose2019cores}, have been shown to not produce dark matter cores. The dependence of feedback-induced core formation on the star formation density threshold has been studied in more detail by \cite{dutton2019impact} and \cite{benitez2019cores}. Both concluded that density thresholds higher than the mean ISM density, which allows for some ISM phase structure and clustered star formation as observed, is necessary in forming feedback-induced cores.

\cite{di2013dependence} studied the relationship between the inner local density slope of dark matter, $\alpha$, and the stellar mass fraction, $M_{\star}/M_{\rm halo}$, of simulated galaxies from the {\footnotesize MUGS} \citep{stinson2010galaxy} and {\footnotesize MaGICC} \citep{brooks2012magicc,stinson2012magicc} simulations for a wide range stellar mass systems, $M_{\star} \simeq 10^{5-11}\ M_{\odot}$. They found that core formation is a strong function the mass-ratio of stars formed to total halo mass and demonstrated that there is a characteristic mass-ratio for efficient core formation $M_{\star}/M_{\rm halo} \simeq 5 \times 10^{-3}$, above and below which galaxy halos approach the cuspy behavior associated with dark matter only simulations. \cite{chan2015impact} used galaxies of stellar masses, $M_{\star} = 10^{3-11}\ M_{\odot}$, from the FIRE-1 suite \citep{hopkins2014galaxies} to study feedback-induced core formation and found similar results. \cite{tollet2016nihao} used the {\footnotesize NIHAO} suite \citep{wang2015nihao1} for a wide range of halo masses, $M_{\rm halo} = 10^{10-12}\ M_{\odot}$ and further confirmed this qualitative phenomena. Recently, \cite{maccio2020nihao} extended the work of \cite{tollet2016nihao} with the inclusion of black hole feedback for galaxies spanning eight orders in magnitude in stellar mass.

The above-mentioned simulation groups agree on a few additional qualitative points. First, feedback typically does not produce significant deviations from cuspy dark matter only predictions in the smallest galaxies: $M_{\star}/M_{\rm halo} < 10^{-4}$ ($M_{\star} \lesssim 10^6 \, M_{\odot}$, typically), as expected on energetic grounds \citep{penarrubia2012coupling,garrison2013}. Second, dark matter halos become more cored as $M_{\star}/M_{\rm halo}$ increases up until $M_{\star}/M_{\rm halo} \simeq 5 \times 10^{-3}$, which is the region of {\it peak} core formation. These halos are not well modeled by cuspy density profiles and must be described by an alternative dark matter profile that has a pronounced flattening in slope at small radii. In higher mass halos, $M_{\rm halo} \simeq 10^{12}\ M_{\odot}$, baryonic contraction actually makes halos denser at the stellar half-mass radius than dark matter only simulations would suggest. However, \cite{chan2015impact} found that within this radius, small cores are often present even within baryonically-contracted $10^{12}\ M_{\odot}$ halos.

The analysis done in \cite{di2014mass} explored a general five-parameter density profile to characterize halos with either cuspy or cored inner density profiles. In addition to a characteristic radius and density, this profile had three shape parameters: $\alpha,\ \beta,\ \text{and}\ \gamma$  \citep{zhao1996models}. They found that the values of the three shape parameters varied regularly as a function of the $M_{\star}/M_{\rm halo}$ and provided fitting functions that captured these trends. Therefore, given $M_{\star}/M_{\rm halo}$, the  \cite{di2014mass} profile reduces to a two free-parameter function that may be used to compare predictions with observations in a fairly straightforward manner.

The $\alpha\beta\gamma$-profile can be regarded as a generalization of the \citet[][NFW]{navarro1997universal} profile, which provides a good fit to dark matter only simulations. Since dark matter only simulations have traditionally been characterized by the NFW profile, there have been attempts to modify the NFW form by allowing for a constant density core radius parameter $r_{c}\equiv r_{\rm core}$. For example, \citet{penarrubia2012coupling} suggested a three-parameter core profile: the classic NFW profile with a core radius in the inner radial regions of the halo. \cite{read2016cNFW} derives a core profile starting with an NFW form by  connecting core formation to features of star-formation efficiency and the stellar half-mass radius. 
More recently, \cite{freundlich2020dekel} used {\footnotesize NIHAO} to explore a constrained version of the $\alpha\beta\gamma$ profile that has three-parameters, the ``Dekel+'' profile \citep{dekel2017profile}, with a variable inner slope and concentration parameter.

\begin{figure}
    \centering
    \includegraphics[width=\columnwidth]{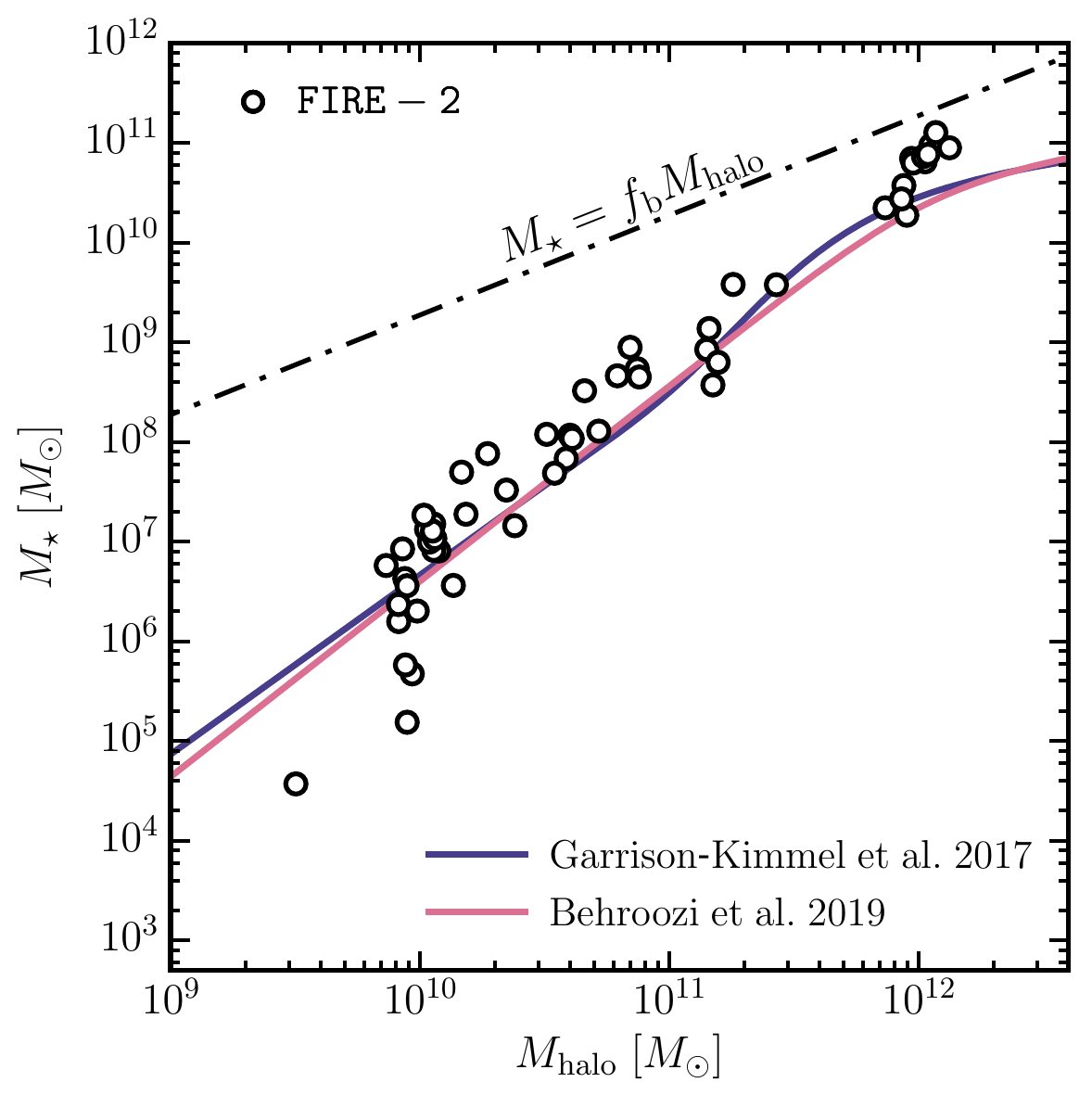}
    \caption{---
        {\bf \emph{Stellar mass to halo mass relations}}. The white points show galaxies from the FIRE-2 simulations studied in this paper.
        The curves are the median abundance matching relations presented in \protect\cite[zero scatter]{sgk2017chaos} (blue) and \protect\cite{behroozi2019um} (pink).
    }
    \label{fig:1}
\end{figure}

In what follows, we revisit the question of dark matter halo density profiles in cosmological galaxy formation simulations using the FIRE-2 feedback model \citep{hopkins2018fire}. The simulations we consider herein allow us to resolve to within $0.5\%$ of the halo virial radius in halos that produce galaxies spanning six orders of magnitude in stellar mass. We introduce a new analytic density profile, the ``core-Einasto'', that extends the \citet{einasto1965influence} form by adding one free parameter, a physical core radius, $r_{c}$. 
It is well known that the two-parameter Einasto profile provides a better fit to dark matter only simulations than the two-parameter NFW \citep{navarro2004inner,wang2019zoom}. Similarly we find that the three-parameter core-Einasto profile provides a better fit to FIRE-2 halos than two popular three-parameter versions of generalized double-power law profiles: the Dekel+ profile \citep{dekel2017profile,freundlich2020dekel} and a cored extension of the NFW 
\citep{penarrubia2012coupling}. We also find that the two-parameter \citet{di2014mass} profile is not a good fit to our feedback-affected halos. 

In addition to providing a better fit to our FIRE-2 halos than other three-parameter profiles, the core-Einasto profile utilizes a physically-meaningful core-radius parameter, $r_c$. The numerical value of $r_c$ matches well to the radius where a visual profile begins to flatten towards a constant density. The combination of accuracy, intuitive parameters, and ease-of-use will hopefully allow our three-parameter core-Einasto profile to become a useful tool for comparing predictions to observations.

This article is structured as follows: Section~\ref{sec:numerical.methods} discusses our sample of high resolution galaxies simulated with FIRE-2 physics along with their relevant properties. We also discuss the numerical intricacies considered for our galaxies. Section~\ref{sec:slope} revisits the analysis of correlations between $\alpha$ and $M_{\star}/M_{\rm halo}$ for our sample of galaxies and dark matter halos. In Section~\ref{sec:cEinasto}, we introduce the cored version of the classic Einasto profile used to model $\Lambda$CDM halos. We use the properties of these profiles to provide constraints on dark matter cores as a function and of $M_{\star}/M_{\rm halo}$. We summarize our results and discuss potential uses for observational and cosmological studies in Section~\ref{sec:conclude}. 
The appendix includes five sections: \ref{sec:cEinasto.Mstar} has expressions for fitting parameters as a function of stellar mass; \ref{sec:analytic.cEinasto} derives analytical expressions for the mass and gravitational potential implied by the core-Einasto profile; \ref{sec:cEinasto.BC} has a four-parameter core-Einasto extension that better accounts for adiabatic contraction in Milky Way size halos; \ref{sec:cNFW} presents comparisons to fits with alternative three-parameter profiles and also presents fits for the five-parameter $\alpha\beta\gamma$ form; and \ref{sec:sim.sample} provides tables that list all halo properties and best-fit profile parameters for each halo in our sample.

\section{Numerical Methodology}
\label{sec:numerical.methods}
In this section, we briefly describe the suite of high-resolution simulations used in our analysis. We discuss the FIRE-2 model for full galaxy formation physics in Section~\ref{sec:FIRE2.model}, the numerical parameters used in our high resolution simulations in Sections~\ref{sec:simulations} and \ref{sec:power.radius}, and present the halo sample used in this analysis in Section~\ref{sec:halo.sample}. The numerical simulations presented here are all part of the Feedback In Realistic Environments (FIRE) project\footnote{The {\footnotesize FIRE} project website: \url{http://fire.northwestern.edu}} and are listed in Table \ref{tab:halo.sample} at the end of this article.

\subsection{The FIRE-2 model} 
\label{sec:FIRE2.model}
Our simulations were run using the multi-method code {\footnotesize GIZMO} \citep{hopkins2015new}, with the second-order mesh-free Lagrangian-Godunov finite mass (MFM) method for hydrodynamics. {\footnotesize GIZMO} utilizes an updated version of the PM+Tree algorithm from {\footnotesize GADGET-3} \citep{springel2005cosmological} to calculate gravity and adopts fully conservative adaptive gravitational softening for gas \citep{price2007energy}. The FIRE-2 model \citep{hopkins2018fire}, which is an updated version of the FIRE-1 feedback scheme from \cite{hopkins2014galaxies}, is used to implement star formation and stellar feedback physics. Gas and gravitational physics implemented are discussed in complete detail in \cite{hopkins2018fire}. Here we discuss in brief detail the feedback physics relevant to core formation.

The simulations presented here tabulate the relevant ionization states and cooling rates from a compilation of {\footnotesize CLOUDY} runs \citep{ferland1998cloudy}, accounting for gas self-shielding. The gas cooling mechanisms follow the cooling rates of $T=10 - 10^{10}$ K; these include metallicity-dependent fine-structure atomic cooling, low temperature molecular cooling, and high temperature metal-line cooling that followed 11 separately tracked species. Gas is heated and ionized throughout cosmic time using the redshift dependent UV background model from \cite{faucher2009new} that ionizes and heats gas in an optically thin approximation and uses an approximate prescription to account for self-shielding of dense gas using a Sobolev/Jeans-length approximation. Stars are formed in Jeans-unstable, molecular gas regions at densities $n_{\rm H} \geq 10^{3}\ \rm cm ^{-3}$, with $100 \%$ instantaneous efficiency per local free-fall time in dense gas. Each star particle is an assumed stellar population with a \cite{kroupa2001variation} IMF that inherits its metallicity from its parent gas particle and has an age determined by its formation time. The stellar feedback implemented includes stellar winds, radiation pressure from young stars, Type II and Type Ia supernovae, photoelectric heating, and photo-heating from ionizing radiation. Feedback event rates, luminosities, energies, mass-loss rates, and other quantities are tabulated directly from stellar evolution models \cite[{\footnotesize STARBURST99
};][]{leitherer1999starburst99}. 

\subsection{Numerical simulations}
\label{sec:simulations}
All simulations in this analysis use a zoom-in technique \citep{onorbe2013zoom} to reach high resolutions in a cosmological environment by constructing a convex-hull region and refining it in progressively higher-resolution shells until the desired resolution is reached in the inner-most region. All initial conditions are generated with {\footnotesize MUSIC} \citep{hahn2013music} and then the simulations are evolved from redshifts $z\approx100$ to $z=0$ assuming a flat $\Lambda$CDM cosmology. We note that the cosmological parameters in each of the simulations vary to some degree, but remain consistent with \cite{ade2016planck}. Across our entire simulation sample: $h = 0.68-0.71$,  $\Omega_{\Lambda} = 1 - \Omega_{m} = 0.69-0.73$,  $\Omega_{b} = 0.0455 - 0.048$, $\sigma_{8} = 0.801 - 0.82$, $n_{s} = 0.961-0.97$.
In post-processing, halos are identified using the phase-space halo finder {\footnotesize ROCKSTAR} \citep{behroozi2012rockstar}, which uses adaptive, hierarchical refinement of the friends-of-friends groups in 6-dimensional phase-space and one time dimension. This results in robust tracking of halos and subhalos \citep{srisawat2013merger}.

\subsection{Halo sample \texorpdfstring{\&}{and} nomenclature}
\label{sec:halo.sample}
Throughout this paper, dark matter halos are defined as spherical systems with virial radius, $r_{\rm vir}$, inside of which the average density is equal to $\Delta_{\rm vir}(z)\rho_{\rm crit}(z)$. Here, $\rho_{\rm crit}(z) := 3 H^{2}(z)/8\pi G$ is the critical density of the universe and $\Delta_{\rm vir}(z)$ is the redshift evolving virial overdensity defined in \cite{bryan1998statistical}. The virial mass of a dark matter halo, denoted by $M_{\rm halo}$, is then defined as the dark matter mass within $r_{\rm vir}$. The stellar mass of the galaxy, $M_{\star}$, is then taken to be the total sum of the stellar particles inside $10 \% \times r_{\rm vir}$. It follows that the three-dimensional stellar-half-mass radius, $r_{1/2}$, is the radius that encloses half of the defined stellar mass. Finally we refer to the ``stellar fraction'' of the halo as the ratio between the quantified stellar mass and the halo mass: $M_{\star}/M_{\rm halo}$.

Fig.~\ref{fig:1} outlines our sample of galaxies, where just the dark matter halo masses (from the FIRE-2 runs) are plotted against $M_{\star}$. We compare our sample with the the abundance matching relations presented in \cite[zero scatter]{sgk2017chaos} and \cite{behroozi2019um} as the blue and pink curves, respectively, showing the best fit median abundance matching relations. Table~\ref{tab:halo.sample} lists all of the halos galaxies in this paper, including their $z=0$ properties from the FIRE-2 runs. Given our large sample, we chose to divide our galaxy sample into four convenient classifications of objects using the convention from \cite{bullock2017small}:\footnote{Note that these classifications are based on galaxies that span specific stellar mass ranges.}

\begin{itemize}
    \item[] {\bf \emph{Ultra-Faint Dwarfs}}: 
    Defined to have stellar masses of $M_{\star} \approx 10^{2-5}\ M_{\odot}$ at $z=0$. These are analogs of galaxies to be detected within limited local volumes around M31 and the Milky Way.
    \vspace{1ex}
    
    \item[] {\bf \emph{Classical Dwarfs}}: 
    Defined to have stellar masses of $M_{\star} \approx 10^{5-7}\ M_{\odot}$ at $z=0$. These are analogs of the faintest galaxies known prior to {\it SDSS}. 
    \vspace{1ex}
    
    \item[] {\bf \emph{Bright Dwarfs}}: 
    Defined to have stellar masses of $M_{\star} \approx 10^{7-10}\ M_{\odot}$ at $z=0$. These are analogs of the faintest galaxies that can be seen in wide-field galaxy surveys.
    \vspace{1ex}
    
    \item[] {\bf \em Milky Way-Mass Halos}: 
    Defined to host spiral galaxies with stellar mass of $M_{\star} \approx 10^{10-11}\ M_{\odot}$ at $z=0$. At the peak of abundance-matching relation, this maps to the generally accepted range in Milky Way-mass halos of $M_{\rm halo} = [0.8 - 2.4] \times 10^{12}$. Hereafter, we abbreviate Milky Way as ``MW''.
\end{itemize}

Lastly, each zoomed-in halo run with full FIRE-2 physics has an analogous dark matter only (DMO) version.  The individual dark matter particle masses in the DMO versions are larger by a factor of $(1 - f_{b})^{-1}$ in these runs, where $f_{b} := \Omega_{b}/\Omega_{m}$ is the cosmic baryon fraction, but the initial conditions are otherwise identical. The density profiles quoted from the DMO simulations have been scaled $m_{\rm p} \rightarrow (1-f_{b})m_{\rm p}$ in order to roughly account for the exclusion of the baryons. Other quantities are also adjusted accordingly: $\rho(r) \rightarrow (1-f_{b})\rho(r)$, $M_{\rm halo}(<r) \rightarrow (1-f_{b})M_{\rm halo}(<r)$ and $V_{\rm circ}(r) \rightarrow \sqrt{1-f_{b}}V_{\rm circ}(r)$, for all of the results analyzed in the DMO runs. This provides a simple comparison set to understand the additional effects of energetic feedback seen in our FIRE-2 runs.

\subsection{Radial profiles}
For each main halo identified by {\footnotesize ROCKSTAR}, the center of the halo is quantified through a ``shrinking spheres'' iteration scheme \citep{power2003inner,navarro2004inner}: the center of mass of particles is computed in a sphere and then has its radius reduced by half and re-centered on the new center of mass. This is done successively until the sphere contains one thousand particles. The final center of mass position is determined at this last iteration. For our galaxies, this is done for the combined star and dark matter particles found inside the virial radius while the center of mass for the DMO analogs are done with only dark matter inside the halo.\footnote{We also compared our results with centers defined as the most bound dark matter particle in the halo determined by {\footnotesize ROCKSTAR}. We find no qualitative differences in our final results.} 
The spherically averaged local density profile, $\rho(r)$, is constructed in 35 logarithmically spaced bins over $[0.005-1]\,\times \, r_{\rm vir}$.
We expected systematic uncertainties in the binned density estimates to be extremely minimal due to large number of particles in each simulation sample. Throughout the entirety of this paper, we refer to these local density profiles as the density profiles for the dark matter halo.

\subsection{Region of numerical convergence}
\label{sec:power.radius}
We expect the innermost regions of our simulated halos to be affected by numerical relaxation. With a variety of galaxies simulated at different resolutions, we must account for resolution differently in each simulation. We do so using the method specified in \cite{power2003inner}, where the effective resolution of cosmological simulations is related to the radius where the two-body relaxation timescale, $t_{\rm relax}$, becomes shorter than the age of the universe, $t_{0}$. Precisely, the radius at which numerical convergence is achieved, $r_{\rm conv}$, is dependent on the  number of enclosed particles, $N(<r)$, as well as the mean density enclosed at the associated radius, $\bar{\rho}(r) = 3M(<r)/4\pi r^{3}$, where $M(<r)$ is the total mass contained within radius $r$. Therefore, $r_{\rm conv}$ is governed by the following equation:
\begin{align}
    \frac{t_{\rm relax}(r)}{t_{0}} = \frac{\sqrt{200}}{8}\frac{N}{\ln N}
    \left[ \frac{\bar{\rho}(r)}{\rho_{\rm crit}} \right]^{-1/2}
    \, .
    \label{eq:power.radius}
\end{align}
A rigorous study of the numerical convergence for DMO halos and the FIRE-2 galaxies (dark matter with baryons) has been discussed in detail in \cite{hopkins2018fire}. There, the convergence has been gauged as a function of mass resolution, force resolution, time resolution, and so on. 

\begin{figure*}
    \centering
    \includegraphics[width=0.9\textwidth]{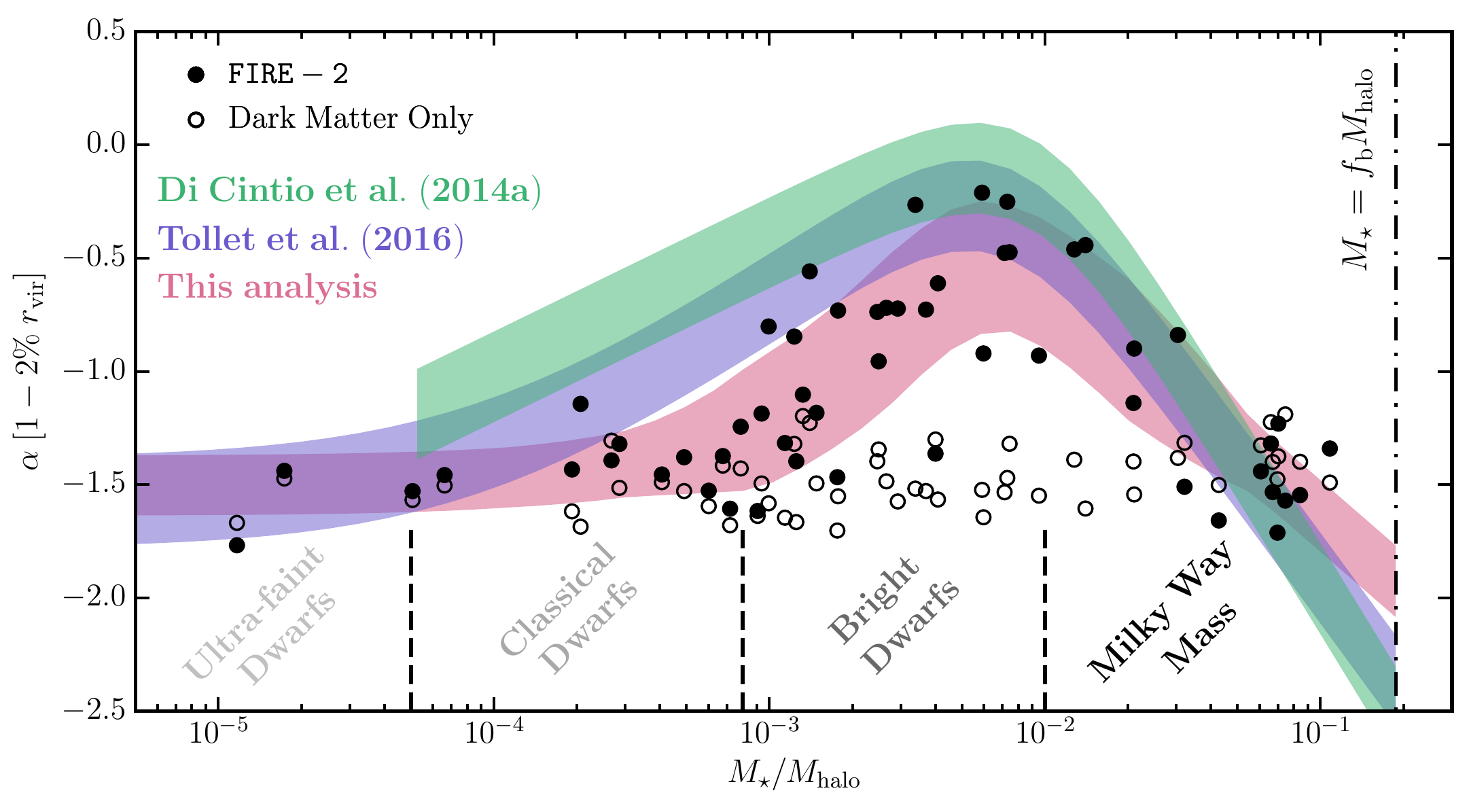} 
    \caption{---
        {\bf \em The impact of feedback physics on the inner dark matter densities}. Shown is the inner dark matter density slope, $\alpha$, averaged over $[1-2\%] \times r_{\rm vir}$, as function of the stellar mass fraction, $M_{\star}/M_{\rm halo}$, at $z=0$. Cored profiles have $\alpha \sim 0$,  while cuspy inner density profiles have lower values of $\alpha \lesssim -1$. The open circles are the DMO analogs, which all have $\alpha \approx -1.5$ as expected from a NFW profile. The pink shaded region shows the $1\sigma$ dispersion about the smoothed binned median. As a comparison, the fits from \protect\cite[green]{di2013dependence} and \protect\cite[blue]{tollet2016nihao} are also plotted using a constant width of $\Delta \alpha = \pm 0.2 $ relative to the mean relation \protect\citep{tollet2016nihao}. 
        The curve from \protect\cite{di2013dependence} was only fitted to down to a stellar mass fraction of $M_{\star}/M_{\rm halo} \simeq 4\times10^{-5}$, so we restrict the curve to that mass limit}. The dispersion in $\alpha$ increases from the stellar mass fraction from $M_{\star}/M_{\rm halo} \gtrsim 10^{-4}$, the regime of classical dwarfs and the brightest dwarfs, to the MW-mass halos with $M_{\star}/M_{\rm halo} \simeq 10^{-1}$. Feedback-induced core formation peaks at $M_{\star}/M_{\rm halo} \simeq 5 \times 10^{-3}$, the regime of the brightest dwarfs. At $M_{\star}/M_{\rm halo} \lesssim 10^{-4}$, the regime of classical dwarfs and ultra-faints, the impact of stellar feedback is negligible. 
    \label{fig:2}
\end{figure*}

For the DMO simulations, convergence was shown to be well resolved to the radius at which the criterion satisfies $t_{\rm relax} > 0.6\ t_{0}$ with $<1\%$ resolution level deviations. This typically equates to ${\sim}2000$ particles and is more conservative for the ranges of resolution levels analyzed in our halo sample. However, even at ${\sim}200$ particles (resulting in a factor ${\sim}2$ smaller radius of convergence), the convergence is good to ${\sim}10\%$ in the density profile. Hereafter, we adopt $t_{\rm relax} = 0.6\ t_{0}$ as our resolution criterion to maintain consistency across all of our simulations. We define $r_{\rm conv} := r_{0.6}^{\rm DM}$ to be the radius at which the resolution criterion is fulfilled for the \textit{dark matter only} analogs of each sample halo, meaning that $r > r_{\rm conv}$ is our best estimate of the numerically converged region. 
In \cite{hopkins2018fire}, convergence for simulations ran with baryons can be much better or worse in comparison to their DMO analogs, but convergence is entirely dominated by the convergence from the baryons. So in the context of our galaxies, the criterion of convergence has much more to do with the star-formation dynamics and converging baryonic physics rather than having to do with the number of particles enclosing a specific region. With this, $r_{\rm conv}$ from the DMO analogs are applied to the galaxies of the FIRE-2 halos throughout this paper as a conservative estimate. For more details regarding the numerical convergence study of FIRE-2 halos, we refer to \cite{hopkins2018fire}.

\begin{figure*}
    \centering
    \includegraphics[width=0.9\textwidth]{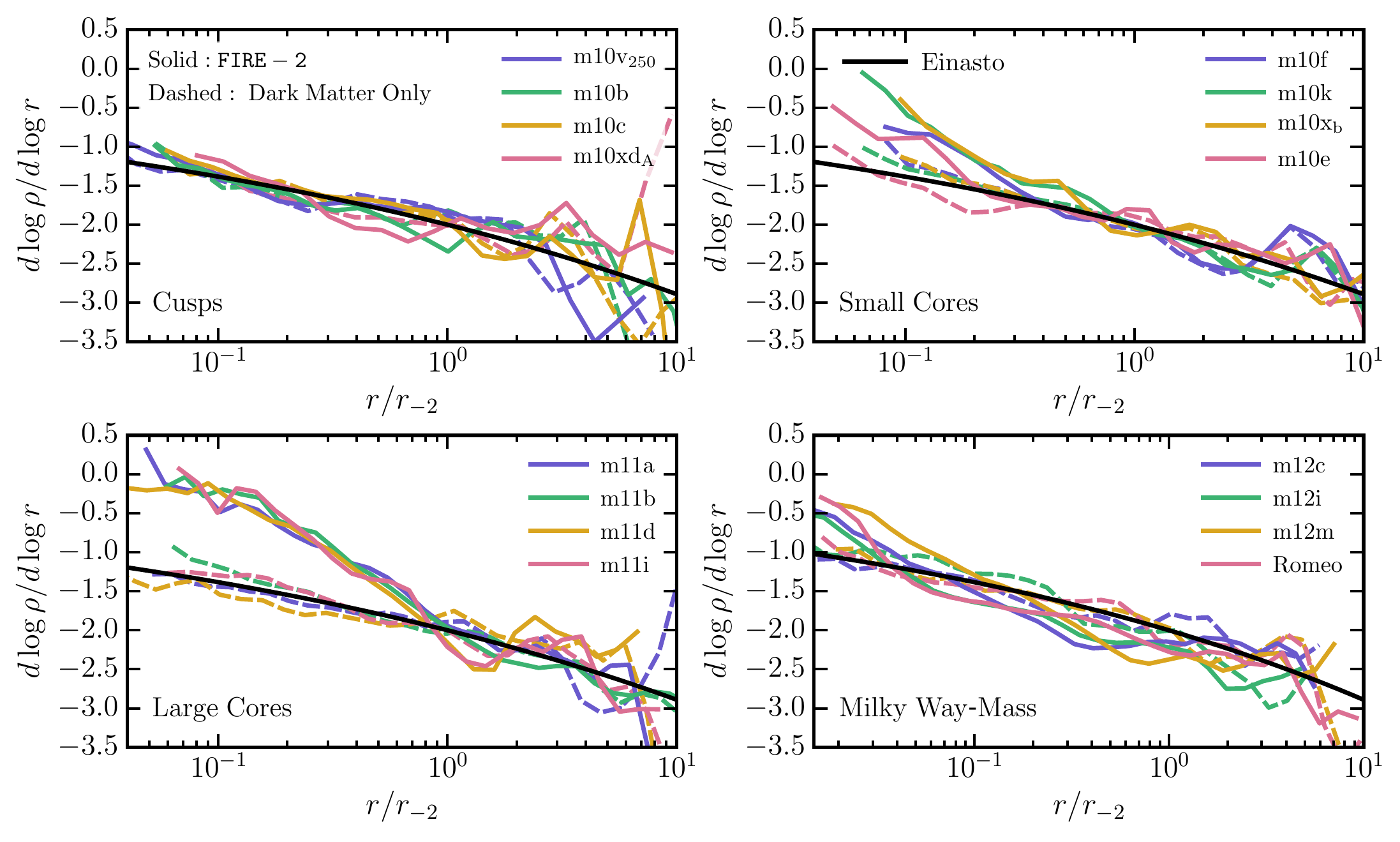} 
    \caption{ ---
        {\bf \em Comparison of the log-slope behaviour}. 
        The four panels show galaxies grouped by the behavior of their inner density profiles: galaxies with cusps, small cores, large cores, and MW-mass halos. The resolved portions of the FIRE-2 galaxies are depicted as the solid lines while the resolved DMO analog profiles are plotted as dashed lines. The solid black line illustrates the slope expected from Eq.~\eqref{eq:einasto.logslope}. All of the radial values are normalized by $r_{-2}$ of the DMO analogs, which are computed by fitting Eq.~\eqref{eq:einasto.density} to each individual dashed curves. As expected, the galaxies with cusps are well described by Eq.~\eqref{eq:einasto.density}. Galaxies with small cores have profiles that start to rise very slowly towards $d\log \rho / d\log r = 0$ at $\sim r_{-2}$. The largest cores in our sample are seen to have slight excesses in the density at around $r_{-2}$ (the ``dip'' in the profile) and begins to rise substantially for decreasing values of $r$. Milky-Way mass halos are the outliers in the trend, in which the galaxies' log-slopes are inconsistent with their dark matter analogs beginning at $r_{-2}$. At radii $r\ll r_{-2}$, the log-slopes are shown to form cores abruptly.
    }
    \label{fig:3}
\end{figure*}

\section{Stellar Fraction Relation with the Inner-Density Slope}
\label{sec:slope}
We begin by comparing our catalog of galaxies with previous results in the literature. The stellar mass fraction, which we define as the ratio between the stellar mass and halo mass, $M_{\star}/M_{\rm halo}$, has a relationship with the slope of the dark matter density profile found at the innermost radii \citep{di2013dependence,chan2015impact,tollet2016nihao}. Following the convention of \citet{di2013dependence}, the effect of feedback on the inner dark matter halo density can be captured by exploring the best-fitting power law for the dark matter density profile over a specific radial range, $\rho(r) \propto r^{\alpha}$. \cite{di2013dependence} suggested using $\alpha$ fitted over the radial range $r \in [1-2 \%\ r_{\rm vir}]$ since the lower limit of $1\%\ r_{\rm vir}$ satisfied the \cite{power2003inner} radius criterion of convergence for the majority of their halo sample.

Fig.~\ref{fig:2} summarizes the relation between $\alpha$ and the stellar mass fraction at $z=0$ for our simulations and compares to results from \protect\cite[][green band]{di2013dependence} and \protect\cite[][blue band]{tollet2016nihao}. 
The analysis performed in \protect\cite[green]{di2013dependence} included only stellar mass fractions down to $M_{\star}/M_{\rm halo} \simeq 4\times10^{-5}$, so we restrict their curve to that limit. 
The differences between the two curves included differences in cosmological models used, as noted in \citep{tollet2016nihao}. The black filled circles are our simulated FIRE-2 galaxies and the black open circles are the results for the DMO simulations (for which we use the stellar mass of their galaxy analogs). For all values of $M_{\star}/M_{\rm halo}$, the DMO analogs are cuspy, with $\alpha \approx -1.5$, which is expected when assuming the behavior of an analytic NFW profile along with scatter induced by the mass-concentration relation \citep[see][]{bullock2017small}. 

The pink band captures our results using the fitting-formula shape suggested by \cite{tollet2016nihao}: 
\begin{align}
    \alpha(x)
    &=
    n - \log_{10}\left[ n_{1}\left( 1+\frac{x}{x_{1}} \right)^{-\beta} + \left( \frac{x}{x_{0}}\right)^{\gamma}\right]
    \, ,
\end{align}
where $x=M_{\star}/M_{\rm halo}$. We find that $n = -1.60$, $n_{1} = 0.80$, $x_{0}=9.18\times 10^{-2}$, $x_{1}=6.54\times 10^{-3}$, $\beta = 5$, and $\gamma = 1.05$ matches our results in the median. The general purpose of this fit is to guide the eye. We also binned by $M_{\star}/M_{\rm halo}$ to compute a rough estimate of the standard deviation found at each stellar fraction. The width of the pink band roughly corresponds to the $1\sigma$ dispersion about the median. The width of the green and blue bands are set at a constant $\Delta\alpha = \pm 0.2 $.

Ultra-faint and classical dwarf galaxies, with low stellar mass fractions of  $M_{\star}/M_{\rm halo} \lesssim 10^{-3}$,  have inner densities slopes of $\alpha \approx -1.5$, the same as their DMO analogs. From there and increasing to $M_{\star}/M_{\rm halo} \simeq 5 \times 10^{-3}$, the inner dark matter densities of the bright dwarf galaxies transition to more cored profiles. At $M_{\star}/M_{\rm halo} \simeq 5 \times 10^{-3}$, our galaxies reach efficient core formation (shown more directly below), with $\alpha \approx -0.25$. The diversity in core strength, as quantified by $\alpha$, is largest from $M_{\star}/M_{\rm halo} \approx 10^{-3}$ to $5 \times 10^{-3}$, with a variance of $\Delta \alpha \approx \pm\ 0.35$ about the median.
Note that one bright dwarf (m11q) at  $M_{\star}/M_{\rm halo} \simeq 4 \times 10^{-3}$ has what appears to be a cuspy central density. We checked the assembly history of this galaxy and verified that it is not particularly unusual, with its last major merger at $z \sim 2$. This galaxy does in fact have a constant density core (see Table~\ref{tab:halo.fits} in the appendix), but at a radius $\sim 850$ pc, which is smaller than $1 \%$ $r_{\rm vir} \sim 1500$ pc, meaning that it is not detected using this $\alpha$ slope measurement.
From the region of efficient core formation to MW masses, $\alpha$ decreases. The scatter in $\alpha$ remains large ($\Delta \alpha \approx \pm\ 0.3$) until $M_{\star}/M_{\rm halo} \approx 6\times 10^{-2}$, which is in the range of the majority of the MW-mass halos. The scatter is minimized at $\Delta \alpha \approx \pm\ 0.15$ for these galaxy masses. 

\begin{figure*}
    \centering
    \includegraphics[width=0.925\textwidth]{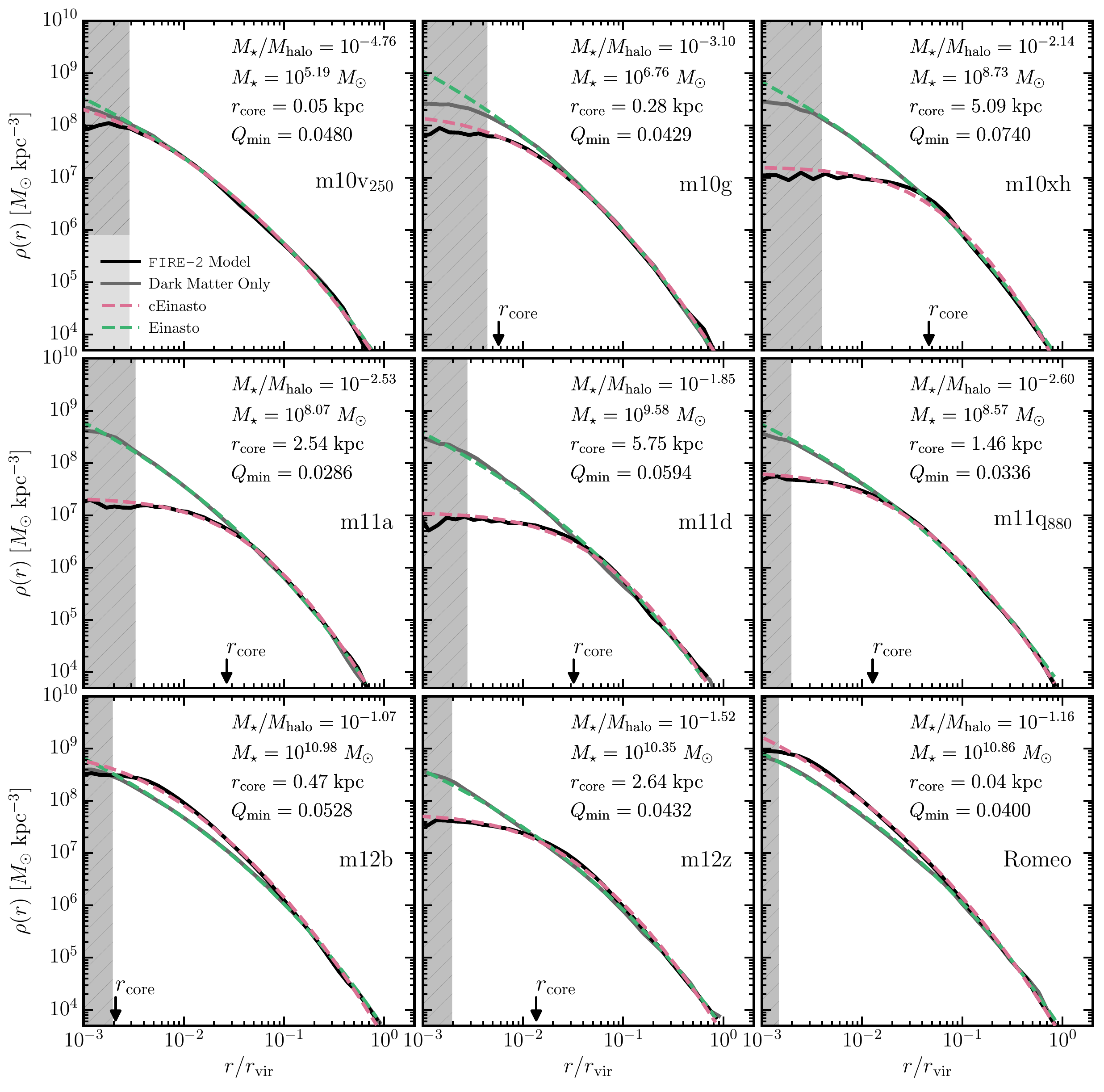}
    \caption{---
        {\bf \emph{Profiles of the local dark matter density}}:
        The $\rho_{\rm cEin}$ fits (pink dashed curves, with $\alpha_\epsilon=0.16$) are plotted along with the FIRE-2 galaxies (black curves) for a sample of galaxy halos. The $\rho_{\rm Ein}$ fits (green dashed curves, with $\alpha_\epsilon=0.16$) to the density profiles of DMO analogs (grey curves) are plotted as well. The vertical grey band encloses the radius where numerical two-body relaxation might effect the halo. Each panel has a list of relevant parameters for each galaxy: the stellar mass fraction ($M_{\star}/M_{\rm halo}$), the stellar mass ($M_{\star}$), the dark matter core radius from the $\rho_{\rm cEin}$ profile fit ($r_{c}$), and the minimum value of the merit function ($Q_{\rm min}$) that indicates the goodness-of-fit. The fitted dark matter core radius, $r_{c}$, is indicated by the black arrow pointing along the radial axis to show its location in units of $r_{\rm vir}$. For most of the depicted galaxies, the $\rho_{\rm cEin}$ profile fits perform exceptionally well in parameterizing the location of $r_{c}$. Note that {\em these examples include the full range of fit quality in our sample} (as measured by $Q_{\rm min}$), including some of the poorest fits, e.g., m10xh in the upper right corner.  
    }
    \label{fig:4}
\end{figure*}

Our findings agree with  previous results in the literature for the region of efficiently peaked core formation: $M_{\star}/M_{\rm halo} \simeq 5 \times 10^{-3}$ \citep{di2013dependence,chan2015impact,tollet2016nihao}. While we do not have a significant sample of ultra-faint dwarfs, we find negligible core formation for $M_{\star}/M_{\rm halo} \lesssim 10^{-4}$. The most significant difference we see with past results are (i) core formation that is less pronounced than previously reported for $M_{\star}/M_{\rm halo} \simeq 10^{-3}$ ($M_\star \simeq 10^7$ $M_\odot$) and (ii) more scatter in $\alpha$ within the regime of the brightest dwarfs, with $\alpha$ ranging from quite cuspy ($\alpha \approx -1.5$) to very cored ($\alpha \approx -0.25$) over the small range $M_{\star}/M_{\rm halo} \simeq [2-5]\times 10^{-3}$. 

While results on $\alpha$ at $r \simeq 1.5\% \, r_{\rm vir}$ have proven useful for characterizing the effectiveness of core formation as a function of stellar mass fraction in dark matter halos in the past, more recent simulations have allowed predictions at even smaller radii. This can potentially lead to small cores being unaccounted for \citep[see][]{chan2015impact,wheeler2019resolved}. For example, while Fig.~\ref{fig:2} gives the impression that MW-mass halos will have density structure similar to the DMO (NFW-like) expectation, this is only because the log-slope at $[1-2 \%] r_{\rm vir}$ does not provide a complete picture. That is, while the log-slope at this radius is similar to that expected in the absence of galaxy formation, the overall density amplitude at ${\sim}1\%$ of the virial radius is higher. In fact, as we will see in the upcoming section, at even smaller radii, our MW-mass halos have cored density profiles.\footnote{Also seen from the implementation of FIRE-1 physics for MW-mass halos in \protect\cite{chan2015impact}.} This motivates a more complete examination into the shapes of profiles of simulated galaxy halos. 

\section{A Density Profile for Feedback-affected Halos}
\label{sec:cEinasto}
In this section, we present a new dark matter density profile that allows for constant-density cores of the type seen in our simulated galaxy halos. The new profile generalizes the \cite{einasto1965influence} profile, which has proven to be an excellent fit for halos formed in DMO simulations. Our ``core-Einasto'' (cEinasto) profile extends its behaviour with one free parameter --- a core radius, $r_{c}$. After demonstrating that this profile does sufficiently well of capturing the density structure for a majority of the FIRE-2 halos, we follow the methodology employed in \cite{di2014mass}, and provide fits for halo fitting parameters as functions of $M_{\star}/M_{\rm halo}$ at $z=0$. In Appendix~\ref{sec:cEinasto.Mstar} we provide profile parametrization as a function of galaxy stellar mass, $M_{\star}$. We note that in the course of this analysis, we explored several different options for analytic cored profiles and found that the core-Einasto form was the best of these fits. In Appendix~\ref{sec:cNFW} we show an example comparison between the core-Einasto profile and the \cite{penarrubia2012coupling} (core-NFW) profile and demonstrate that core-Einasto provides a superior fit with the same number of free parameters.

\begin{figure*}
    \centering
    \includegraphics[width=\columnwidth]{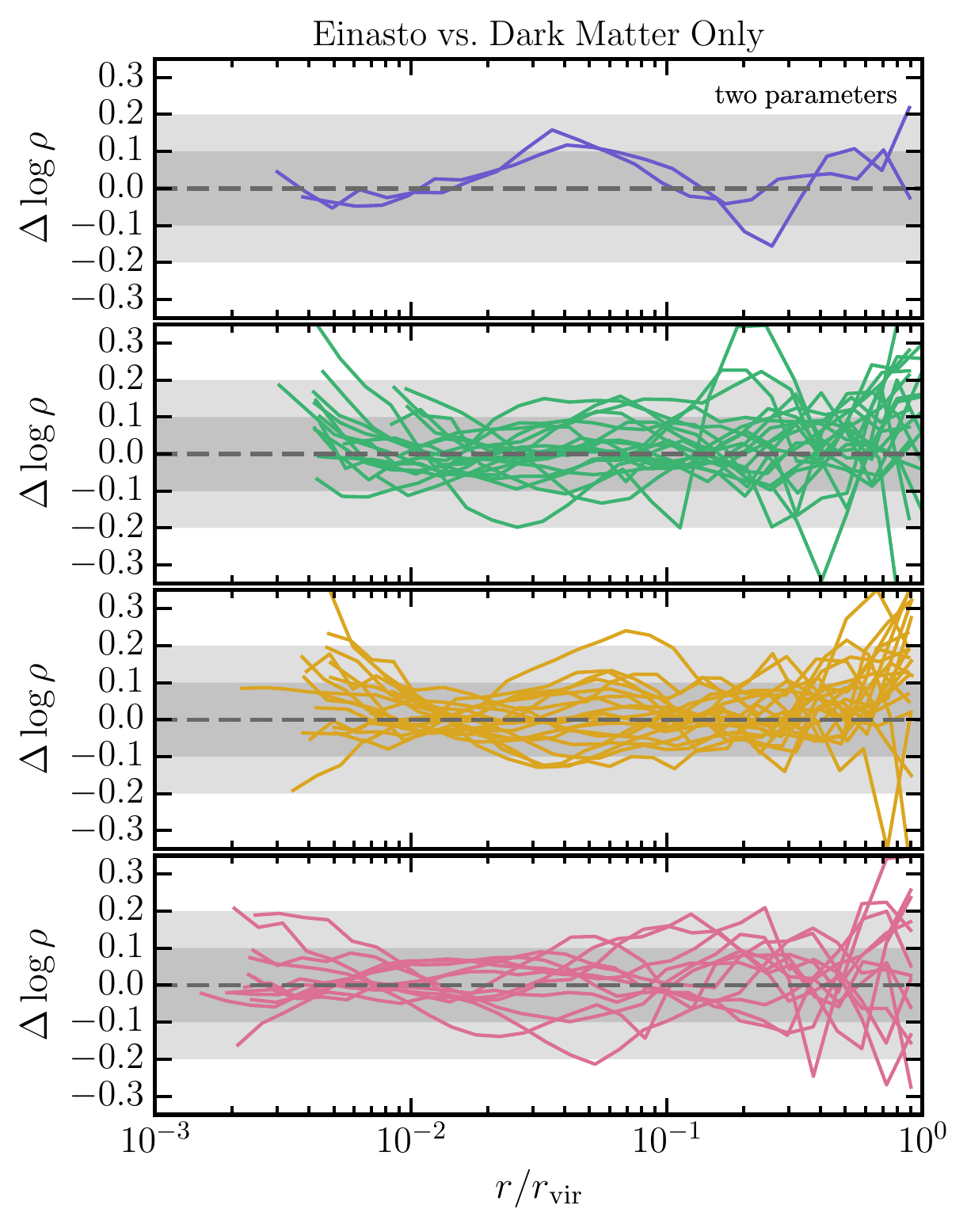}
    \includegraphics[width=\columnwidth]{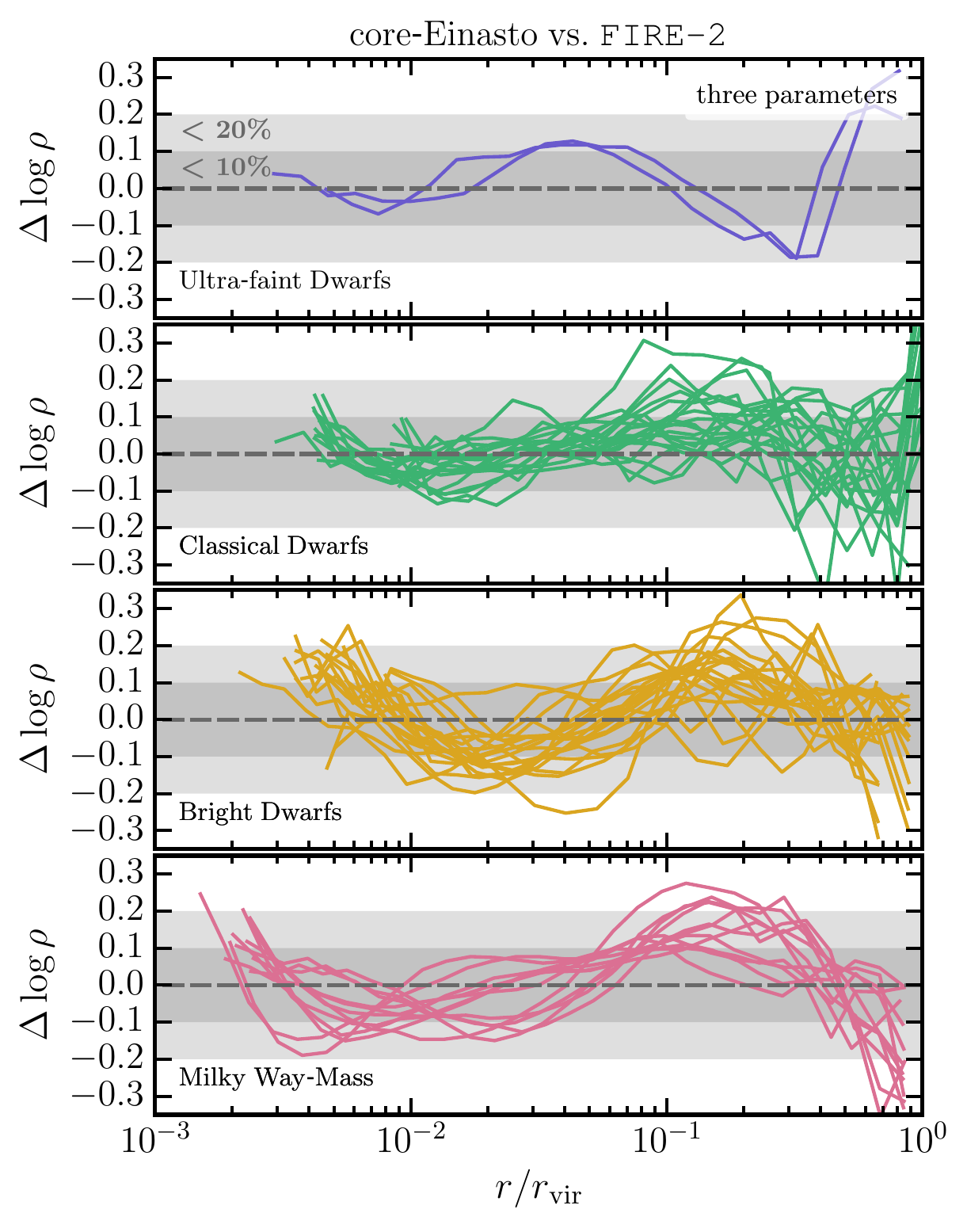}
    \caption{---
        {\bf \em Profile residuals}:
        Deviation from the best profile fits for each individual halo (fit subtracted from simulation).
        The left column shows residuals for fits to our DMO analogs using Einasto profiles with $\alpha_{\epsilon} = 0.16$. The right column shows residuals for the hydrodynamic simulations of the same halos fit using  the core-Einasto profile with $\alpha_{\epsilon} = 0.16$. For clarity, we have grouped halos by the four classification groups discussed in Section~\ref{sec:numerical.methods} in each  row: ultra-faint dwarfs, classical dwarfs, bright dwarfs, and MW-mass halos. Residuals are computed from the inner-most resolved radius, $r_{\rm conv}$, out the virial radius of each halo. The darker and lighter shaded gray enclose residuals of 10$\%$ and $20\%$, respectively. The core-Einasto fits to the full physics runs are almost as good as the Einasto fits are for the DMO halos. The offsets are less than $15\%$ in the inner regions of classical dwarfs and most bright dwarfs. Several of MW-size halos show worse fits, with offsets as large as $20 \%$, which is a result of both baryonic contraction and feedback-induced  dark matter cores.
    }
    \label{fig:5}
\end{figure*}

\subsection{Profiles for dark matter only halos}
\label{sec:profile.method}
Dark matter halos in $\Lambda$CDM are fairly well-described by the Navarro-Frank-White \citep[][NFW]{navarro1997universal} double-power law profile. While power laws are robust for understanding and are analytically friendly to work with, it has been made apparent that dark matter density profiles are not perfectly captured by the power-law construction. \cite{navarro2004inner,navarro2010diversity} demonstrated that higher resolution dark matter density profiles have log-slopes\footnote{We refer ``log-slope'' as the logarithmic derivative of the local density profile: $d\log{\rho}/d\log{r}$.} that decrease monotonically as $r$ approaches the center, which is not captured by the NFW at small $r$. This indicates that the innermost regions of CDM halos are shallower than an NFW.
Their study suggested a different radial profile for DMO halos, starting with the log-slope relation:
\begin{align}
    \frac{d\log \rho}{d\log r}(r)
    &=
    -2\left( \frac{r}{r_{-2}} \right)^{\alpha_{\epsilon}}
    \, .
    \label{eq:einasto.logslope}
\end{align}
This results in the three-parameter Einasto profile 
\begin{align}
    \log 
    \left[\frac{\rho_{\rm Ein}(r)}{\rho_{-2}}\right]
    &=
    - \frac{2}{\alpha_{\epsilon}} \left[ \left(\frac{r}{r_{-2}}\right)^{\alpha_{\epsilon}} - 1 \right]
    \, ,
    \label{eq:einasto.density}
\end{align}
where $\alpha_{\epsilon}$ is the so-called {\it shape parameter} that tunes how slow or fast the slope changes with radius, and $r_{-2}$ (as well as $\rho_{-2} := \rho(r_{-2})$) is the radius (density) at which the logarithmic slope of the density profile is equal to $-2$, i.e. $d\log \rho/d\log r\rvert_{r=r_{-2}} = -2$. 

The shape parameter, $\alpha_{\epsilon}$, is a key component of Eq.~\eqref{eq:einasto.density}. When obtained from Einasto profile fits to dark matter halos of cosmological simulations, it has been shown to correlate with the overdensity peak height of the dark matter halo and is calibrated based on the cosmology \citep[e.g.][]{gao2008redshift,dutton2014cold,klypin2016multidark}. Fixing $\alpha_{\epsilon} \simeq 0.16$ has been shown to provide a good fit for DMO halos throughout the literature \citep[][]{prada2006far,merritt2006empirical,gao2008redshift}. With this choice, $\rho_{\rm Ein}$ becomes a two-parameter function, one that still provides a better fit to DMO simulations than the two-parameter NFW profile.\footnote{Of course, one can acquire even better density profile fits to as good as $5-10\%$ for halos in our mass range when leaving $\alpha_{\epsilon}$ as a free parameter, as this value tailors to each shape to the dark matter halo. This however, leaves ambiguity in the value of $r_{-2}$, as this is now dependent on $\alpha_{\epsilon}$.} Recently, \cite{wang2019zoom} have shown that the two-parameter version of $\rho_{\rm Ein}$ provides a adequate fit for DMO halos over 30 orders of magnitude in halo mass.  We fix $\alpha_{\epsilon} = 0.16$ in what follows.

\begin{figure*}
    \centering
    \includegraphics[width=0.925\textwidth]{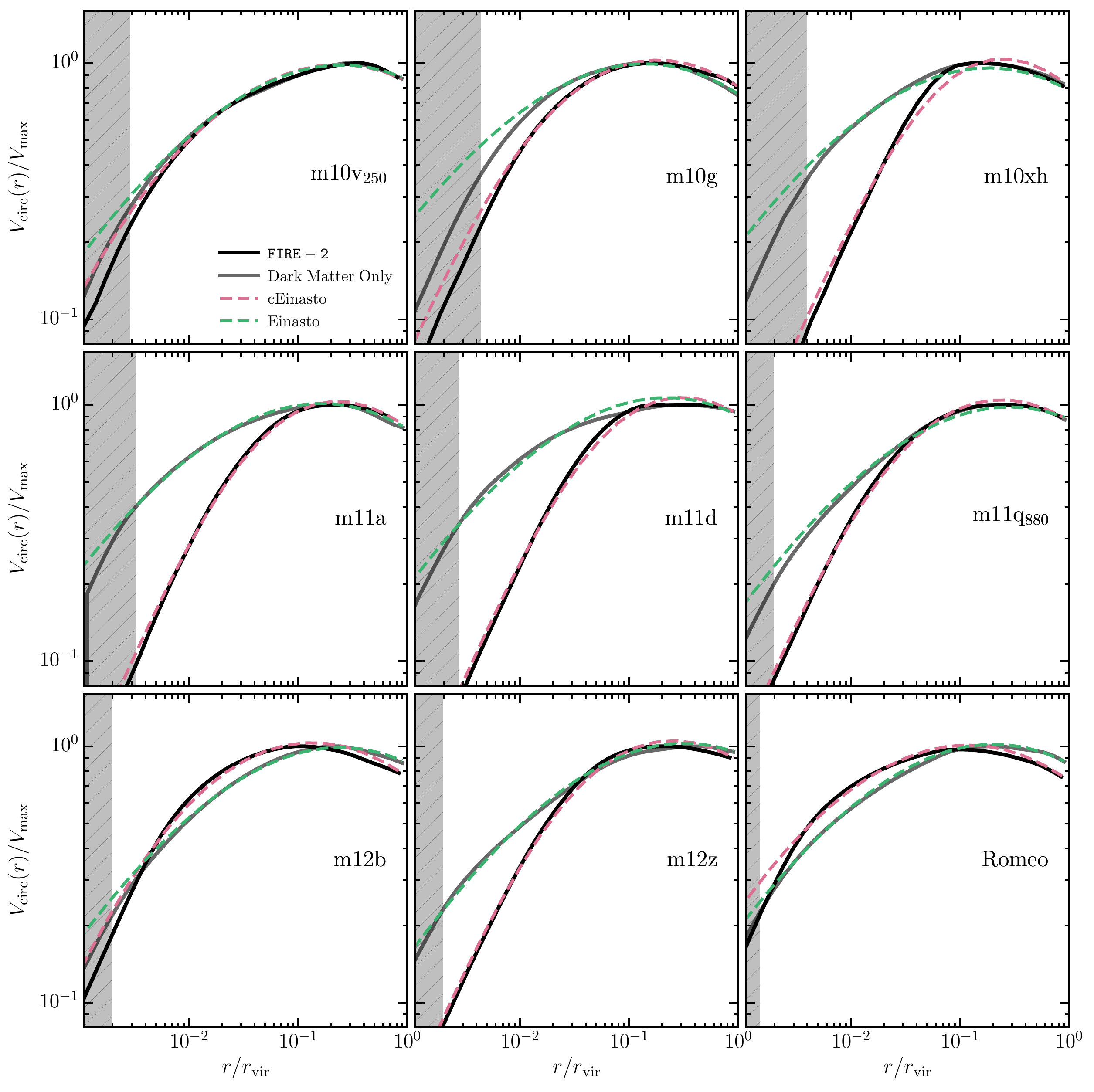}
    \caption{---
        {\bf \em Dark matter circular velocity curves}.
        Shown are the dark matter components of the circular velocity curves, $V_{\rm circ}(r) = \sqrt{GM(<r)/r}$, of the same halos presented in Fig.~\ref{fig:4}. The dashed pink and green curves are plotted using the analytical forms of Eqs.~\eqref{eq:cored.einasto.integrated.mass}~and~\eqref{eq:einasto.integrated.mass}, respectively. Curves of $V_{\rm cEin}$ and $V_{\rm Ein}$ are normalized by $V_{\rm max}$ of the galaxy and DMO analog, respectively. Analytical fits are able to capture the density normalization of the simulated halos robustly for all of the dwarf galaxies, even while it can under-estimate or over-estimate the integrated mass in at the outer radii.   
    }
    \label{fig:6}
\end{figure*}

\subsection{Cored profile for feedback-affected CDM halos}
We follow \citet{navarro2004inner} and consider the behaviour of the log-slope of the density profiles for our galaxy halos as a function of radius. Fig.~\ref{fig:3} shows log-slope profiles for four classifications of halos in our full-physics runs: ``cusps'', ``small cores'', ``large cores'', and ``Milky Way-mass halos''. The halos simulated with FIRE-2 physics are plotted as colored solid curves while their respective DMO analogs are shown as dashed lines with the same color. Starting with the upper-left panel, low-mass dwarfs tend to be hosted by cuspy dark matter halos. Similarly, halos with small cores tend to host higher-mass classical dwarfs. Halos with the largest cores correspond the brightest dwarf galaxies, which we have seen previously in Fig~\ref{fig:2}, while MW-mass galaxies have dark matter halo profiles that are more complicated (and are discussed further below). For reference, the solid black line shows the log-slope of the Einasto profile, Eq.~\eqref{eq:einasto.logslope}. The galaxies and DMO analogs have their radii normalized by $r_{-2}$ from the DMO runs.

As expected, Eq.~\eqref{eq:einasto.logslope} captures the log-slope trend of the DMO halos. The same is true for FIRE-2 runs with low stellar mass fraction (``cusps'' in this case). Halos labeled ``small cores'' tend to slightly deviate from Eq.~\eqref{eq:einasto.logslope}, with upturns in the log-slope trend for $r \lesssim 0.03 \times r_{-2}$. The lower left panel contains galaxy halos (solid lines) that approach $d\log \rho / d\log r = 0$ at small radii -- that is, a true core. This behavior never occurs beyond $r_{-2}$ of the analogous DMO profiles, and cores are only see at $r \ll r_{-2}$. MW-mass halos have more complicated profiles. Their log-slopes tend to lie below the log-slope of DMO analogs from $r \simeq [0.1 - 1]\times r_{-2}$; this is a consequence of baryonic contraction. However, we see that at $r \ll r_{-2}$, the log-slopes begin to rise towards 0, indicating that small cores can form in our MW sample.

In order to capture the behavior illustrated in Fig.~\ref{fig:3}, we start by writing a more general form of Eq.~\eqref{eq:einasto.logslope} that allows the log-slope to increase more sharply within a physical core radius, $r_{c}$:
\begin{align}
    \frac{d\log \rho}{d\log r}(r)
    &=
    -2\left( \frac{r}{\tilde{r}_{s}} \right)^{\alpha_{\epsilon}}
    \widetilde{\mathcal{C}}(r|r_{c})
    \label{eq:modified.einasto.logslope}
    \, .
\end{align}
Implemented here is a radially-dependent damping function, $\widetilde{C}(r|r_{c})$, which is designed to control the rate of which the profile dampens within $r_{c}$. The variable $\tilde{r}_{s}$ plays a similar role as $r_{-2}$ in Eq.~\eqref{eq:einasto.logslope}, but will no longer be the radius where the log-slope is equal to $-2$ owing to the presence of $r_{c}$. We demand that the behavior of the damping function satisfies the limiting cases of $\widetilde{C} \rightarrow 1$ and $\tilde{r}_{s} \rightarrow r_{-2}$ as $r_{c} \rightarrow 0$ in order to (i) capture the qualitative expectations of cores that can substantially vary in size and (ii) revert back to the form of $\rho_{\rm Ein}$ in the absence of a core. 

We adopt the following form:
\begin{align}
    \widetilde{\mathcal{C}}(r|r_{c})
    &=
    \bigg( 1 + \frac{r_{c}}{r} \bigg)^{\alpha_{\epsilon} - 1} 
    \, ,
\end{align}
such that
\begin{align}
    \frac{d\log \rho}{d\log r}(r)
    &=
    -2\left( \frac{r}{\tilde{r}_{s}} \right)^{\alpha_{\epsilon}}
    \bigg( 1 + \frac{r_{c}}{r} \bigg)^{\alpha_{\epsilon} - 1}
    \label{eq:cored.einasto.logslope}
    \, .
\end{align}
In particular, the log-slope of the density profile approaches zero more quickly for larger values of $r_{c}$. Integrating out Eq.~\eqref{eq:cored.einasto.logslope} gives us a cored counterpart of $\rho_{\rm Ein}$, the {\bf \em core-Einasto} profile:
\begin{align}
    \log 
    \left[\frac{\rho_{\rm cEin}(r)}{\tilde\rho_{s}}\right]
    &=
    - \frac{2}{\alpha_{\epsilon}} \left[ \left(\frac{r + r _{c}}{\tilde{r}_{s}}\right)^{\alpha_{\epsilon}} - 1 \right]
    \, .
    \label{eq:cored.einasto.density}
\end{align}
Here, $\tilde{\rho}_{s}$ is a density free parameter in the fit. In what follows we set $\alpha_{\epsilon}=0.16$, which reduces the expression to a three-parameter profile. In the limiting case of $r_{c} \rightarrow 0$, we re-acquire $\rho_{\rm Ein}$, where now $\tilde{\rho}_{s} \rightarrow \rho_{-2}$. Note that the central density with the presence of a core, $\rho_{0} := \rho_{\rm cEin}(r=0)$, is parametrized as
\begin{align}
    \rho_{0} 
    &=
    \tilde{\rho}_{s}  
    \exp\left\{-\frac{2}{\alpha_{\epsilon}} \left[ \left( \frac{r_{c}}{\tilde{r}_{s}} \right)^{\alpha_{\epsilon}} -1 \right] \right\}
    \label{eq:cored.einasto.central.density}
    \, .
\end{align}
Alternatively, we can reparameterize $\tilde{\rho}_{s}$ by mapping to $\tilde{\rho}_{-2}:=\rho_{\rm cEin}(r_{-2})$, the density (and radius) where the log-slope is equal to $-2$. This allows us to re-express Eq.~\eqref{eq:cored.einasto.density} as
\begin{align}
    \log 
    \left[\frac{\rho_{\rm cEin}(r)}{\tilde{\rho}_{-2}}\right]
    &=
    - \frac{2}{\alpha_{\epsilon}} \left[ \left(\frac{r + r _{c}}{\tilde{r}_{s}}\right)^{\alpha_{\epsilon}} - \left(\frac{r_{-2} + r _{c}}{\tilde{r}_{s}}\right)^{\alpha_{\epsilon}} \right]
    \, ,
\end{align}
which certainly work in our zero core limit to re-acquire Eq.~\eqref{eq:einasto.density}. However, this expression now introduces an additional free parameter, $r_{-2}$, that can likely lead to degenerate results in acquiring $r_{c}$ and $\tilde{r}_{s}$. With that, we prefer to adopt the form of Eq.~\eqref{eq:cored.einasto.density} for our analysis hereinafter.
Analytic expressions for the mass profile, gravitational potential, and energy for the core-Einasto profile are presented in Appendix~\ref{sec:analytic.cEinasto}.

\begin{figure*}
    \centering
    \includegraphics[width=\columnwidth]{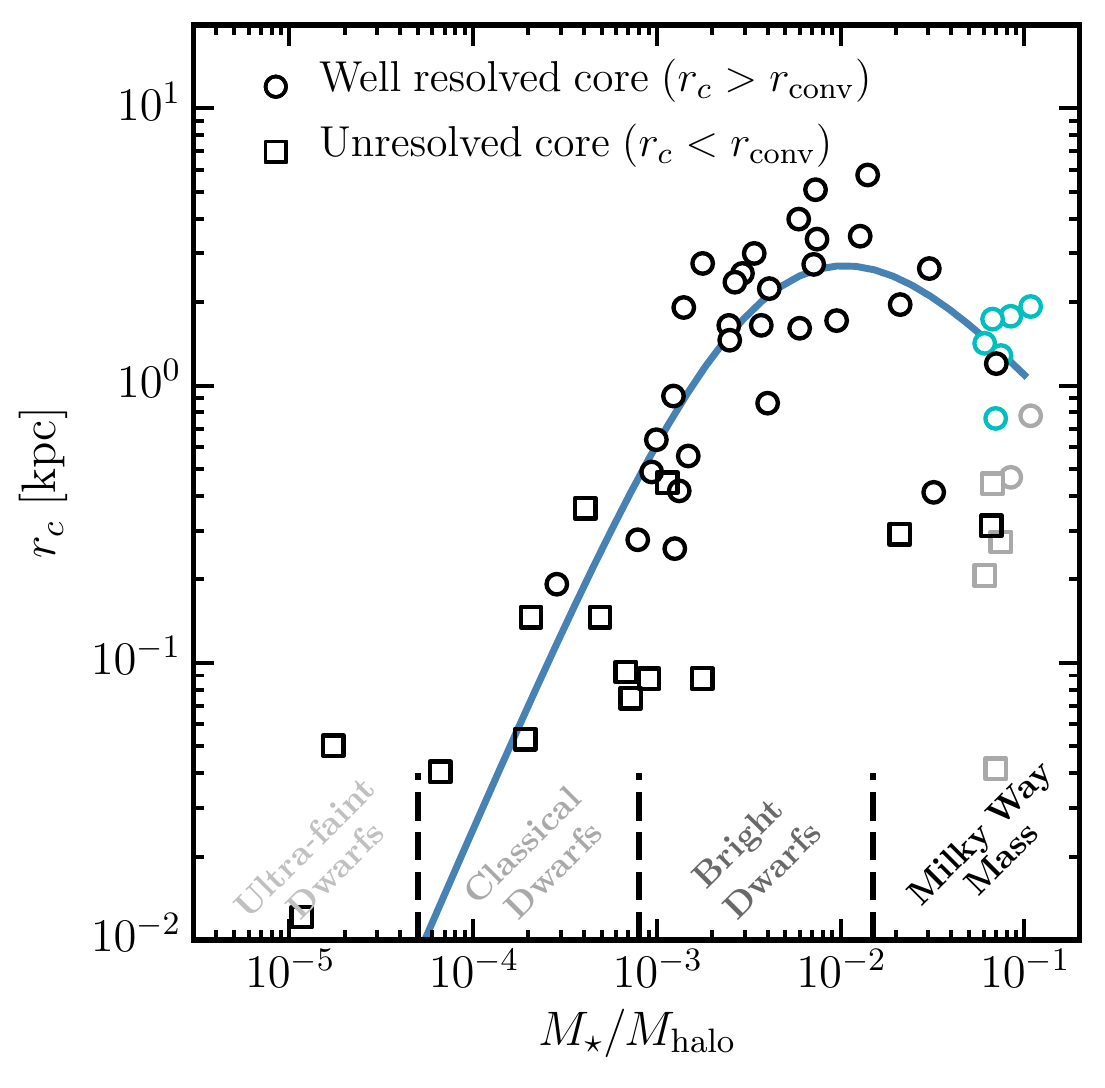}
    \includegraphics[width=\columnwidth]{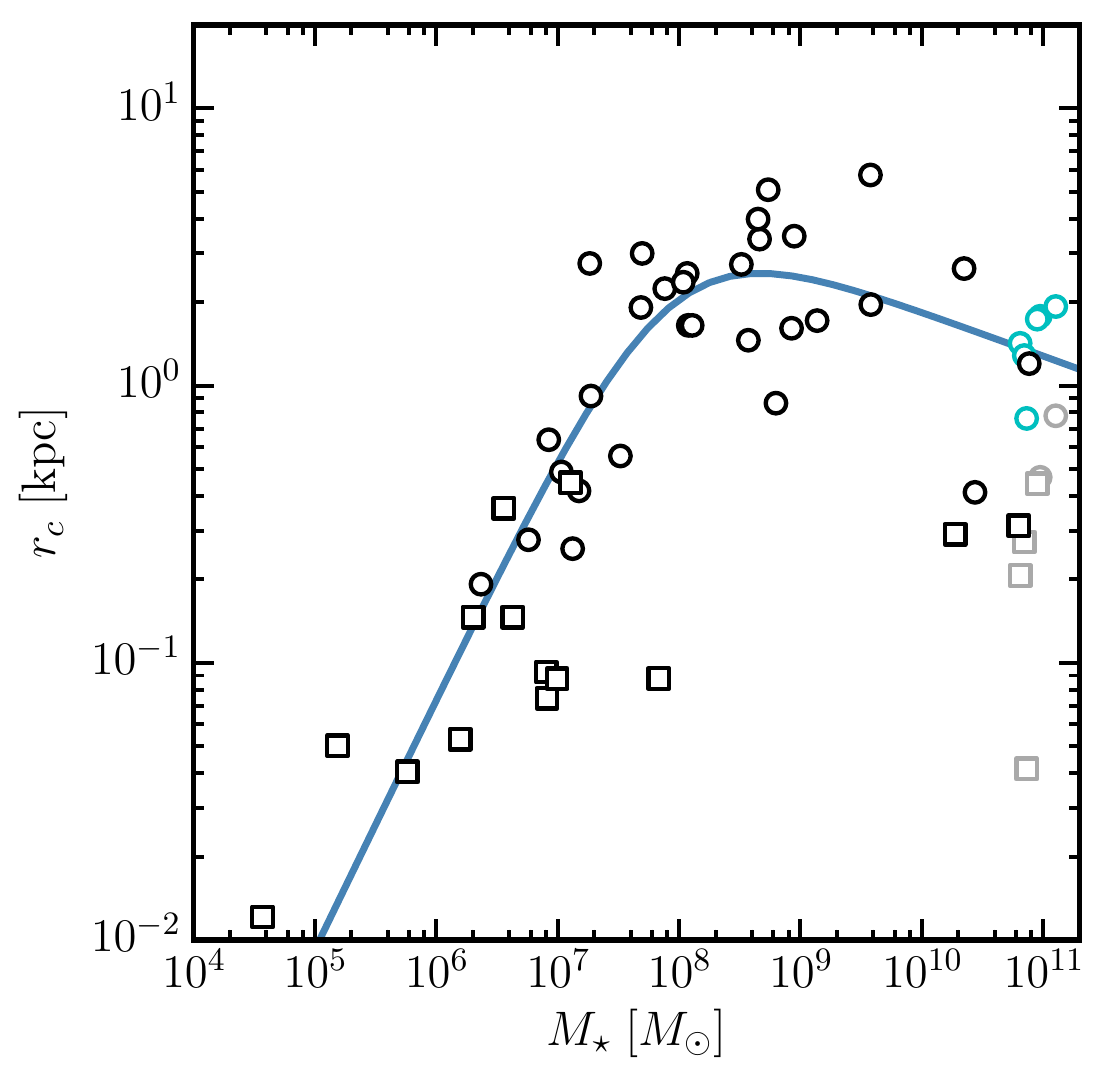}
    \caption{---
        {\bf \em Feedback-induced core formation}. 
        Circles show core radii that are larger than the convergence radius of the simulation ($r_{c}>r_{\rm conv}$) while squares are values smaller than the convergence radius ($r_{c}<r_{\rm conv}$). MW halos with significant baryonic contraction, which are therefore not as well fit by the $\rho_{\rm cEin}$ function, are shown in light grey. The cyan points show $r_{c}$ values for MW-mass galaxies returned from a four-parameter ``baryonic contracted cored-Einasto'' profile, $\rho_{\rm cEin,BC}$, introduced in Appendix~\ref{sec:cEinasto.BC}, in order to better account for baryonic contraction.
        {\bf \em Left}: Core radius as a function of stellar to halo mass ratio. The solid blue curve is a fit to the dark black and cyan points using Eq.~\protect\eqref{eq:rcore.fit}, with the best fit parameters given in Table~\ref{tab:rcore.fit}.  We note that this trend mirrors results shown in Fig.~\ref{fig:1}, with the largest core radii values occuring in the ``Bright Dwarfs'' regime.
        {\bf \em Right}:
        Dark matter core radius as a function of $M_{\star}$.
        Peak core formation, while scattered, appears around $M_{\star} = 10^{8-9}\ M_{\odot}$. The solid blue curve is our best fitting line using Eq.~\eqref{eq:rcore.fit} and best-fit parameters from Table~\ref{tab:rcore.fit} for $x=M_{\star}$.
    }
    \label{fig:7}
\end{figure*}

\subsection{Resulting profile fits}
All functional fits are performed using the Levenberg-Marquart minimization algorithm. We restrict our radial density profile fits to the radial range of $r_{\rm conv}$ to $r_{\rm vir}$. Best-fit models are obtained by simultaneously adjusting the parameters of the analytical density profiles in order to minimize a figure-of-merit function, defined by
\begin{align}
    Q^{2} 
    = 
    \frac{1}{N_{\rm bins}} \sum_{i}^{N_{\rm bins}}
    \left[
    \log_{10}\rho_{i} - \log_{10}\rho_{i}^{\rm model}
    \right]^{2}
    \, ,
    \label{eq:merit.function}
\end{align}
which weights all the logarithmic radial bins equally and, for a given radial range, is fairly independent of the number of bins used \citep{navarro2010diversity}. That is, the minimum figure-of-merit, denoted as $Q_{\rm min}$, quantifies the residuals of the true profile from the model caused by shape differences induced in the fitting routine. 

\begin{figure*}
    \centering
    \includegraphics[width=0.975\textwidth]{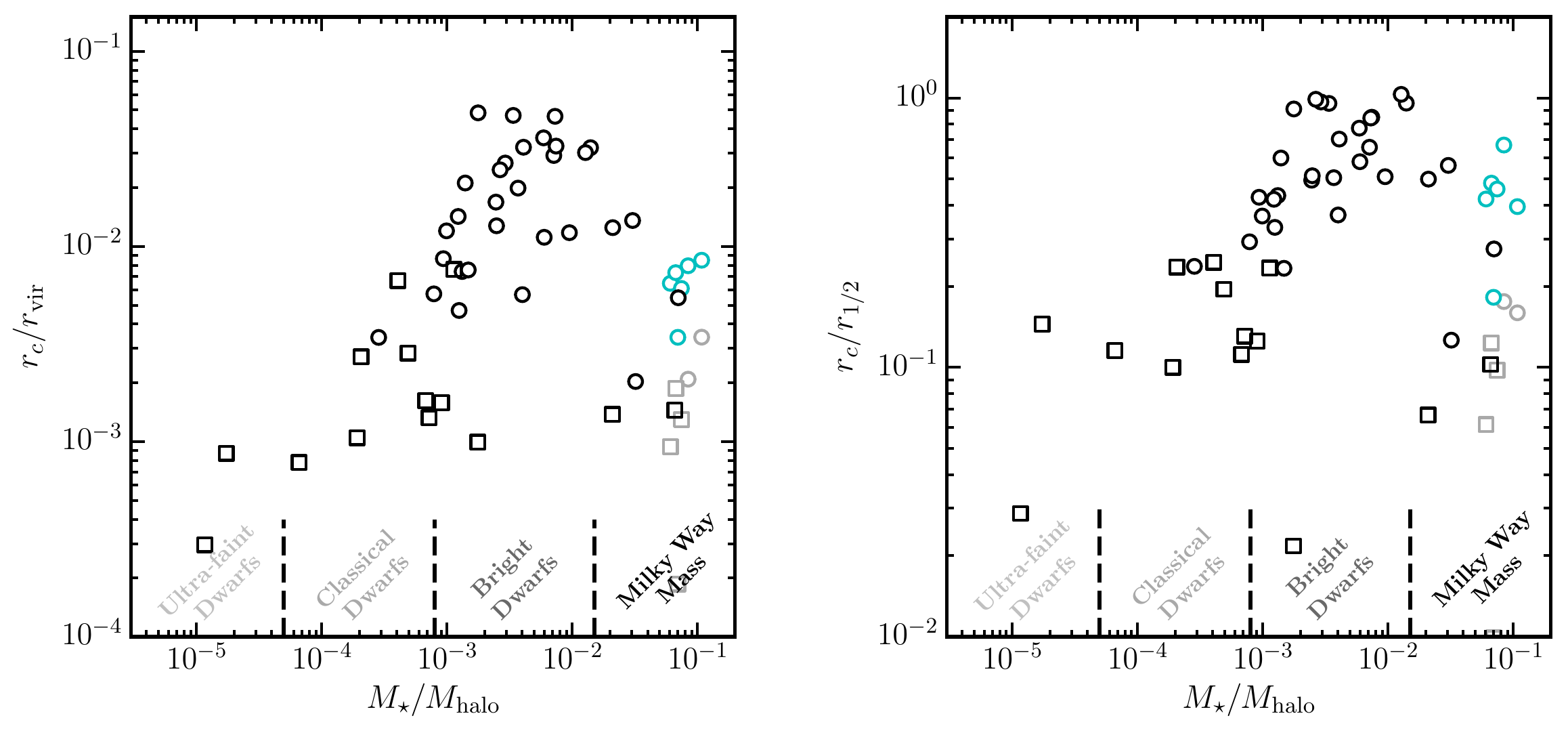}
    \caption{---
        {\bf \em Core radius relative to the halo and galaxy size}.
        Similar to Fig.~\ref{fig:7}, except with the core radii scaled by the virial radius of the dark matter halos (left) and stellar-half-mass radius of the galaxies (right). {\bf\em Left}: The fractional size of cores rises toward the regime of peak core formation, where  $r_{c} \simeq 0.05\, r_{\rm vir}$. MW-mass halos have  $r_{c}/r_{\rm vir}$ values comparable to those of dwarf galaxies with  $M_{\star}/M_{\rm halo}\sim 10^{-3}$. {\bf\em Right}: All resolved cores are constrained to a lower bound of $r_{c} \gtrsim 0.1\, r_{1/2}$. At peak core formation, $r_{c} \simeq r_{1/2}$ for some of the the brightest dwarfs.
        }
    \label{fig:8}
\end{figure*}

\subsubsection{Local dark matter density}
Fig.~\ref{fig:4} provides example fits for a sample of dark matter density profiles. Dark matter halos simulated using FIRE-2 (black curves) are fitted with $\rho_{\rm cEin}$ (pink dashed) while the DMO analogs (grey line) are fitted with $\rho_{\rm Ein}$ (dashed green). In each panel, we list the galaxy's stellar mass fraction ($M_{\star}/M_{\rm halo}$), stellar mass ($M_{\star}$), dark matter core radius ($r_{c}$) given by fitting $\rho_{\rm cEin}$, and the goodness-of-fit ($Q_{\rm min}$) from fitting $\rho_{\rm cEin}$. The location of the best-fit dark matter core radius, scaled by the virial radius, is indicated by the black arrow in each panel. Table~\ref{tab:halo.fits} lists the fit results for all of our galaxies, including the fit parameters and the $Q_{\rm min}$ values. We can see that the value $r_{c}$ is effectively determined for a wide range of galaxy sizes. For even the worst profile fits (e.g. m10xh with $Q_{\rm min} = 0.074$; top-right panel), the value of $r_{c}$ is still identified at the location where one's eye might pick out a dark matter core in the local density profile. 

As a way of examining the robustness of Eq.~\eqref{eq:cored.einasto.density}, we fit core-Einasto to the DMO analogs and found that in every case the best-fit core-radii were either zero or smaller than the radius of convergence. This provides confidence that this profile does not force or impose cores that do not exist in the resolved regions of the halo.  However, it does suggest that $r_c$ values smaller than the convergence limit should not be taken as robust indications for the existence of real cores. For example, the upper left panel of Fig.~\ref{fig:4} shows an $\rho_{\rm cEin}$ fit to m10v$_{250}$ (baryon simulated), a profile that is unaltered by feedback in the resolved region owing to its small stellar mass. The best-fit core radius ($r_{c}\simeq50\ \rm pc$) is much smaller than the radius of convergence ($r_{\rm conv} \simeq 160\ \rm pc$) in this case. 

While we find success in characterizing dwarf galaxies with $\rho_{\rm cEin}$, almost all of the MW-mass halos have cored regions that are more sharply pronounced than enabled by the  $\rho_{\rm cEin}$ profile. As one can see (e.g. m12b and Romeo), the values of $r_{c}$ from the fits do not coincide with the locations of the bend seen in the simulated profiles.\footnote{The core radius of Romeo from the $\rho_{\rm cEin}$ fit does not appear in Fig.~\ref{fig:4} (bottom right panel) since the fitted value of $r_{c}$ is located inside the region of numerical convergence ($r_{c}/r_{\rm vir} < 10^{-3}$).} Based on our entire sample of MW-mass halos, we find that the $\rho_{\rm cEin}$ profile performs less well for MW-mass halos that have both a small central dark matter core and baryonic contraction in the inner densities. On the other hand, MW-mass halos with little evidence of either baryonic contraction (e.g. m12z) or a core are successfully characterized by $\rho_{\rm cEin}$. MW-mass halos with no core, but with only baryonic contraction, are also well-modeled by $\rho_{\rm Ein}$. In Appendix~\ref{sec:cEinasto.BC}, we formulate a more general core profile with one additional free parameter that captures the behavior for baryonic contracted halos with cores. This allows us to accurately quantify the core radii for the rest of our MW-mass halos. 

\subsubsection{Density profile residuals}
Profile residuals of the local dark matter density are presented in Fig.~\ref{fig:5} for DMO analog fitted with the Einasto profile (left) and to the dark matter halos of the FIRE-2 physics runs fit to core-Einasto (right). Results are split into the four galaxy classifications defined in Section~\ref{sec:numerical.methods}. The residuals for the left and right columns are comparable, which is remarkable given that the right-hand fits have only one additional free parameter to account for the full impact of complex galaxy formation physics.  Notice that the largest deviations are present large radii ($r \gtrsim 0.3 r_{\rm vir}$). This behavior has been seen in the past for DMO halos, where the outer regions may not be fully relaxed \citep[e.g.][]{ludlow2010pseudo,ludlow2016relations}, and may contain large substructures. 

While we have only two ultra-faint galaxies (blue curves) in our sample, both galaxies are well described to $10 \%$ for a majority of the radii. This is unsurprising, as these halos lack the requisite star formation to induce cores; the core-Einasto fit is therefore effectively the same as a standard Einasto fit, with $r_c$ values that are smaller than the convergence radius. Almost all of the classical dwarf galaxies (green curves) have excellent core-Einasto fits, with deviations in the range $10-15 \%$ at worst. At small radii $(r \lesssim 0.1\times r_{\rm vir})$, core-Einasto is shown to be sufficient in fitting the FIRE-2 halos compared to their DMO analogs in the same radial regions. For a majority of the brightest dwarfs in our sample, deviations are constrained within $15 \%$. For MW-mass halos, the quality of the fit can range from quite good to as bad as $20 \%$. As mentioned previously, the worst fits are for the MW-mass halos impacted by both baryonic contraction and feedback-induced core formation at small radii. We find deviations of $10-15\%$ in the inner-most regions for profiles of MWs with just cores (e.g. m12z in Fig.~\ref{fig:4}) or just having baryonic contraction with no cores.

\begin{figure}
    \centering
    \includegraphics[width=\columnwidth]{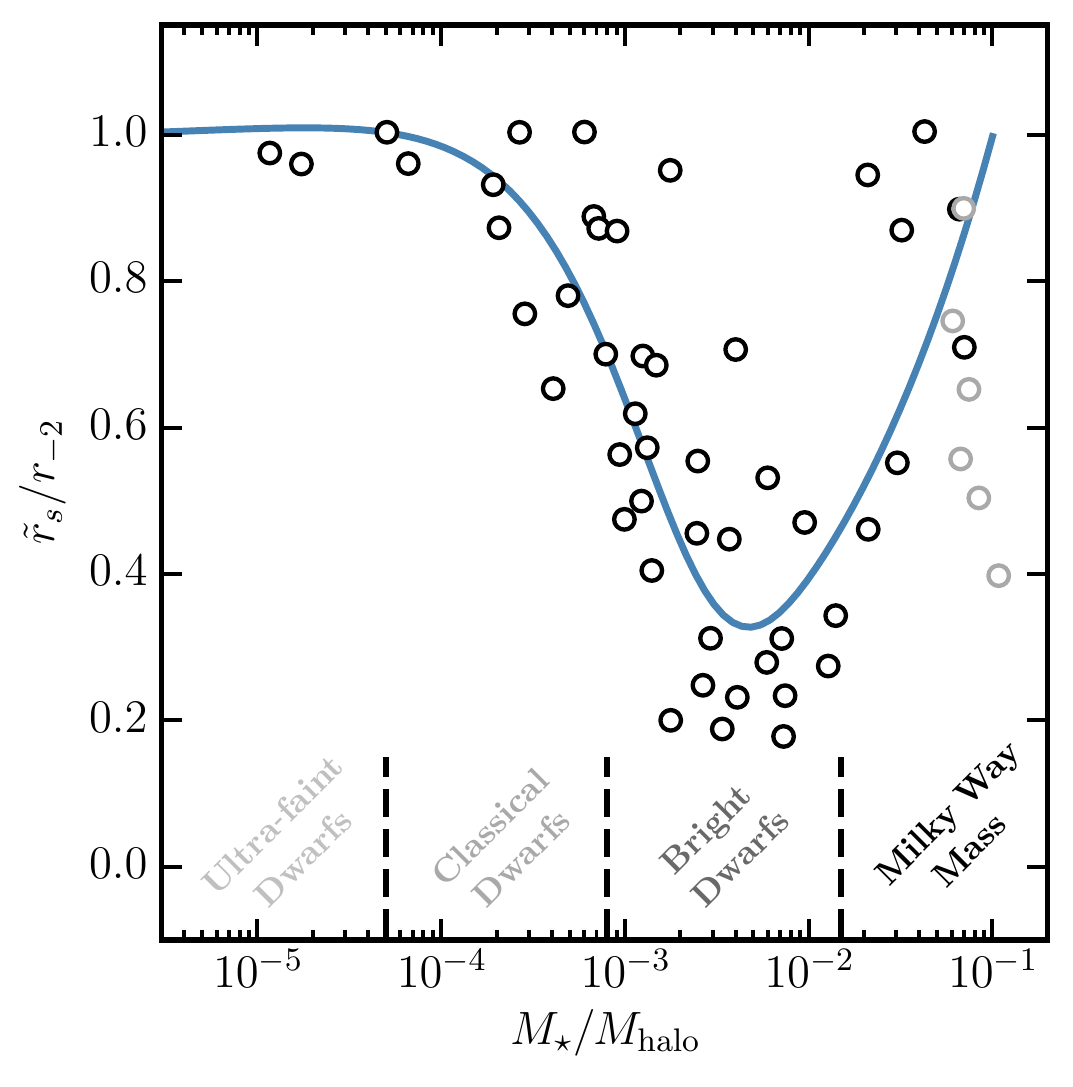}
    \caption{---
        {\bf \em Effects of core formation on the global structure of the dark matter halo}.
        A trend can be seen between the fitting parameter, $\tilde{r}_{s}$, and the interpolated scale radius, $r_{-2}$, from the $\rho_{\rm cEin}$ profile. We again show the results not well fitted with $\rho_{\rm cEin}$, which are highlighted in light grey like in Fig~\ref{fig:7}.
        As the core radius of the halos becomes larger, i.e. as we increase $M_{\star}/M_{\rm halo} \sim 5 \times 10^{-3}$, the physical interpretation of $\tilde{r}_{s}$ changes. This result shows that the formation of a core (found most prominently in the regime of the bright dwarfs) results in a change to the dark matter halo at larger scales (as parametrized by $r_{-2}$). 
        The solid blue curve is our best fits using Eq.~\protect\eqref{eq:rs.fit}.
    }
    \label{fig:9}
\end{figure}

In both columns, there are are hints of a sinusoidal feature in the residuals. This behavior is not unusual when simplified fits are compared to detailed dark matter halo profiles
\citep[e.g][]{griffen2016caterpillar}. Reducing the residual behavior even more would require more free parameters in the form of $\widetilde{\mathcal{C}}$ in Eq.~\eqref{eq:modified.einasto.logslope} and/or allowing the value of $\alpha_{\epsilon}$ to vary from halo-to-halo. However, given that the gross residuals for our core-Einasto fits to the FIRE-2 runs are close to those of Einasto fits to DMO runs, we are satisfied that the given parameterization provides a useful balance between simplicity and accuracy. 
In Appendix \ref{sec:cNFW} we do find that our halos are modeled better by the three-parameter core-Einasto profile than two alternative three-parameter profiles: the core extension for the NFW from \cite{penarrubia2012coupling} and Dekel+ from \cite{dekel2017profile} and \cite{freundlich2020dekel}.

\subsubsection{Dark matter circular velocity}
Fig.~\ref{fig:6} provides an alternative view of the results shown in Fig.~\ref{fig:4}: it shows the circular velocity curves of the dark matter component,\footnote{For the analysis of observed galaxies, spherically averaged rotation curves are typically presented using their total mass, i.e., their combined baryonic and dark matter components. We chose to show just the dark matter components here to compare with our core-Einasto model.} $V_{\rm circ}(r) = \sqrt{GM(<r)/r}$, for the same halos presented in Fig.~\ref{fig:4}, each normalized by  $V_{\rm max} := \mathrm{max}[V_{\rm circ}(r)]$ of the dark matter curve. The analytical profiles for $V_{\rm cEin}$ and $V_{\rm Ein}$ are plotted using Eqs.~\eqref{eq:cored.einasto.integrated.mass}~and~\eqref{eq:einasto.integrated.mass}, respectively, for the values obtained from the fits shown in Fig.~\ref{fig:4}. These analytical curves are normalized by the $V_{\rm max}$ values of the simulated halos to which they are fitted. For profile fits over-estimating (or under-estimating) the mass found in the simulated profiles by $15-20\%$ (e.g., m10xh and m11d), the most substantial effects can seen at the outer radii, near where $V_{\rm max}$ is attained. However, even for the worst profile fits in our sample, {\it the central density normalization is well-captured for dwarf galaxies of varying stellar mass fractions}.

\subsection{Parametrization of the physical core radius}
\label{sec:rcore.param}
For the left plot in Fig.~\ref{fig:7}, we show the relationship between $M_{\star}/M_{\rm halo}$ and the fitted values of $r_{c}$. Circular points denote the values of $r_{c}$ that we verify as resolved cores (with $r_{c} > r_{\rm conv}$ for the local dark matter density profiles). This sample includes the MW-mass core radii fit using using the four parameter function $\rho_{\rm cEin,BC}$ (cyan highlights) described in Appendix~\ref{sec:cEinasto.BC} instead of their $r_{c}$ values from $\rho_{\rm cEin}$ (shown by  gray points for reference). Squares denote best-fit core radii that have values smaller the numerical convergence region ($r_{c} < r_{\rm conv}$). It is important to note that in some cases, we obtain fit values of $r_{c}$ that are formally smaller than $r_{\rm conv}$ yet large enough that the halo is not well-described by the standard $\rho_{\rm Ein}$ form. This comes about because dark matter halos impacted by stellar feedback produce dark matter profiles that are no longer self-similar in nature, meaning the core-Einasto fit balances $\tilde{r}_{s}$ and $r_{c}$ to accommodate the shape of the density profile. 

We see that our robustly-determined $r_{c}$ values ($r_{c}>r_{\rm conv}$), begin to appear at the higher mass end for the classical dwarf galaxy regime,  $M_{\star}/M_{\rm halo} \gtrsim 7 \times 10^{-3}$, with values that are physically quite small, $r_{c} \simeq [0.2-0.3]\ \rm kpc$. As the stellar mass fraction increases toward the region of bright dwarf galaxies, $M_{\star}/M_{\rm halo} \simeq [10^{-3} - 10^{-2}]$, the sizes of the core radii, $r_{c}$, increase with $M_{\star}/M_{\rm halo}$. Importantly, the largest dark matter cores, $r_{c} \simeq [5-6]\ \rm kpc$, coincide with the stellar mass fraction at the peak core formation that we have seen previously ($M_{\star}/M_{\rm halo}\simeq 5 \times 10^{-3}$). A majority of the galaxies at the MW-mass scale have dark matter cores  as $r_{c} \simeq 1-2\ \rm kpc$, though two remain fairly cuspy (m12r and m12w). To provide further insight into observations of real galaxies comparable to the simulations analyzed here, the right plot in Fig.~\ref{fig:7} shows the trend of $r_{c}$ with $M_{\star}$. The largest cores tend to form in galaxies with $M_{\star} \simeq 10^{8-9}\ M_{\odot}$. Notably, a significant amount of scatter is seen for fixed value of $r_{c} \simeq 2-3$ kpc, which tends to be apparent for galaxies with $M_{\star} \simeq 10^{8-11}\ M_{\odot}$.

The formation of small cores for MW-mass halos using FIRE-1 was discussed in \cite{chan2015impact}, where they found that small cores for MW-size galaxies tend form in the low-mass galaxy progenitors at $z\sim 2$, which have stellar-to-halo mass ratios suitable for core formation. These progenitors have their resulting innermost dark matter profile amplified at $z=0$ due to baryonic contraction. This phenomena also drives out old stars formed in situ in MW-like galaxies \citep{el2018ancient}. Other simulation groups have not reported the existence of small cores at the MW-mass regime. This could however be due to differences in numerical resolution. For example, the NIHAO simulations presented in \cite{tollet2016nihao} and \cite{maccio2020nihao} study MW-mass halos at a lower resolution than ours, with  convergence down to $r_{\rm conv}\simeq 1.25$ kpc compared 330-500 pc in our runs (see Table \ref{tab:halo.sample}). The MW cores in our simulations are $\sim 1$ kpc in size. Such cores would be difficult to form without having a convergence radius smaller than this limit. 


\begin{table}
    \setlength{\tabcolsep}{5.55pt}
    \centering
    \captionsetup{justification=centering}
    \caption{Best-fit parameters for the physical core radius, $r_{\rm core}$. For complete data set: $-3.54 \lesssim \log_{10}(M_{\star}/M_{\rm halo}) \lesssim -0.97$ and 
    $6.37 \lesssim \log_{10}(M_{\star}/M_{\odot}) \lesssim 11.10 $.
    }
    \begin{tabular}{SSSSSSS} 
    \toprule
    \toprule
    {Parameter} & 
    {$\mathcal{A}_{1}$} & {$\mathcal{A}_{2}$} &
    {$x^{*}_{1}$} & {$x^{*}_{2}$} & 
    {$\beta_{1}$} & {$\gamma_{1}$} 
    \\ 
    \midrule
    {$M_{\star}/M_{\rm halo}$} &
    {1.21} & {0.71} &
    {$7.2 \times 10^{-3}$} & {0.011 } &
    {2.31} & {1.55} 
    \\
    {$M_{\star}/M_{\odot}$} &
    {1.33} & {$4.3 \times 10^{7}$} &
    {1.93} & {0.55} &
    {1.06} & {0.90} 
    \\
    \bottomrule
    \bottomrule
    \vspace{-1ex}
    \\
    \multicolumn{6}{l}{\textbf{Note.} Use Eq. \protect\eqref{eq:rcore.fit} for either $x=M_{\star}/M_{\rm halo}$ or $x=M_{\star}/M_{\odot}$.}\\
    \end{tabular}
\label{tab:rcore.fit}
\end{table}


We find that the relationship between $r_{c}$ and $x = M_{\star}/M_{\rm halo}$ (and $x = M_{\star}/M_{\odot}$)  can be captured as a double-power law
\begin{align}
    r_{c}\left( x \right)
    &=
    10^{\mathcal{A}_{1}}
    \left(\mathcal{A}_{2} + \frac{x}{x^{*}_{1}} \right)^{-\beta_{1}}
    \left(  \frac{x}{x^{*}_{2}} \right)^{ \gamma_{1}}
    \ \mathrm{kpc} 
    \, ,
    \label{eq:rcore.fit}
\end{align}
where $\{ \beta_{1}, \gamma_{1}\}$ are free parameter slopes that control the transition of $x$.  The quantities $\{ x_{1}^{*},x_{2}^{*}\}$ are normalization parameters associated with both slopes, and $\{\mathcal{A}_{1},\mathcal{A}_{2}\}$ are constants of the fit. Best-fit parameters for $x = M_{\star}/M_{\rm halo}$ and $M_\star/M_\odot$ are given in Table~\ref{tab:rcore.fit}. The trend for our plotted data for $r_{c}$ as a function of $M_{\star}/M_{\rm halo}$ and $M_\star/M_\odot$ is shown by the blue curves in the left and right plots in Fig.~\ref{fig:7}, respectively.

\begin{figure*}
    \centering
    \includegraphics[width=0.95\columnwidth]{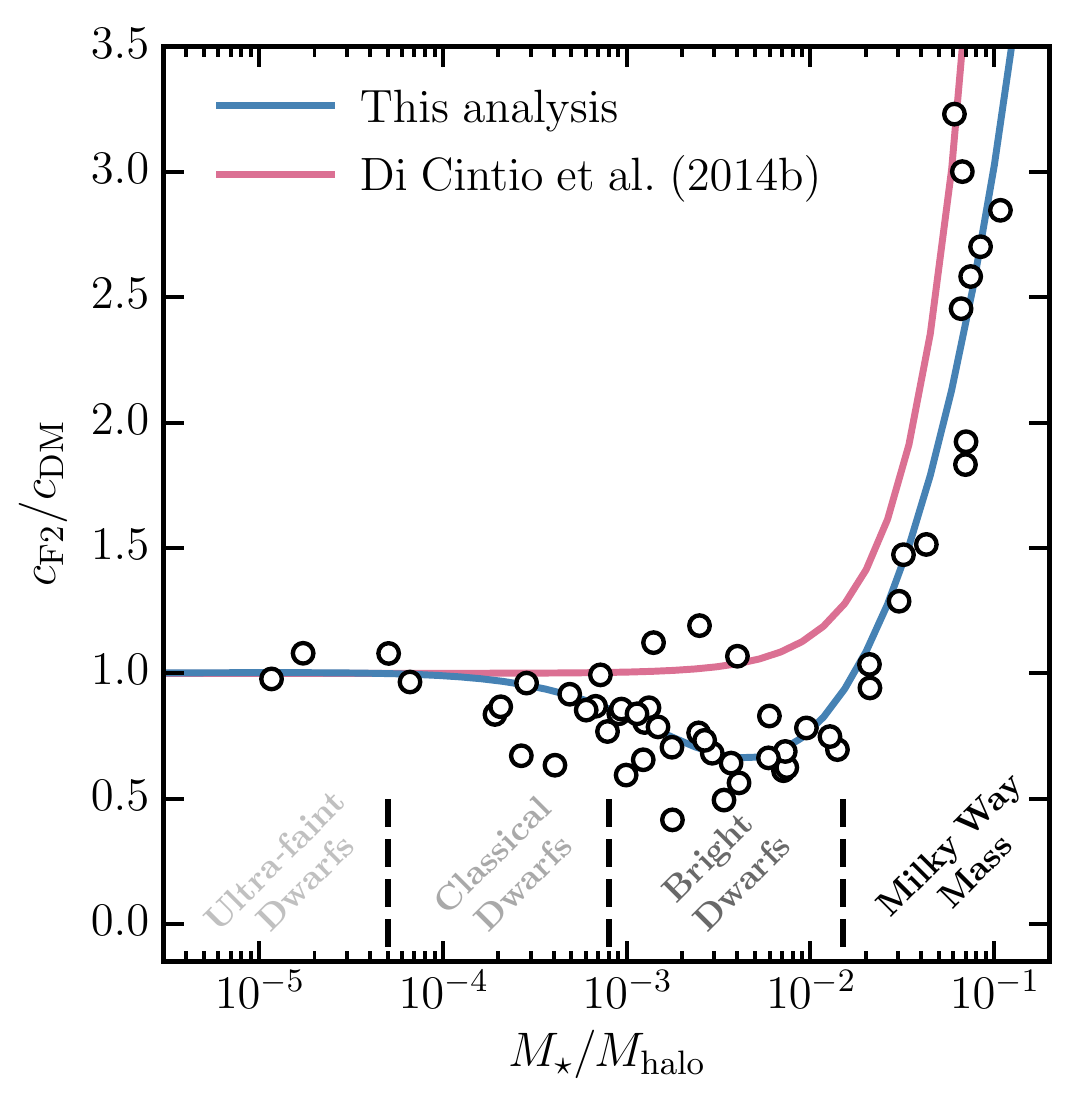}
    \qquad
    \includegraphics[width=0.935\columnwidth]{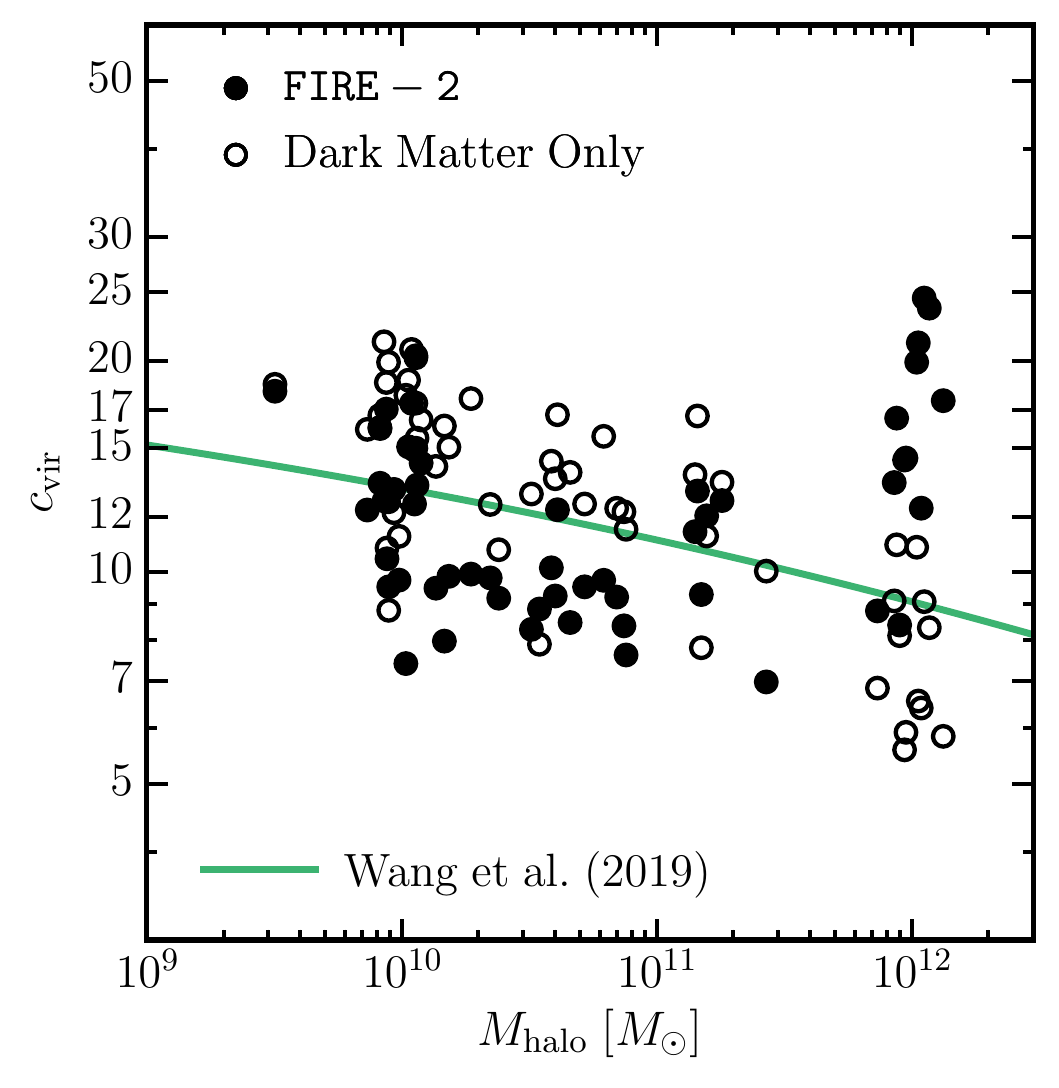}
    \caption{---
        {\bf \em The impact of feedback on halo concentration}. 
       Halo concentration defined by $c_{\rm vir} := r_{\rm vir}/r_{-2}$. 
       {\bf \em Left}: 
       FIRE-2 halo concentrations ($c_{\rm F2}$) using $r_{-2}$ interpolated from the best fits of $\rho_{\rm cEin}$ and best fits of $\rho_{\rm cEin, BC}$ from Appendix~\ref{sec:cEinasto.BC}. For the halo concentrations of the DMO analogs ($c_{\rm DM}$), $r_{-2}$ is taken from the free parameter fit of $\rho_{\rm Ein}$.
       Galaxies with have the largest cores (the brightest dwarfs) have their halo concentrations lowered by a factor of two compared to DMO analog concentrations. The solid blue curve is our best fit of the scatter points using Eq.~\protect\eqref{eq:conc.fit}. Also plotted is the fit from \protect\cite{di2014mass} as the solid pink curve for comparison.
       {\bf \em Right}: 
       Concentration as a function of dark matter halo mass. Galaxies and DMO analogs are denoted by the filled and unfilled black circles, respectively. The solid green curve is the concentration relation from \protect\cite{wang2019zoom}, which was also calibrated using the $c_{\rm vir}$ from the Einasto profile (including the same shape considered here, $\alpha_{\epsilon}=0.16$).
    }
    \label{fig:10}
\end{figure*}

Fig.~\ref{fig:8} is similar to left plot in Fig.~\ref{fig:7} except with the values of $r_{c}$ normalized by the size of the dark matter halo virial radius ($r_{\rm vir}$; left plot) or the half-stellar-mass radius of the galaxy it hosts ($r_{1/2}$; right plot) as a function of $M_{\star}/M_{\rm halo}$. Notably, the  normalization for each plot roughly follows the same trend that we have seen in previous figures: as $M_{\star}/M_{\rm halo}$ increases from $10^{-4}$ to $10^{-2}$, galaxies have larger cores, {\it even relative to the size of the dark matter halo or its central galaxy}. The trend peaks at the mass scale of robust core formation. At this peak, the brightest galaxies tend have cores of $r_{c} \sim 0.04 \, r_{\rm vir}$ (albeit with large scatter) and $r_{c}\sim r_{1/2}$. Interestingly, most MW-mass halos have $r_{c}/r_{\rm vir}$ values similar to dwarfs with stellar fractions that are 100 times lower and $r_{c}/r_{1/2}$ values comparable to many of the brightest dwarfs.   

\subsection{Parametrization of \texorpdfstring{\boldmath$\tilde{r}_{s}$}{rs}}
We wish to quantify how the free parameter, $\tilde{r}_{s}$, is related to $r_{-2}$ from using $\rho_{\rm cEin}$, the radius at which the log-slope of the local dark matter density is equal to $-2$, in the presence of a dark matter core. Unfortunately, the relation between $\tilde{r}_{s}$ and $r_{-2}$ for the FIRE-2 dark matter halos cannot be solved analytically as the additional power of $\alpha_{\epsilon}$ means they are non-linearly related. However, we can paramterize the covariance between $\tilde{r}_{s}$ and $r_{-2}$ from introducing $r_{c}$. Fig.~\ref{fig:9} shows the ratio of $\tilde{r}_{s}$ to $r_{-2}$ as a function of $M_{\star}/M_{\rm halo}$ for the FIRE-2 halos. Here, $r_{-2}$ is interpolated from only the $\rho_{\rm cEin}$ fits. As expected, dwarf galaxies with no cores (or cores small enough to effectively be approximated as $r_c=0$) have $\tilde{r}_{s} \simeq r_{-2}$. As we transition towards the region of peak core formation, $\tilde{r}_{s}$ gradually decreases relative to $r_{-2}$. We then see a sudden upturn at the MW-mass scale, which is a consequence of baryonic contraction. The relation for $\tilde{r}_{s}$ to $r_{-2}$ as a function of $M_{\star}$ is also discussed in Appendix~\ref{sec:cEinasto.Mstar}. 


\begin{table}
    \centering
    \setlength{\tabcolsep}{7.75pt}
    \captionsetup{justification=centering}
    \caption{Best-fit parameters for $\tilde{r}_{s}/r_{-2}$. 
    }
    \begin{tabular}{SSSSSS} 
    \toprule
    \toprule
    {Parameter} & 
    {$\mathcal{B}$} & {$x^{*}_{3}$} & {$x^{*}_{4}$} & {$\beta_{2}$} & {$\gamma_{2}$} 
    \\ 
    \midrule
    {$M_{\star}/M_{\rm halo}$} &
    {1.51} & {$0.044$} & {0.28} & 
    {31.79} & {0.40} 
    \\
    {$M_{\star}/M_{\odot}$} &
    {$0.098$} & {$5.1 \times 10^{6}$} & {$1.4 \times 10^{6}$} & 
    {0.57} & {0.20}
    \\
    \bottomrule
    \bottomrule
    \vspace{-1ex}
    \\
    \multicolumn{6}{l}{\textbf{Note.} Use Eq. \protect\eqref{eq:rs.fit} for either $x=M_{\star}/M_{\rm halo}$ or $x=M_{\star}/M_{\odot}$.}\\
    \end{tabular}
\label{tab:rsr2.fit}
\end{table}


The relationship between $\tilde{r}_{s}/r_{-2}$ and either $x = M_{\star}/M_{\rm halo}$ (or $x = M_{\star}/M_{\odot}$) can be captured as a double-power law:
\begin{align}
    \left[\tilde{r}_{s}/r_{-2}\right]\left(x\right)
    &=
    \left(1 + \frac{x}{x^{*}_{3}} \right)^{-\beta_{2}}
    +
    \mathcal{B} 
    \left( \frac{x}{x^{*}_{4}} \right)^{ \gamma_{2}}
    \, ,
    \label{eq:rs.fit}
\end{align}
where  $\{ \beta_{2}, \gamma_{2}\}$ are free parameter slopes that control the transition, the quantities $\{ x_{3}^{*},x_{4}^{*}\}$ are normalization values associated with these slopes, and $\mathcal{B}$ is a constant. The best fit parameters for $x= M_{\star}/M_{\rm halo}$ are given in Table \ref{tab:rsr2.fit}. The trend for our data is plotted as the blue curve in Fig.~\ref{fig:9}.

\subsection{Parametrization of the halo concentration}
The stellar feedback in dark matter halos also affects the halo concentration through the gravitational coupling of dark matter to the rapidly changing central gravitational potential. We adopt the halo concentration parameter $c_{\rm vir} := r_{\rm vir}/r_{-2}$. This definition of $c_{\rm vir}$ will be applied for the established results modeled by $\rho_{\rm cEin}$, $\rho_{\rm cEin,BC}$, and $\rho_{\rm Ein}$.\footnote{For the FIRE-2 halos fitted well with $\rho_{\rm cEin}$ and the MWs fitted with $\rho_{\rm cEin,BC}$ in Appendix~\ref{sec:cEinasto.BC}, the value of $r_{-2}$ is interpolated from the analytical profile fits, while for the DMO halos, $r_{-2}$ is taken from the free parameter fit of $\rho_{\rm Ein}$.} Ratios of the concentration parameter between the FIRE-2 halos, $c_{\rm F2}$, and their DMO analogs, $c_{\rm DM}$, are shown in the left panel of Fig.~\ref{fig:10} as a function of $M_{\star}/M_{\rm halo}$. The result from \cite{di2014mass} is plotted as the pink curve. We also extend this discussion with the parametrization done for $M_{\star}$ in Appendix~\ref{sec:cEinasto.Mstar}.

Galaxies with lower stellar mass fraction limit ($M_{\star}/M_{\rm halo} \lesssim 10^{-4}$) have values of $c_{\rm vir}$ comparable to their DMO analogs. Noticeable differences of the concentrations become apparent as $M_{\star}/M_{\rm halo}$ starts to increase towards the classical dwarf and bright galaxy regime. Importantly, as $M_{\star}/M_{\rm halo}$ approaches the peak of sufficient core formation, the halo concentrations for the FIRE-2 galaxies are conspicuously smaller -- by 30-50\% -- than the halo concentrations of their DMO analogs. This could mean that the strength of stellar feedback, which we can also probe by the size $r_{\rm c}$, in these halos has been strong enough to affect the density structure out to $r_{-2}$, an effect not seen previously (e.g., compare with the pink curve from \cite{di2014mass}). However, the relation from \cite{di2014mass} used the parameters obtained from fitting the $\alpha\beta\gamma$-profile to acquire $r_{-2}$ while we numerically interpolated from our resulting profile fits. 
We explore the differences in concentration that arise for the same halos when fitting different profiles in Appendix~\ref{sec:cNFW}. We find that when using the $\alpha\beta\gamma$-profile, the concentration can shift somewhat, but there is a tendency to be lower in the bright dwarf regime, following a similar qualitative trend shown in Fig.~\ref{fig:10}.

As stellar fractions reach the the MW regime, we see the opposite effect: the concentrations of our galaxy halos are significantly larger than their DMO analogs because of baryonic contraction. 

The relationship between the concentration parameters of our galaxy halos can be parameterized as a double power law:
\begin{align}
    [c_{\rm F2}/c_{\rm DM}]\left(x \right)
    &=
    \left(1 + \frac{x}{x^{*}_{5}} \right)^{-\beta_{3}}
    +
    \mathcal{C} 
    \left( \frac{x}{x^{*}_{5}} \right)^{ \gamma_{3}}
    \label{eq:conc.fit}
    \, ,
\end{align}
where either $x = M_{\star}/M_{\rm halo}$ or $x = M_{\star}/M_{\odot}$, $\{ \beta_{3}, \gamma_{3}\}$ are slopes, and $ x_{5}^{*}$ is a free normalization value to anchor the transition between slopes, and $\mathcal{C}$ is a constant. Best fit parameters for $x = M_{\star}/M_{\rm halo}$ are given in Table \ref{tab:conc.fit}. The trend for our data is plotted as the blue curve in the left plot of Fig.~\ref{fig:10}.


\begin{table}
    \centering
    \setlength{\tabcolsep}{8.75pt}
    \captionsetup{justification=centering}
    \caption{Best-fit parameters for $c_{\rm F2}/c_{\rm DM}$. 
    }
    \begin{tabular}{SSSSS} 
    \toprule
    \toprule
    {Parameter} & 
    {$\mathcal{C}$} & {$x^{*}_{5}$} & {$\beta_{3}$} & {$\gamma_{3}$} 
    \\ 
    \midrule
    {$M_{\star}/M_{\rm halo}$} &
    {0.374} & {$4.28 \times 10^{-3}$} & {1.80} & {0.66}
    \\ 
    {$M_{\star}/M_{\odot}$} &
    {$6.39 \times 10^{-4}$} & {$1.77 \times 10^{5}$} & { 0.057} & {0.62}
    \\
    \bottomrule
    \bottomrule
    \vspace{-1ex}
    \\
    \multicolumn{5}{l}{\textbf{Note.} Use Eq. \protect\eqref{eq:conc.fit} for either $x=M_{\star}/M_{\rm halo}$ or $x=M_{\star}/M_{\odot}$.}\\
    \end{tabular}
\label{tab:conc.fit}
\end{table}


The right plot in Fig.~\ref{fig:10} shows the dark matter halo concentration directly: $c_{\rm vir}$ as a function of the dark matter halo mass, $M_{\rm halo}$. Black filled circles are the results for the FIRE-2 halos while open circles are the DMO analogs. The solid green curve traces the recent results of the concentration-mass relation from \cite{wang2019zoom}, which extends to masses all way down to the Earth mass dark matter halos. Note that \cite{wang2019zoom} uses the same concentration definition as we do as well. Additionally, they also fit halos with an Einasto profile same shape parameter we adopted ( $\alpha_{\epsilon} = 0.16$). The DMO analogs in our halo mass range follow the \cite{wang2019zoom} relation with significant scatter about the median. Interestingly, galaxy halos with $M_{\rm halo} = 10^{10-11}\ M_{\odot}$ all have about the same concentrations of $c_{\rm vir} \simeq 9$, with small scatter. In the $M_{\rm halo} = 10^{12}\ M_{\odot}$ region, baryonic contraction of the galaxy can increase the halo concentration significantly, to $c_{\rm vir} \simeq 15-25$). 
Observational measurements of the MW's halo concentration, which usually assume an NFW profile, have often found values typical of those we find here for our FIRE-2 halos ($c_{\rm vir} \approx 15-25$) -- well above the expectation for DMO halos of that mass ($c_{\rm vir} \sim 9$) \citep{battaglia2005radial,catena2010novel,deason2012broken,nesti2013dark}.
This then also suggests that for real galaxies, the predictions from \cite{wang2019zoom} will be an underestimated.

\section{Summary and concluding remarks}
\label{sec:conclude}
In this paper, we studied and modeled the $z=0$ dark matter density profiles of 54 zoom-in galaxy simulations run using the FIRE-2 feedback model. Our sample includes galaxies with stellar masses ranging from ultra-faint dwarfs to MW-mass galaxies, a factor of around 7 decades in stellar mass and 3 decades in halo mass. Details on these simulated halos, as well as parameter fits for each dark matter halo, are provided in Appendix \ref{sec:sim.sample}. 

The most significant contribution of this paper has been the introduction of the ``core-Einasto'': a new, three-parameter analytic density profile that provides a good fit to our FIRE-2 galaxy halos by allowing for a prominent constant density core, Eq.~\eqref{eq:cored.einasto.density}. Specifically, our main conclusions are as follows:
\begin{enumerate}
    \item 
    We find that feedback creates prominent cores in the centers of dark matter halos that have galaxy stellar masses $M_{\star}/M_{\rm halo} \simeq 5\times 10^{-3}$ or $M_\star \sim 10^9\ M_{\odot}$, roughly comparable to the stellar masses spanning the mass ranges of the SMC and the LMC (Fig.~\ref{fig:2}, \ref{fig:6}, and \ref{fig:7}). This mass regime is in agreement with previous studies \citep[e.g.][]{di2013dependence,tollet2016nihao}. Feedback-induced core formation becomes less important for galaxies with larger and smaller stellar masses. 
    
    \item 
    We find no evidence that feedback alters the density structure of halos that host galaxies smaller than $M_{\star} \simeq 10^6\, M_{\odot}$ or $M_{\star}/M_{\rm halo} \lesssim 10^{-4}$ down to radii $\sim 0.5\%\, r_{\rm vir}$ ($\sim 100$ pc; see also \citealt{fitts2017fire}). This in turn results in concentration values matching those seen in DMO analogs (Fig.~\ref{fig:10}). However, in FIRE-2 simulations with higher resolution, feedback may produce cores ${\sim} 100\ \rm pc$ in such galaxies \citep[see][]{wheeler2019resolved}.

    \item The core-Einasto profile, Eq.~\eqref{eq:cored.einasto.density}, takes the Einasto profile, Eq.~\eqref{eq:einasto.density}, and adds one additional parameter, a core radius $r_{c}$. The profile returns to the standard Einasto form as $r_{c} \rightarrow 0$. With a fixed $\alpha_\epsilon = 0.16$, we find that the three-parameter core-Einasto profile is able to characterize the majority of our feedback-impacted dark matter halos almost as well as the standard two-parameter Einasto profile does for DMO halos (Figs.~\ref{fig:4} -- \ref{fig:5}). In Appendix~\ref{sec:cNFW} we compare fits using the core-Einasto profile to two other three-parameter profiles (core-NFW and Dekel+) and show that the core-Einasto provides a better fit to FIRE-2 halos. 
    
    \item Fitted core radii are the largest ($r_{c} \simeq 1-5$ kpc) for  bright dwarf galaxies of $M_\star/M_{\rm halo} \simeq 5 \times 10^{-3}$ (or $M_{\star} \sim 10^{9} \, M_{\odot}$; Figs.~\ref{fig:7} -- \ref{fig:7}). Fitted core radii become smaller as the stellar to halo mass ratio moves away from this value (or equivalently, at both higher and lower stellar masses).
    The physical core radius is found to never be much larger than the stellar half-light radius, $r_{c} \lesssim r_{1/2}$, and only approaches $r_{1/2}$ in galaxies of the characteristic mass for core formation, $M_{\star} \sim 10^{9} \, M_{\odot}$ (Fig.~\ref{fig:8}).
    
    \item 
    Feedback and galaxy formation alters the global structure of dark matter halos well beyond the core region (Figs.~\ref{fig:9} -- \ref{fig:10}). Halos that host bright dwarf galaxies are often less concentrated than their DMO analogs, with $c_{\rm vir}$ values $30\%$ smaller. This differs slightly from the results in \cite{di2014mass}, who found no change in concentration at this mass scale. At higher masses, approaching the MW scale, the trend reverses and halos become much more concentrated owing to baryonic contraction. 
    
    \item 
    While baryonic contraction makes halos more concentrated and denser at the stellar half-light radius for MW size galaxies, we find that feedback can still produce small dark matter cores of ${\sim}\, 0.5-2$ kpc in size at this mass scale. The formation of cores in MW-size halos was previously discussed in \citet{chan2015impact}.  The combination of core-formation and baryonic contraction makes the resultant profiles complicated enough that Eq.~\eqref{eq:cored.einasto.density} does less well at capturing the full shape (with $\approx 20\%$ residuals, Fig.~\ref{fig:5}). To accommodate these features, we introduce a four-parameter contracted core profile in Appendix~\ref{sec:cEinasto.BC} (see Fig.~\ref{fig:C1}). The presence of dark matter cores in MW-size galaxies might be supported by dynamical modeling of MW data. \citet{portail2017dynamical} find evidence for a dark matter core comparable in size to what we quantify our feedback-affected MW-mass halos. 
\end{enumerate}

Though our results for core-Einasto and $r_{c}$ relations have focused on halos at $z=0$, the evolution of $r_{c}$ throughout cosmic time would provide an interesting future avenue of study, one that could provide further insight on the energy budget needed to transform cusps to cores in $\Lambda$CDM throughout cosmic time. Similarly, the methodology implemented and discussed in our analysis may be beneficial for a variety of studies in galaxy formation with alternative dark matter models. That is, our methods can be applicable in constraining characteristics of dark matter halos formed in other dark matter models. For example, dwarf galaxies simulated in self-interacting dark matter have characteristic central densities that are proportional to the interaction cross-section \citep[see][]{rocha2013sidm}. Preliminary results indicate that cores in self-interacting dark matter halos are ``sharper'' than those in feedback-affected CDM halos, perhaps indicating a path for differentiating between the two models in the presence of exquisite data (M.~Straight et al., in preparation).  

Perhaps the most exciting direction for future work will involve direct comparisons and modeling of observational data. In order to enable comparisons with observations, we provide fitting functions for $r_{\rm c}$ and other profile fit parameters as a function of $M_{\star}/M_{\rm halo}$ (see Eqs.~(\ref{eq:rcore.fit} -- \ref{eq:conc.fit}) and Tables \ref{tab:rcore.fit} -- \ref{tab:conc.fit}). Appendix~\ref{sec:cEinasto.Mstar} provides fits as a function of $M_{\star}$. Best fit parameters for all 54 of our galaxies are listed in Table~\ref{tab:halo.sample}. Resulting core-Einasto parameters can be utilized with analytic expressions for the mass profile, gravitational potential, and energy as presented in Appendix~\ref{sec:analytic.cEinasto}.

We have also shown that the dark matter rotation curves are well-captured by the core-Einasto fits in our simulations in Fig.~\ref{fig:6}, which motivates a comparison to current rotation curve data, such as the that from the {THINGS} survey \citep{walter2008things,oh2015little} or {SPARC} \citep{lelli2016psarc}. For examples of modeling with analytical profiles, we refer to the reader to analysis conducted by, but not limited to, \cite{kamada2017diverse,katz2017testing,ren2019diversity,kaplinghat2019spiral,robles2019sfdm,li2020sparc}. With the advent of future astrometric data being collected by {\it Gaia} \citep{brown2016gaia,prusti2016gaia,brown2018gaia,helmi2018gaia}, our model can also be combined with the central density normalizations obtainable in \cite{lazar2020mass} from the proper motions of dispersion-supported galaxies in order to constrain possible core radii and central densities via Eq.~\eqref{eq:cored.einasto.central.density}.

\section*{Acknowledgements}
We would like to thank the referee for helpful comments on the earlier versions of this article.
AL and JSB was supported by the National Science Foundation (NSF) grant AST-1910965.
MBK acknowledges support from NSF CAREER award AST-1752913, NSF grant AST-1910346, NASA grant NNX17AG29G, and HST-AR-14282, HST-AR-14554, HST-AR-15006, HST-GO-14191, and HST-GO-15658 from the Space Telescope Science Institute, which is operated by AURA, Inc., under NASA contract NAS5-26555. 
TKC is supported by STFC astronomy consolidated grant ST/T000244
ASG is supported by the McDonald Observatory at the University of Texas at Austin, through the Harlan J. Smith fellowship.
AW received support from NASA through ATP grant 80NSSC18K1097 and HST grants GO-14734, AR-15057, AR-15809, and GO-15902 from STScI; the Heising-Simons Foundation; and a Hellman Fellowship.
KE is supported by an NSF graduate research fellowship.
Support for CW was provided by NASA through the NASA Hubble Fellowship grant \#10938 awarded by STScI.
DK acknowledges support from NSF grant AST-1715101 and the Cottrell Scholar Award from the Research Corporation for Science Advancement.  Simulations presented in this work utilized resources granted by the Extreme Science and Engineering Discovery Environment (XSEDE), which is supported by National Science Foundation grant no. OCI-1053575.
CAFG was supported by NSF through grants AST-1517491, AST-1715216, and CAREER award AST-1652522; by NASA through grant 17-ATP17-0067; and by a Cottrell Scholar Award from the Research Corporation for Science Advancement. 
The analysis in this paper depended on the python packages {\footnotesize NumPy} \citep{van2011numpy}, {\footnotesize SciPy} \citep{oliphant2007python}, and {\footnotesize Matplotlib} \citep{hunter2007matplotlib}; We are thankful to the developers of these tools. This research has made intensive use of NASA's Astrophysics Data System (\url{http://ui.adsabs.harvard.edu/}) and the arXiv eprint service (\url{http://arxiv.org}).

\bibliography{references}

\begin{thebibliography}{}
\makeatletter
\relax
\def\mn@urlcharsother{\let\do\@makeother \do\$\do\&\do\#\do\^\do\_\do\%\do\~}
\def\mn@doi{\begingroup\mn@urlcharsother \@ifnextchar [ {\mn@doi@}
  {\mn@doi@[]}}
\def\mn@doi@[#1]#2{\def\@tempa{#1}\ifx\@tempa\@empty \href
  {http://dx.doi.org/#2} {doi:#2}\else \href {http://dx.doi.org/#2} {#1}\fi
  \endgroup}
\def\mn@eprint#1#2{\mn@eprint@#1:#2::\@nil}
\def\mn@eprint@arXiv#1{\href {http://arxiv.org/abs/#1} {{\tt arXiv:#1}}}
\def\mn@eprint@dblp#1{\href {http://dblp.uni-trier.de/rec/bibtex/#1.xml}
  {dblp:#1}}
\def\mn@eprint@#1:#2:#3:#4\@nil{\def\@tempa {#1}\def\@tempb {#2}\def\@tempc
  {#3}\ifx \@tempc \@empty \let \@tempc \@tempb \let \@tempb \@tempa \fi \ifx
  \@tempb \@empty \def\@tempb {arXiv}\fi \@ifundefined
  {mn@eprint@\@tempb}{\@tempb:\@tempc}{\expandafter \expandafter \csname
  mn@eprint@\@tempb\endcsname \expandafter{\@tempc}}}

\bibitem[\protect\citeauthoryear{{Battaglia} et~al.,}{{Battaglia}
  et~al.}{2005}]{battaglia2005radial}
{Battaglia} G.,  et~al., 2005, \mn@doi [\mnras]
  {10.1111/j.1365-2966.2005.09367.x}, \href
  {https://ui.adsabs.harvard.edu/abs/2005MNRAS.364..433B} {364, 433}

\bibitem[\protect\citeauthoryear{{Behroozi}, {Wechsler}  \& {Wu}}{{Behroozi}
  et~al.}{2013}]{behroozi2012rockstar}
{Behroozi} P.~S.,  {Wechsler} R.~H.,   {Wu} H.-Y.,  2013, \mn@doi [\apj]
  {10.1088/0004-637X/762/2/109}, \href
  {https://ui.adsabs.harvard.edu/abs/2013ApJ...762..109B} {762, 109}

\bibitem[\protect\citeauthoryear{{Behroozi}, {Wechsler}, {Hearin}  \&
  {Conroy}}{{Behroozi} et~al.}{2019}]{behroozi2019um}
{Behroozi} P.,  {Wechsler} R.~H.,  {Hearin} A.~P.,   {Conroy} C.,  2019,
  \mn@doi [\mnras] {10.1093/mnras/stz1182}, \href
  {https://ui.adsabs.harvard.edu/abs/2019MNRAS.488.3143B} {488, 3143}

\bibitem[\protect\citeauthoryear{{Ben{\'\i}tez-Llambay}, {Frenk}, {Ludlow}  \&
  {Navarro}}{{Ben{\'\i}tez-Llambay} et~al.}{2019}]{benitez2019cores}
{Ben{\'\i}tez-Llambay} A.,  {Frenk} C.~S.,  {Ludlow} A.~D.,   {Navarro} J.~F.,
  2019, \mn@doi [\mnras] {10.1093/mnras/stz1890}, \href
  {https://ui.adsabs.harvard.edu/abs/2019MNRAS.488.2387B} {488, 2387}

\bibitem[\protect\citeauthoryear{{Binney} \& {Tremaine}}{{Binney} \&
  {Tremaine}}{2008}]{binney2011galactic}
{Binney} J.,  {Tremaine} S.,  2008, {Galactic Dynamics: Second Edition}

\bibitem[\protect\citeauthoryear{{Blumenthal}, {Faber}, {Flores}  \&
  {Primack}}{{Blumenthal} et~al.}{1986}]{blumenthal1986contraction}
{Blumenthal} G.~R.,  {Faber} S.~M.,  {Flores} R.,   {Primack} J.~R.,  1986,
  \mn@doi [\apj] {10.1086/163867}, \href
  {https://ui.adsabs.harvard.edu/abs/1986ApJ...301...27B} {301, 27}

\bibitem[\protect\citeauthoryear{{Bose} et~al.,}{{Bose}
  et~al.}{2019}]{bose2019cores}
{Bose} S.,  et~al., 2019, \mn@doi [\mnras] {10.1093/mnras/stz1168}, \href
  {https://ui.adsabs.harvard.edu/abs/2019MNRAS.486.4790B} {486, 4790}

\bibitem[\protect\citeauthoryear{{Boylan-Kolchin}, {Bullock}  \&
  {Kaplinghat}}{{Boylan-Kolchin} et~al.}{2011}]{boylan2011too}
{Boylan-Kolchin} M.,  {Bullock} J.~S.,   {Kaplinghat} M.,  2011, \mn@doi
  [\mnras] {10.1111/j.1745-3933.2011.01074.x}, \href
  {https://ui.adsabs.harvard.edu/abs/2011MNRAS.415L..40B} {415, L40}

\bibitem[\protect\citeauthoryear{{Brook} \& {Di Cintio}}{{Brook} \& {Di
  Cintio}}{2015}]{brook2015local}
{Brook} C.~B.,  {Di Cintio} A.,  2015, \mn@doi [\mnras] {10.1093/mnras/stv864},
  \href {https://ui.adsabs.harvard.edu/abs/2015MNRAS.450.3920B} {450, 3920}

\bibitem[\protect\citeauthoryear{{Brook}, {Stinson}, {Gibson}, {Wadsley}  \&
  {Quinn}}{{Brook} et~al.}{2012}]{brooks2012magicc}
{Brook} C.~B.,  {Stinson} G.,  {Gibson} B.~K.,  {Wadsley} J.,   {Quinn} T.,
  2012, \mn@doi [\mnras] {10.1111/j.1365-2966.2012.21306.x}, \href
  {https://ui.adsabs.harvard.edu/abs/2012MNRAS.424.1275B} {424, 1275}

\bibitem[\protect\citeauthoryear{{Brooks} \& {Zolotov}}{{Brooks} \&
  {Zolotov}}{2014}]{brooks2014why}
{Brooks} A.~M.,  {Zolotov} A.,  2014, \mn@doi [\apj]
  {10.1088/0004-637X/786/2/87}, \href
  {https://ui.adsabs.harvard.edu/abs/2014ApJ...786...87B} {786, 87}

\bibitem[\protect\citeauthoryear{{Bryan} \& {Norman}}{{Bryan} \&
  {Norman}}{1998}]{bryan1998statistical}
{Bryan} G.~L.,  {Norman} M.~L.,  1998, \mn@doi [\apj] {10.1086/305262}, \href
  {https://ui.adsabs.harvard.edu/abs/1998ApJ...495...80B} {495, 80}

\bibitem[\protect\citeauthoryear{{Bullock} \& {Boylan-Kolchin}}{{Bullock} \&
  {Boylan-Kolchin}}{2017}]{bullock2017small}
{Bullock} J.~S.,  {Boylan-Kolchin} M.,  2017, \mn@doi [\araa]
  {10.1146/annurev-astro-091916-055313}, \href
  {https://ui.adsabs.harvard.edu/abs/2017ARA&A..55..343B} {55, 343}

\bibitem[\protect\citeauthoryear{{Catena} \& {Ullio}}{{Catena} \&
  {Ullio}}{2010}]{catena2010novel}
{Catena} R.,  {Ullio} P.,  2010, \mn@doi [\jcap]
  {10.1088/1475-7516/2010/08/004}, \href
  {https://ui.adsabs.harvard.edu/abs/2010JCAP...08..004C} {2010, 004}

\bibitem[\protect\citeauthoryear{{Chan}, {Kere{\v{s}}}, {O{\~n}orbe},
  {Hopkins}, {Muratov}, {Faucher-Gigu{\`e}re}  \& {Quataert}}{{Chan}
  et~al.}{2015}]{chan2015impact}
{Chan} T.~K.,  {Kere{\v{s}}} D.,  {O{\~n}orbe} J.,  {Hopkins} P.~F.,  {Muratov}
  A.~L.,  {Faucher-Gigu{\`e}re} C.~A.,   {Quataert} E.,  2015, \mn@doi [\mnras]
  {10.1093/mnras/stv2165}, \href
  {https://ui.adsabs.harvard.edu/abs/2015MNRAS.454.2981C} {454, 2981}

\bibitem[\protect\citeauthoryear{{Chan}, {Kere{\v{s}}}, {Wetzel}, {Hopkins},
  {Faucher-Gigu{\`e}re}, {El-Badry}, {Garrison-Kimmel}  \&
  {Boylan-Kolchin}}{{Chan} et~al.}{2018}]{chan2018origin}
{Chan} T.~K.,  {Kere{\v{s}}} D.,  {Wetzel} A.,  {Hopkins} P.~F.,
  {Faucher-Gigu{\`e}re} C.~A.,  {El-Badry} K.,  {Garrison-Kimmel} S.,
  {Boylan-Kolchin} M.,  2018, \mn@doi [\mnras] {10.1093/mnras/sty1153}, \href
  {https://ui.adsabs.harvard.edu/abs/2018MNRAS.478..906C} {478, 906}

\bibitem[\protect\citeauthoryear{{Cole}, {Dehnen}  \& {Wilkinson}}{{Cole}
  et~al.}{2011}]{cole2011weakening}
{Cole} D.~R.,  {Dehnen} W.,   {Wilkinson} M.~I.,  2011, \mn@doi [\mnras]
  {10.1111/j.1365-2966.2011.19110.x}, \href
  {https://ui.adsabs.harvard.edu/abs/2011MNRAS.416.1118C} {416, 1118}

\bibitem[\protect\citeauthoryear{{Deason}, {Belokurov}, {Evans}  \&
  {An}}{{Deason} et~al.}{2012}]{deason2012broken}
{Deason} A.~J.,  {Belokurov} V.,  {Evans} N.~W.,   {An} J.,  2012, \mn@doi
  [\mnras] {10.1111/j.1745-3933.2012.01283.x}, \href
  {https://ui.adsabs.harvard.edu/abs/2012MNRAS.424L..44D} {424, L44}

\bibitem[\protect\citeauthoryear{{Dekel}, {Ishai}, {Dutton}  \&
  {Maccio}}{{Dekel} et~al.}{2017}]{dekel2017profile}
{Dekel} A.,  {Ishai} G.,  {Dutton} A.~A.,   {Maccio} A.~V.,  2017, \mn@doi
  [\mnras] {10.1093/mnras/stx486}, \href
  {https://ui.adsabs.harvard.edu/abs/2017MNRAS.468.1005D} {468, 1005}

\bibitem[\protect\citeauthoryear{{Di Cintio}, {Brook}, {Macci{\`o}}, {Stinson},
  {Knebe}, {Dutton}  \& {Wadsley}}{{Di Cintio}
  et~al.}{2014a}]{di2013dependence}
{Di Cintio} A.,  {Brook} C.~B.,  {Macci{\`o}} A.~V.,  {Stinson} G.~S.,  {Knebe}
  A.,  {Dutton} A.~A.,   {Wadsley} J.,  2014a, \mn@doi [\mnras]
  {10.1093/mnras/stt1891}, \href
  {https://ui.adsabs.harvard.edu/abs/2014MNRAS.437..415D} {437, 415}

\bibitem[\protect\citeauthoryear{{Di Cintio}, {Brook}, {Dutton}, {Macci{\`o}},
  {Stinson}  \& {Knebe}}{{Di Cintio} et~al.}{2014b}]{di2014mass}
{Di Cintio} A.,  {Brook} C.~B.,  {Dutton} A.~A.,  {Macci{\`o}} A.~V.,
  {Stinson} G.~S.,   {Knebe} A.,  2014b, \mn@doi [\mnras]
  {10.1093/mnras/stu729}, \href
  {https://ui.adsabs.harvard.edu/abs/2014MNRAS.441.2986D} {441, 2986}

\bibitem[\protect\citeauthoryear{{Dubinski} \& {Carlberg}}{{Dubinski} \&
  {Carlberg}}{1991}]{dubinkski1991}
{Dubinski} J.,  {Carlberg} R.~G.,  1991, \mn@doi [\apj] {10.1086/170451}, \href
  {https://ui.adsabs.harvard.edu/abs/1991ApJ...378..496D} {378, 496}

\bibitem[\protect\citeauthoryear{{Dutton} \& {Macci{\`o}}}{{Dutton} \&
  {Macci{\`o}}}{2014}]{dutton2014cold}
{Dutton} A.~A.,  {Macci{\`o}} A.~V.,  2014, \mn@doi [\mnras]
  {10.1093/mnras/stu742}, \href
  {https://ui.adsabs.harvard.edu/abs/2014MNRAS.441.3359D} {441, 3359}

\bibitem[\protect\citeauthoryear{{Dutton}, {Macci{\`o}}, {Buck}, {Dixon},
  {Blank}  \& {Obreja}}{{Dutton} et~al.}{2019}]{dutton2019impact}
{Dutton} A.~A.,  {Macci{\`o}} A.~V.,  {Buck} T.,  {Dixon} K.~L.,  {Blank} M.,
  {Obreja} A.,  2019, \mn@doi [\mnras] {10.1093/mnras/stz889}, \href
  {https://ui.adsabs.harvard.edu/abs/2019MNRAS.486..655D} {486, 655}

\bibitem[\protect\citeauthoryear{{Einasto}}{{Einasto}}{1965}]{einasto1965influence}
{Einasto} J.,  1965, Trudy Astrofizicheskogo Instituta Alma-Ata, \href
  {https://ui.adsabs.harvard.edu/abs/1965TrAlm...5...87E} {5, 87}

\bibitem[\protect\citeauthoryear{{El-Badry}, {Wetzel}, {Geha}, {Hopkins},
  {Kere{\v{s}}}, {Chan}  \& {Faucher-Gigu{\`e}re}}{{El-Badry}
  et~al.}{2016}]{el2016breathing}
{El-Badry} K.,  {Wetzel} A.,  {Geha} M.,  {Hopkins} P.~F.,  {Kere{\v{s}}} D.,
  {Chan} T.~K.,   {Faucher-Gigu{\`e}re} C.-A.,  2016, \mn@doi [\apj]
  {10.3847/0004-637X/820/2/131}, \href
  {https://ui.adsabs.harvard.edu/abs/2016ApJ...820..131E} {820, 131}

\bibitem[\protect\citeauthoryear{{El-Badry} et~al.,}{{El-Badry}
  et~al.}{2018a}]{el2017gas}
{El-Badry} K.,  et~al., 2018a, \mn@doi [\mnras] {10.1093/mnras/stx2482}, \href
  {https://ui.adsabs.harvard.edu/abs/2018MNRAS.473.1930E} {473, 1930}

\bibitem[\protect\citeauthoryear{{El-Badry} et~al.,}{{El-Badry}
  et~al.}{2018b}]{el2018ancient}
{El-Badry} K.,  et~al., 2018b, \mn@doi [\mnras] {10.1093/mnras/sty1864}, \href
  {https://ui.adsabs.harvard.edu/abs/2018MNRAS.480..652E} {480, 652}

\bibitem[\protect\citeauthoryear{{El-Zant}, {Shlosman}  \& {Hoffman}}{{El-Zant}
  et~al.}{2001}]{elzant2001cusp}
{El-Zant} A.,  {Shlosman} I.,   {Hoffman} Y.,  2001, \mn@doi [\apj]
  {10.1086/322516}, \href
  {https://ui.adsabs.harvard.edu/abs/2001ApJ...560..636E} {560, 636}

\bibitem[\protect\citeauthoryear{{Faucher-Gigu{\`e}re}, {Lidz}, {Zaldarriaga}
  \& {Hernquist}}{{Faucher-Gigu{\`e}re} et~al.}{2009}]{faucher2009new}
{Faucher-Gigu{\`e}re} C.-A.,  {Lidz} A.,  {Zaldarriaga} M.,   {Hernquist} L.,
  2009, \mn@doi [\apj] {10.1088/0004-637X/703/2/1416}, \href
  {https://ui.adsabs.harvard.edu/abs/2009ApJ...703.1416F} {703, 1416}

\bibitem[\protect\citeauthoryear{{Ferland}, {Korista}, {Verner}, {Ferguson},
  {Kingdon}  \& {Verner}}{{Ferland} et~al.}{1998}]{ferland1998cloudy}
{Ferland} G.~J.,  {Korista} K.~T.,  {Verner} D.~A.,  {Ferguson} J.~W.,
  {Kingdon} J.~B.,   {Verner} E.~M.,  1998, \mn@doi [\pasp] {10.1086/316190},
  \href {https://ui.adsabs.harvard.edu/abs/1998PASP..110..761F} {110, 761}

\bibitem[\protect\citeauthoryear{{Fitts} et~al.,}{{Fitts}
  et~al.}{2017}]{fitts2017fire}
{Fitts} A.,  et~al., 2017, \mn@doi [\mnras] {10.1093/mnras/stx1757}, \href
  {https://ui.adsabs.harvard.edu/abs/2017MNRAS.471.3547F} {471, 3547}

\bibitem[\protect\citeauthoryear{{Flores} \& {Primack}}{{Flores} \&
  {Primack}}{1994}]{flores1994observational}
{Flores} R.~A.,  {Primack} J.~R.,  1994, \mn@doi [\apjl] {10.1086/187350},
  \href {https://ui.adsabs.harvard.edu/abs/1994ApJ...427L...1F} {427, L1}

\bibitem[\protect\citeauthoryear{{Freundlich} et~al.,}{{Freundlich}
  et~al.}{2020}]{freundlich2020dekel}
{Freundlich} J.,  et~al., 2020, arXiv e-prints, \href
  {https://ui.adsabs.harvard.edu/abs/2020arXiv200408395F} {p. arXiv:2004.08395}

\bibitem[\protect\citeauthoryear{{Gaia Collaboration} et~al.,}{{Gaia
  Collaboration} et~al.}{2016a}]{prusti2016gaia}
{Gaia Collaboration} et~al., 2016a, \mn@doi [\aap]
  {10.1051/0004-6361/201629272}, \href
  {https://ui.adsabs.harvard.edu/abs/2016A&A...595A...1G} {595, A1}

\bibitem[\protect\citeauthoryear{{Gaia Collaboration} et~al.,}{{Gaia
  Collaboration} et~al.}{2016b}]{brown2016gaia}
{Gaia Collaboration} et~al., 2016b, \mn@doi [\aap]
  {10.1051/0004-6361/201629512}, \href
  {https://ui.adsabs.harvard.edu/abs/2016A&A...595A...2G} {595, A2}

\bibitem[\protect\citeauthoryear{{Gaia Collaboration} et~al.,}{{Gaia
  Collaboration} et~al.}{2018a}]{brown2018gaia}
{Gaia Collaboration} et~al., 2018a, \mn@doi [\aap]
  {10.1051/0004-6361/201833051}, \href
  {https://ui.adsabs.harvard.edu/abs/2018A&A...616A...1G} {616, A1}

\bibitem[\protect\citeauthoryear{{Gaia Collaboration} et~al.,}{{Gaia
  Collaboration} et~al.}{2018b}]{helmi2018gaia}
{Gaia Collaboration} et~al., 2018b, \mn@doi [\aap]
  {10.1051/0004-6361/201832698}, \href
  {https://ui.adsabs.harvard.edu/abs/2018A&A...616A..12G} {616, A12}

\bibitem[\protect\citeauthoryear{{Gao}, {Navarro}, {Cole}, {Frenk}, {White},
  {Springel}, {Jenkins}  \& {Neto}}{{Gao} et~al.}{2008}]{gao2008redshift}
{Gao} L.,  {Navarro} J.~F.,  {Cole} S.,  {Frenk} C.~S.,  {White} S. D.~M.,
  {Springel} V.,  {Jenkins} A.,   {Neto} A.~F.,  2008, \mn@doi [\mnras]
  {10.1111/j.1365-2966.2008.13277.x}, \href
  {https://ui.adsabs.harvard.edu/abs/2008MNRAS.387..536G} {387, 536}

\bibitem[\protect\citeauthoryear{{Garrison-Kimmel}, {Rocha}, {Boylan-Kolchin},
  {Bullock}  \& {Lally}}{{Garrison-Kimmel} et~al.}{2013}]{garrison2013}
{Garrison-Kimmel} S.,  {Rocha} M.,  {Boylan-Kolchin} M.,  {Bullock} J.~S.,
  {Lally} J.,  2013, \mn@doi [\mnras] {10.1093/mnras/stt984}, \href
  {https://ui.adsabs.harvard.edu/abs/2013MNRAS.433.3539G} {433, 3539}

\bibitem[\protect\citeauthoryear{{Garrison-Kimmel}, {Boylan-Kolchin}, {Bullock}
   \& {Kirby}}{{Garrison-Kimmel} et~al.}{2014}]{garrison2014too}
{Garrison-Kimmel} S.,  {Boylan-Kolchin} M.,  {Bullock} J.~S.,   {Kirby} E.~N.,
  2014, \mn@doi [\mnras] {10.1093/mnras/stu1477}, \href
  {https://ui.adsabs.harvard.edu/abs/2014MNRAS.444..222G} {444, 222}

\bibitem[\protect\citeauthoryear{{Garrison-Kimmel}, {Bullock}, {Boylan-Kolchin}
   \& {Bardwell}}{{Garrison-Kimmel} et~al.}{2017}]{sgk2017chaos}
{Garrison-Kimmel} S.,  {Bullock} J.~S.,  {Boylan-Kolchin} M.,   {Bardwell} E.,
  2017, \mn@doi [\mnras] {10.1093/mnras/stw2564}, \href
  {https://ui.adsabs.harvard.edu/abs/2017MNRAS.464.3108G} {464, 3108}

\bibitem[\protect\citeauthoryear{{Garrison-Kimmel} et~al.,}{{Garrison-Kimmel}
  et~al.}{2019}]{garrison2018local}
{Garrison-Kimmel} S.,  et~al., 2019, \mn@doi [\mnras] {10.1093/mnras/stz1317},
  \href {https://ui.adsabs.harvard.edu/abs/2019MNRAS.487.1380G} {487, 1380}

\bibitem[\protect\citeauthoryear{{Gentile}, {Salucci}, {Klein}, {Vergani}  \&
  {Kalberla}}{{Gentile} et~al.}{2004}]{gentile2004cored}
{Gentile} G.,  {Salucci} P.,  {Klein} U.,  {Vergani} D.,   {Kalberla} P.,
  2004, \mn@doi [\mnras] {10.1111/j.1365-2966.2004.07836.x}, \href
  {https://ui.adsabs.harvard.edu/abs/2004MNRAS.351..903G} {351, 903}

\bibitem[\protect\citeauthoryear{{Gnedin}, {Kravtsov}, {Klypin}  \&
  {Nagai}}{{Gnedin} et~al.}{2004}]{gnedin2004response}
{Gnedin} O.~Y.,  {Kravtsov} A.~V.,  {Klypin} A.~A.,   {Nagai} D.,  2004,
  \mn@doi [\apj] {10.1086/424914}, \href
  {https://ui.adsabs.harvard.edu/abs/2004ApJ...616...16G} {616, 16}

\bibitem[\protect\citeauthoryear{{Goerdt}, {Moore}, {Read}  \&
  {Stadel}}{{Goerdt} et~al.}{2010}]{goerdt2010core}
{Goerdt} T.,  {Moore} B.,  {Read} J.~I.,   {Stadel} J.,  2010, \mn@doi [\apj]
  {10.1088/0004-637X/725/2/1707}, \href
  {https://ui.adsabs.harvard.edu/abs/2010ApJ...725.1707G} {725, 1707}

\bibitem[\protect\citeauthoryear{{Governato} et~al.,}{{Governato}
  et~al.}{2010}]{governato2010bulgeless}
{Governato} F.,  et~al., 2010, \mn@doi [\nat] {10.1038/nature08640}, \href
  {https://ui.adsabs.harvard.edu/abs/2010Natur.463..203G} {463, 203}

\bibitem[\protect\citeauthoryear{{Governato} et~al.,}{{Governato}
  et~al.}{2012}]{governato2012cuspy}
{Governato} F.,  et~al., 2012, \mn@doi [\mnras]
  {10.1111/j.1365-2966.2012.20696.x}, \href
  {https://ui.adsabs.harvard.edu/abs/2012MNRAS.422.1231G} {422, 1231}

\bibitem[\protect\citeauthoryear{{Graus} et~al.,}{{Graus}
  et~al.}{2019}]{graus2019predicted}
{Graus} A.~S.,  et~al., 2019, \mn@doi [\mnras] {10.1093/mnras/stz2649}, \href
  {https://ui.adsabs.harvard.edu/abs/2019MNRAS.490.1186G} {490, 1186}

\bibitem[\protect\citeauthoryear{{Griffen}, {Ji}, {Dooley}, {G{\'o}mez},
  {Vogelsberger}, {O'Shea}  \& {Frebel}}{{Griffen}
  et~al.}{2016}]{griffen2016caterpillar}
{Griffen} B.~F.,  {Ji} A.~P.,  {Dooley} G.~A.,  {G{\'o}mez} F.~A.,
  {Vogelsberger} M.,  {O'Shea} B.~W.,   {Frebel} A.,  2016, \mn@doi [\apj]
  {10.3847/0004-637X/818/1/10}, \href
  {https://ui.adsabs.harvard.edu/abs/2016ApJ...818...10G} {818, 10}

\bibitem[\protect\citeauthoryear{{Hahn} \& {Abel}}{{Hahn} \&
  {Abel}}{2011}]{hahn2013music}
{Hahn} O.,  {Abel} T.,  2011, \mn@doi [\mnras]
  {10.1111/j.1365-2966.2011.18820.x}, \href
  {https://ui.adsabs.harvard.edu/abs/2011MNRAS.415.2101H} {415, 2101}

\bibitem[\protect\citeauthoryear{{Hopkins}}{{Hopkins}}{2015}]{hopkins2015new}
{Hopkins} P.~F.,  2015, \mn@doi [\mnras] {10.1093/mnras/stv195}, \href
  {https://ui.adsabs.harvard.edu/abs/2015MNRAS.450...53H} {450, 53}

\bibitem[\protect\citeauthoryear{{Hopkins}, {Kere{\v{s}}}, {O{\~n}orbe},
  {Faucher-Gigu{\`e}re}, {Quataert}, {Murray}  \& {Bullock}}{{Hopkins}
  et~al.}{2014}]{hopkins2014galaxies}
{Hopkins} P.~F.,  {Kere{\v{s}}} D.,  {O{\~n}orbe} J.,  {Faucher-Gigu{\`e}re}
  C.-A.,  {Quataert} E.,  {Murray} N.,   {Bullock} J.~S.,  2014, \mn@doi
  [\mnras] {10.1093/mnras/stu1738}, \href
  {https://ui.adsabs.harvard.edu/abs/2014MNRAS.445..581H} {445, 581}

\bibitem[\protect\citeauthoryear{{Hopkins} et~al.,}{{Hopkins}
  et~al.}{2018}]{hopkins2018fire}
{Hopkins} P.~F.,  et~al., 2018, \mn@doi [\mnras] {10.1093/mnras/sty1690}, \href
  {https://ui.adsabs.harvard.edu/abs/2018MNRAS.480..800H} {480, 800}

\bibitem[\protect\citeauthoryear{{Hunter}}{{Hunter}}{2007}]{hunter2007matplotlib}
{Hunter} J.~D.,  2007, \mn@doi [Computing in Science and Engineering]
  {10.1109/MCSE.2007.55}, \href
  {https://ui.adsabs.harvard.edu/abs/2007CSE.....9...90H} {9, 90}

\bibitem[\protect\citeauthoryear{{Kamada}, {Kaplinghat}, {Pace}  \&
  {Yu}}{{Kamada} et~al.}{2017}]{kamada2017diverse}
{Kamada} A.,  {Kaplinghat} M.,  {Pace} A.~B.,   {Yu} H.-B.,  2017, \mn@doi
  [\prl] {10.1103/PhysRevLett.119.111102}, \href
  {https://ui.adsabs.harvard.edu/abs/2017PhRvL.119k1102K} {119, 111102}

\bibitem[\protect\citeauthoryear{{Kaplinghat}, {Ren}  \& {Yu}}{{Kaplinghat}
  et~al.}{2019}]{kaplinghat2019spiral}
{Kaplinghat} M.,  {Ren} T.,   {Yu} H.-B.,  2019, arXiv e-prints, \href
  {https://ui.adsabs.harvard.edu/abs/2019arXiv191100544K} {p. arXiv:1911.00544}

\bibitem[\protect\citeauthoryear{{Katz}, {Lelli}, {McGaugh}, {Di Cintio},
  {Brook}  \& {Schombert}}{{Katz} et~al.}{2017}]{katz2017testing}
{Katz} H.,  {Lelli} F.,  {McGaugh} S.~S.,  {Di Cintio} A.,  {Brook} C.~B.,
  {Schombert} J.~M.,  2017, \mn@doi [\mnras] {10.1093/mnras/stw3101}, \href
  {https://ui.adsabs.harvard.edu/abs/2017MNRAS.466.1648K} {466, 1648}

\bibitem[\protect\citeauthoryear{{Klypin}, {Yepes}, {Gottl{\"o}ber}, {Prada}
  \& {He{\ss}}}{{Klypin} et~al.}{2016}]{klypin2016multidark}
{Klypin} A.,  {Yepes} G.,  {Gottl{\"o}ber} S.,  {Prada} F.,   {He{\ss}} S.,
  2016, \mn@doi [\mnras] {10.1093/mnras/stw248}, \href
  {https://ui.adsabs.harvard.edu/abs/2016MNRAS.457.4340K} {457, 4340}

\bibitem[\protect\citeauthoryear{{Kroupa}}{{Kroupa}}{2001}]{kroupa2001variation}
{Kroupa} P.,  2001, \mn@doi [\mnras] {10.1046/j.1365-8711.2001.04022.x}, \href
  {https://ui.adsabs.harvard.edu/abs/2001MNRAS.322..231K} {322, 231}

\bibitem[\protect\citeauthoryear{{Lazar} \& {Bullock}}{{Lazar} \&
  {Bullock}}{2020}]{lazar2020mass}
{Lazar} A.,  {Bullock} J.~S.,  2020, \mn@doi [\mnras] {10.1093/mnras/staa692},
  \href {https://ui.adsabs.harvard.edu/abs/2020MNRAS.tmp..665L} {493, 5825}

\bibitem[\protect\citeauthoryear{{Leitherer} et~al.,}{{Leitherer}
  et~al.}{1999}]{leitherer1999starburst99}
{Leitherer} C.,  et~al., 1999, \mn@doi [\apjs] {10.1086/313233}, \href
  {https://ui.adsabs.harvard.edu/abs/1999ApJS..123....3L} {123, 3}

\bibitem[\protect\citeauthoryear{{Lelli}, {McGaugh}  \& {Schombert}}{{Lelli}
  et~al.}{2016}]{lelli2016psarc}
{Lelli} F.,  {McGaugh} S.~S.,   {Schombert} J.~M.,  2016, \mn@doi [\aj]
  {10.3847/0004-6256/152/6/157}, \href
  {https://ui.adsabs.harvard.edu/abs/2016AJ....152..157L} {152, 157}

\bibitem[\protect\citeauthoryear{{Li}, {Lelli}, {McGaugh}  \& {Schombert}}{{Li}
  et~al.}{2020}]{li2020sparc}
{Li} P.,  {Lelli} F.,  {McGaugh} S.,   {Schombert} J.,  2020, \mn@doi [\apjs]
  {10.3847/1538-4365/ab700e}, \href
  {https://ui.adsabs.harvard.edu/abs/2020ApJS..247...31L} {247, 31}

\bibitem[\protect\citeauthoryear{{Ludlow}, {Navarro}, {Springel},
  {Vogelsberger}, {Wang}, {White}, {Jenkins}  \& {Frenk}}{{Ludlow}
  et~al.}{2010}]{ludlow2010pseudo}
{Ludlow} A.~D.,  {Navarro} J.~F.,  {Springel} V.,  {Vogelsberger} M.,  {Wang}
  J.,  {White} S. D.~M.,  {Jenkins} A.,   {Frenk} C.~S.,  2010, \mn@doi
  [\mnras] {10.1111/j.1365-2966.2010.16678.x}, \href
  {https://ui.adsabs.harvard.edu/abs/2010MNRAS.406..137L} {406, 137}

\bibitem[\protect\citeauthoryear{{Ludlow}, {Bose}, {Angulo}, {Wang},
  {Hellwing}, {Navarro}, {Cole}  \& {Frenk}}{{Ludlow}
  et~al.}{2016}]{ludlow2016relations}
{Ludlow} A.~D.,  {Bose} S.,  {Angulo} R.~E.,  {Wang} L.,  {Hellwing} W.~A.,
  {Navarro} J.~F.,  {Cole} S.,   {Frenk} C.~S.,  2016, \mn@doi [\mnras]
  {10.1093/mnras/stw1046}, \href
  {https://ui.adsabs.harvard.edu/abs/2016MNRAS.460.1214L} {460, 1214}

\bibitem[\protect\citeauthoryear{{Macci{\`o}}, {Crespi}, {Blank}  \&
  {Kang}}{{Macci{\`o}} et~al.}{2020}]{maccio2020nihao}
{Macci{\`o}} A.~V.,  {Crespi} S.,  {Blank} M.,   {Kang} X.,  2020, \mn@doi
  [\mnras] {10.1093/mnrasl/slaa058}, \href
  {https://ui.adsabs.harvard.edu/abs/2020MNRAS.tmpL..54M} {}

\bibitem[\protect\citeauthoryear{{Madau}, {Shen}  \& {Governato}}{{Madau}
  et~al.}{2014}]{madau2014core}
{Madau} P.,  {Shen} S.,   {Governato} F.,  2014, \mn@doi [\apjl]
  {10.1088/2041-8205/789/1/L17}, \href
  {https://ui.adsabs.harvard.edu/abs/2014ApJ...789L..17M} {789, L17}

\bibitem[\protect\citeauthoryear{{Mashchenko}, {Couchman}  \&
  {Wadsley}}{{Mashchenko} et~al.}{2006}]{mashchenko2006removal}
{Mashchenko} S.,  {Couchman} H.~M.~P.,   {Wadsley} J.,  2006, \mn@doi [\nat]
  {10.1038/nature04944}, \href
  {https://ui.adsabs.harvard.edu/abs/2006Natur.442..539M} {442, 539}

\bibitem[\protect\citeauthoryear{{Merritt}, {Graham}, {Moore}, {Diemand }  \&
  {Terzi{\'c}}}{{Merritt} et~al.}{2006}]{merritt2006empirical}
{Merritt} D.,  {Graham} A.~W.,  {Moore} B.,  {Diemand } J.,   {Terzi{\'c}} B.,
  2006, \mn@doi [\aj] {10.1086/508988}, \href
  {https://ui.adsabs.harvard.edu/abs/2006AJ....132.2685M} {132, 2685}

\bibitem[\protect\citeauthoryear{{Moore}}{{Moore}}{1994}]{moore1994evidence}
{Moore} B.,  1994, \mn@doi [\nat] {10.1038/370629a0}, \href
  {https://ui.adsabs.harvard.edu/abs/1994Natur.370..629M} {370, 629}

\bibitem[\protect\citeauthoryear{{Munshi} et~al.,}{{Munshi}
  et~al.}{2013}]{munshi2013stellar}
{Munshi} F.,  et~al., 2013, \mn@doi [\apj] {10.1088/0004-637X/766/1/56}, \href
  {https://ui.adsabs.harvard.edu/abs/2013ApJ...766...56M} {766, 56}

\bibitem[\protect\citeauthoryear{{Navarro}, {Eke}  \& {Frenk}}{{Navarro}
  et~al.}{1996}]{navarro1996cores}
{Navarro} J.~F.,  {Eke} V.~R.,   {Frenk} C.~S.,  1996, \mn@doi [\mnras]
  {10.1093/mnras/283.3.L72}, \href
  {https://ui.adsabs.harvard.edu/abs/1996MNRAS.283L..72N} {283, L72}

\bibitem[\protect\citeauthoryear{{Navarro}, {Frenk}  \& {White}}{{Navarro}
  et~al.}{1997}]{navarro1997universal}
{Navarro} J.~F.,  {Frenk} C.~S.,   {White} S. D.~M.,  1997, \mn@doi [\apj]
  {10.1086/304888}, \href
  {https://ui.adsabs.harvard.edu/abs/1997ApJ...490..493N} {490, 493}

\bibitem[\protect\citeauthoryear{{Navarro} et~al.,}{{Navarro}
  et~al.}{2004}]{navarro2004inner}
{Navarro} J.~F.,  et~al., 2004, \mn@doi [\mnras]
  {10.1111/j.1365-2966.2004.07586.x}, \href
  {https://ui.adsabs.harvard.edu/abs/2004MNRAS.349.1039N} {349, 1039}

\bibitem[\protect\citeauthoryear{{Navarro} et~al.,}{{Navarro}
  et~al.}{2010}]{navarro2010diversity}
{Navarro} J.~F.,  et~al., 2010, \mn@doi [\mnras]
  {10.1111/j.1365-2966.2009.15878.x}, \href
  {https://ui.adsabs.harvard.edu/abs/2010MNRAS.402...21N} {402, 21}

\bibitem[\protect\citeauthoryear{{Nesti} \& {Salucci}}{{Nesti} \&
  {Salucci}}{2013}]{nesti2013dark}
{Nesti} F.,  {Salucci} P.,  2013, \mn@doi [\jcap]
  {10.1088/1475-7516/2013/07/016}, \href
  {https://ui.adsabs.harvard.edu/abs/2013JCAP...07..016N} {2013, 016}

\bibitem[\protect\citeauthoryear{{O{\~n}orbe}, {Garrison-Kimmel}, {Maller},
  {Bullock}, {Rocha}  \& {Hahn}}{{O{\~n}orbe} et~al.}{2014}]{onorbe2013zoom}
{O{\~n}orbe} J.,  {Garrison-Kimmel} S.,  {Maller} A.~H.,  {Bullock} J.~S.,
  {Rocha} M.,   {Hahn} O.,  2014, \mn@doi [\mnras] {10.1093/mnras/stt2020},
  \href {https://ui.adsabs.harvard.edu/abs/2014MNRAS.437.1894O} {437, 1894}

\bibitem[\protect\citeauthoryear{{O{\~n}orbe}, {Boylan-Kolchin}, {Bullock},
  {Hopkins}, {Kere{\v{s}}}, {Faucher-Gigu{\`e}re}, {Quataert}  \&
  {Murray}}{{O{\~n}orbe} et~al.}{2015}]{onorbe2015forged}
{O{\~n}orbe} J.,  {Boylan-Kolchin} M.,  {Bullock} J.~S.,  {Hopkins} P.~F.,
  {Kere{\v{s}}} D.,  {Faucher-Gigu{\`e}re} C.-A.,  {Quataert} E.,   {Murray}
  N.,  2015, \mn@doi [\mnras] {10.1093/mnras/stv2072}, \href
  {https://ui.adsabs.harvard.edu/abs/2015MNRAS.454.2092O} {454, 2092}

\bibitem[\protect\citeauthoryear{{Oh}, {Brook}, {Governato}, {Brinks}, {Mayer},
  {de Blok}, {Brooks}  \& {Walter}}{{Oh} et~al.}{2011}]{oh2011central}
{Oh} S.-H.,  {Brook} C.,  {Governato} F.,  {Brinks} E.,  {Mayer} L.,  {de Blok}
  W.~J.~G.,  {Brooks} A.,   {Walter} F.,  2011, \mn@doi [\aj]
  {10.1088/0004-6256/142/1/24}, \href
  {https://ui.adsabs.harvard.edu/abs/2011AJ....142...24O} {142, 24}

\bibitem[\protect\citeauthoryear{{Oh} et~al.,}{{Oh}
  et~al.}{2015}]{oh2015little}
{Oh} S.-H.,  et~al., 2015, \mn@doi [\aj] {10.1088/0004-6256/149/6/180}, \href
  {https://ui.adsabs.harvard.edu/abs/2015AJ....149..180O} {149, 180}

\bibitem[\protect\citeauthoryear{{Oliphant}}{{Oliphant}}{2007}]{oliphant2007python}
{Oliphant} T.~E.,  2007, \mn@doi [Computing in Science and Engineering]
  {10.1109/MCSE.2007.58}, \href
  {https://ui.adsabs.harvard.edu/abs/2007CSE.....9c..10O} {9, 10}

\bibitem[\protect\citeauthoryear{{Papastergis}, {Giovanelli}, {Haynes}  \&
  {Shankar}}{{Papastergis} et~al.}{2015}]{papastergis2015too}
{Papastergis} E.,  {Giovanelli} R.,  {Haynes} M.~P.,   {Shankar} F.,  2015,
  \mn@doi [\aap] {10.1051/0004-6361/201424909}, \href
  {https://ui.adsabs.harvard.edu/abs/2015A&A...574A.113P} {574, A113}

\bibitem[\protect\citeauthoryear{{Pe{\~n}arrubia}, {Pontzen}, {Walker}  \&
  {Koposov}}{{Pe{\~n}arrubia} et~al.}{2012}]{penarrubia2012coupling}
{Pe{\~n}arrubia} J.,  {Pontzen} A.,  {Walker} M.~G.,   {Koposov} S.~E.,  2012,
  \mn@doi [\apjl] {10.1088/2041-8205/759/2/L42}, \href
  {https://ui.adsabs.harvard.edu/abs/2012ApJ...759L..42P} {759, L42}

\bibitem[\protect\citeauthoryear{{Planck Collaboration} et~al.,}{{Planck
  Collaboration} et~al.}{2016}]{ade2016planck}
{Planck Collaboration} et~al., 2016, \mn@doi [\aap]
  {10.1051/0004-6361/201525830}, \href
  {https://ui.adsabs.harvard.edu/abs/2016A&A...594A..13P} {594, A13}

\bibitem[\protect\citeauthoryear{{Pontzen} \& {Governato}}{{Pontzen} \&
  {Governato}}{2012}]{pontzen2012supernova}
{Pontzen} A.,  {Governato} F.,  2012, \mn@doi [\mnras]
  {10.1111/j.1365-2966.2012.20571.x}, \href
  {https://ui.adsabs.harvard.edu/abs/2012MNRAS.421.3464P} {421, 3464}

\bibitem[\protect\citeauthoryear{{Portail}, {Gerhard}, {Wegg}  \&
  {Ness}}{{Portail} et~al.}{2017}]{portail2017dynamical}
{Portail} M.,  {Gerhard} O.,  {Wegg} C.,   {Ness} M.,  2017, \mn@doi [\mnras]
  {10.1093/mnras/stw2819}, \href
  {https://ui.adsabs.harvard.edu/abs/2017MNRAS.465.1621P} {465, 1621}

\bibitem[\protect\citeauthoryear{{Power}, {Navarro}, {Jenkins}, {Frenk},
  {White}, {Springel}, {Stadel}  \& {Quinn}}{{Power}
  et~al.}{2003}]{power2003inner}
{Power} C.,  {Navarro} J.~F.,  {Jenkins} A.,  {Frenk} C.~S.,  {White} S.~D.~M.,
   {Springel} V.,  {Stadel} J.,   {Quinn} T.,  2003, \mn@doi [\mnras]
  {10.1046/j.1365-8711.2003.05925.x}, \href
  {https://ui.adsabs.harvard.edu/abs/2003MNRAS.338...14P} {338, 14}

\bibitem[\protect\citeauthoryear{{Prada}, {Klypin}, {Simonneau},
  {Betancort-Rijo}, {Patiri}, {Gottl{\"o}ber}  \& {Sanchez-Conde}}{{Prada}
  et~al.}{2006}]{prada2006far}
{Prada} F.,  {Klypin} A.~A.,  {Simonneau} E.,  {Betancort-Rijo} J.,  {Patiri}
  S.,  {Gottl{\"o}ber} S.,   {Sanchez-Conde} M.~A.,  2006, \mn@doi [\apj]
  {10.1086/504456}, \href
  {https://ui.adsabs.harvard.edu/abs/2006ApJ...645.1001P} {645, 1001}

\bibitem[\protect\citeauthoryear{{Price} \& {Monaghan}}{{Price} \&
  {Monaghan}}{2007}]{price2007energy}
{Price} D.~J.,  {Monaghan} J.~J.,  2007, \mn@doi [\mnras]
  {10.1111/j.1365-2966.2006.11241.x}, \href
  {https://ui.adsabs.harvard.edu/abs/2007MNRAS.374.1347P} {374, 1347}

\bibitem[\protect\citeauthoryear{{Read} \& {Gilmore}}{{Read} \&
  {Gilmore}}{2005}]{read2005shallow}
{Read} J.~I.,  {Gilmore} G.,  2005, \mn@doi [\mnras]
  {10.1111/j.1365-2966.2004.08424.x}, \href
  {https://ui.adsabs.harvard.edu/abs/2005MNRAS.356..107R} {356, 107}

\bibitem[\protect\citeauthoryear{{Read}, {Agertz}  \& {Collins}}{{Read}
  et~al.}{2016}]{read2016cNFW}
{Read} J.~I.,  {Agertz} O.,   {Collins} M.~L.~M.,  2016, \mn@doi [\mnras]
  {10.1093/mnras/stw713}, \href
  {https://ui.adsabs.harvard.edu/abs/2016MNRAS.459.2573R} {459, 2573}

\bibitem[\protect\citeauthoryear{{Relatores} et~al.,}{{Relatores}
  et~al.}{2019}]{relatores2019inner}
{Relatores} N.~C.,  et~al., 2019, \mn@doi [\apj] {10.3847/1538-4357/ab5305},
  \href {https://ui.adsabs.harvard.edu/abs/2019ApJ...887...94R} {887, 94}

\bibitem[\protect\citeauthoryear{{Ren}, {Kwa}, {Kaplinghat}  \& {Yu}}{{Ren}
  et~al.}{2019}]{ren2019diversity}
{Ren} T.,  {Kwa} A.,  {Kaplinghat} M.,   {Yu} H.-B.,  2019, \mn@doi [Physical
  Review X] {10.1103/PhysRevX.9.031020}, \href
  {https://ui.adsabs.harvard.edu/abs/2019PhRvX...9c1020R} {9, 031020}

\bibitem[\protect\citeauthoryear{{Robles}, {Bullock}  \&
  {Boylan-Kolchin}}{{Robles} et~al.}{2019}]{robles2019sfdm}
{Robles} V.~H.,  {Bullock} J.~S.,   {Boylan-Kolchin} M.,  2019, \mn@doi
  [\mnras] {10.1093/mnras/sty3190}, \href
  {https://ui.adsabs.harvard.edu/abs/2019MNRAS.483..289R} {483, 289}

\bibitem[\protect\citeauthoryear{{Rocha}, {Peter}, {Bullock}, {Kaplinghat},
  {Garrison-Kimmel}, {O{\~n}orbe}  \& {Moustakas}}{{Rocha}
  et~al.}{2013}]{rocha2013sidm}
{Rocha} M.,  {Peter} A. H.~G.,  {Bullock} J.~S.,  {Kaplinghat} M.,
  {Garrison-Kimmel} S.,  {O{\~n}orbe} J.,   {Moustakas} L.~A.,  2013, \mn@doi
  [\mnras] {10.1093/mnras/sts514}, \href
  {https://ui.adsabs.harvard.edu/abs/2013MNRAS.430...81R} {430, 81}

\bibitem[\protect\citeauthoryear{{Romano-D{\'\i}az}, {Shlosman}, {Hoffman}  \&
  {Heller}}{{Romano-D{\'\i}az} et~al.}{2008}]{romano2008erasing}
{Romano-D{\'\i}az} E.,  {Shlosman} I.,  {Hoffman} Y.,   {Heller} C.,  2008,
  \mn@doi [\apjl] {10.1086/592687}, \href
  {https://ui.adsabs.harvard.edu/abs/2008ApJ...685L.105R} {685, L105}

\bibitem[\protect\citeauthoryear{{Salucci} \& {Burkert}}{{Salucci} \&
  {Burkert}}{2000}]{salucci2000dark}
{Salucci} P.,  {Burkert} A.,  2000, \mn@doi [\apjl] {10.1086/312747}, \href
  {https://ui.adsabs.harvard.edu/abs/2000ApJ...537L...9S} {537, L9}

\bibitem[\protect\citeauthoryear{{Samuel} et~al.,}{{Samuel}
  et~al.}{2020}]{samuel2020profile}
{Samuel} J.,  et~al., 2020, \mn@doi [\mnras] {10.1093/mnras/stz3054}, \href
  {https://ui.adsabs.harvard.edu/abs/2020MNRAS.491.1471S} {491, 1471}

\bibitem[\protect\citeauthoryear{{Spekkens}, {Giovanelli}  \&
  {Haynes}}{{Spekkens} et~al.}{2005}]{spekkens2005cusp}
{Spekkens} K.,  {Giovanelli} R.,   {Haynes} M.~P.,  2005, \mn@doi [\aj]
  {10.1086/429592}, \href
  {https://ui.adsabs.harvard.edu/abs/2005AJ....129.2119S} {129, 2119}

\bibitem[\protect\citeauthoryear{{Springel}}{{Springel}}{2005}]{springel2005cosmological}
{Springel} V.,  2005, \mn@doi [\mnras] {10.1111/j.1365-2966.2005.09655.x},
  \href {https://ui.adsabs.harvard.edu/abs/2005MNRAS.364.1105S} {364, 1105}

\bibitem[\protect\citeauthoryear{{Srisawat} et~al.,}{{Srisawat}
  et~al.}{2013}]{srisawat2013merger}
{Srisawat} C.,  et~al., 2013, \mn@doi [\mnras] {10.1093/mnras/stt1545}, \href
  {https://ui.adsabs.harvard.edu/abs/2013MNRAS.436..150S} {436, 150}

\bibitem[\protect\citeauthoryear{{Stinson}, {Bailin}, {Couchman}, {Wadsley},
  {Shen}, {Nickerson}, {Brook}  \& {Quinn}}{{Stinson}
  et~al.}{2010}]{stinson2010galaxy}
{Stinson} G.~S.,  {Bailin} J.,  {Couchman} H.,  {Wadsley} J.,  {Shen} S.,
  {Nickerson} S.,  {Brook} C.,   {Quinn} T.,  2010, \mn@doi [\mnras]
  {10.1111/j.1365-2966.2010.17187.x}, \href
  {https://ui.adsabs.harvard.edu/abs/2010MNRAS.408..812S} {408, 812}

\bibitem[\protect\citeauthoryear{{Stinson} et~al.,}{{Stinson}
  et~al.}{2012}]{stinson2012magicc}
{Stinson} G.~S.,  et~al., 2012, \mn@doi [\mnras]
  {10.1111/j.1365-2966.2012.21522.x}, \href
  {https://ui.adsabs.harvard.edu/abs/2012MNRAS.425.1270S} {425, 1270}

\bibitem[\protect\citeauthoryear{{Swaters}, {Madore}, {van den Bosch}  \&
  {Balcells}}{{Swaters} et~al.}{2003}]{swaters2003central}
{Swaters} R.~A.,  {Madore} B.~F.,  {van den Bosch} F.~C.,   {Balcells} M.,
  2003, \mn@doi [\apj] {10.1086/345426}, \href
  {https://ui.adsabs.harvard.edu/abs/2003ApJ...583..732S} {583, 732}

\bibitem[\protect\citeauthoryear{{Teyssier}, {Pontzen}, {Dubois}  \&
  {Read}}{{Teyssier} et~al.}{2013}]{teyssier2013cusp}
{Teyssier} R.,  {Pontzen} A.,  {Dubois} Y.,   {Read} J.~I.,  2013, \mn@doi
  [\mnras] {10.1093/mnras/sts563}, \href
  {https://ui.adsabs.harvard.edu/abs/2013MNRAS.429.3068T} {429, 3068}

\bibitem[\protect\citeauthoryear{{Tollerud}, {Boylan-Kolchin}  \&
  {Bullock}}{{Tollerud} et~al.}{2014}]{tollerud2014}
{Tollerud} E.~J.,  {Boylan-Kolchin} M.,   {Bullock} J.~S.,  2014, \mn@doi
  [\mnras] {10.1093/mnras/stu474}, \href
  {https://ui.adsabs.harvard.edu/abs/2014MNRAS.440.3511T} {440, 3511}

\bibitem[\protect\citeauthoryear{{Tollet} et~al.,}{{Tollet}
  et~al.}{2016}]{tollet2016nihao}
{Tollet} E.,  et~al., 2016, \mn@doi [\mnras] {10.1093/mnras/stv2856}, \href
  {https://ui.adsabs.harvard.edu/abs/2016MNRAS.456.3542T} {456, 3542}

\bibitem[\protect\citeauthoryear{{Tonini}, {Lapi}  \& {Salucci}}{{Tonini}
  et~al.}{2006}]{tonini2006angular}
{Tonini} C.,  {Lapi} A.,   {Salucci} P.,  2006, \mn@doi [\apj]
  {10.1086/506431}, \href
  {https://ui.adsabs.harvard.edu/abs/2006ApJ...649..591T} {649, 591}

\bibitem[\protect\citeauthoryear{{Walter}, {Brinks}, {de Blok}, {Bigiel},
  {Kennicutt}, {Thornley}  \& {Leroy}}{{Walter}
  et~al.}{2008}]{walter2008things}
{Walter} F.,  {Brinks} E.,  {de Blok} W.~J.~G.,  {Bigiel} F.,  {Kennicutt}
  Robert~C. J.,  {Thornley} M.~D.,   {Leroy} A.,  2008, \mn@doi [\aj]
  {10.1088/0004-6256/136/6/2563}, \href
  {https://ui.adsabs.harvard.edu/abs/2008AJ....136.2563W} {136, 2563}

\bibitem[\protect\citeauthoryear{{Wang}, {Dutton}, {Stinson}, {Macci{\`o}},
  {Penzo}, {Kang}, {Keller}  \& {Wadsley}}{{Wang}
  et~al.}{2015}]{wang2015nihao1}
{Wang} L.,  {Dutton} A.~A.,  {Stinson} G.~S.,  {Macci{\`o}} A.~V.,  {Penzo} C.,
   {Kang} X.,  {Keller} B.~W.,   {Wadsley} J.,  2015, \mn@doi [\mnras]
  {10.1093/mnras/stv1937}, \href
  {https://ui.adsabs.harvard.edu/abs/2015MNRAS.454...83W} {454, 83}

\bibitem[\protect\citeauthoryear{{Wang}, {Bose}, {Frenk}, {Gao}, {Jenkins},
  {Springel}  \& {White}}{{Wang} et~al.}{2019}]{wang2019zoom}
{Wang} J.,  {Bose} S.,  {Frenk} C.~S.,  {Gao} L.,  {Jenkins} A.,  {Springel}
  V.,   {White} S. D.~M.,  2019, arXiv e-prints, \href
  {https://ui.adsabs.harvard.edu/abs/2019arXiv191109720W} {p. arXiv:1911.09720}

\bibitem[\protect\citeauthoryear{{Wetzel}, {Hopkins}, {Kim},
  {Faucher-Gigu{\`e}re}, {Kere{\v{s}}}  \& {Quataert}}{{Wetzel}
  et~al.}{2016}]{wetzel2016reconciling}
{Wetzel} A.~R.,  {Hopkins} P.~F.,  {Kim} J.-h.,  {Faucher-Gigu{\`e}re} C.-A.,
  {Kere{\v{s}}} D.,   {Quataert} E.,  2016, \mn@doi [\apjl]
  {10.3847/2041-8205/827/2/L23}, \href
  {https://ui.adsabs.harvard.edu/abs/2016ApJ...827L..23W} {827, L23}

\bibitem[\protect\citeauthoryear{{Wheeler} et~al.,}{{Wheeler}
  et~al.}{2019}]{wheeler2019resolved}
{Wheeler} C.,  et~al., 2019, \mn@doi [\mnras] {10.1093/mnras/stz2887}, \href
  {https://ui.adsabs.harvard.edu/abs/2019MNRAS.490.4447W} {490, 4447}

\bibitem[\protect\citeauthoryear{{Zhao}}{{Zhao}}{1996}]{zhao1996models}
{Zhao} H.,  1996, \mn@doi [\mnras] {10.1093/mnras/278.2.488}, \href
  {https://ui.adsabs.harvard.edu/abs/1996MNRAS.278..488Z} {278, 488}

\bibitem[\protect\citeauthoryear{{van der Walt}, {Colbert}  \&
  {Varoquaux}}{{van der Walt} et~al.}{2011}]{van2011numpy}
{van der Walt} S.,  {Colbert} S.~C.,   {Varoquaux} G.,  2011, \mn@doi
  [Computing in Science and Engineering] {10.1109/MCSE.2011.37}, \href
  {https://ui.adsabs.harvard.edu/abs/2011CSE....13b..22V} {13, 22}

\makeatother
\end{thebibliography}

\appendix
\section{Stellar Mass Parameterization of the Core-Einasto}
\label{sec:cEinasto.Mstar}
\begin{figure*}
    \centering
    \includegraphics[width=\columnwidth]{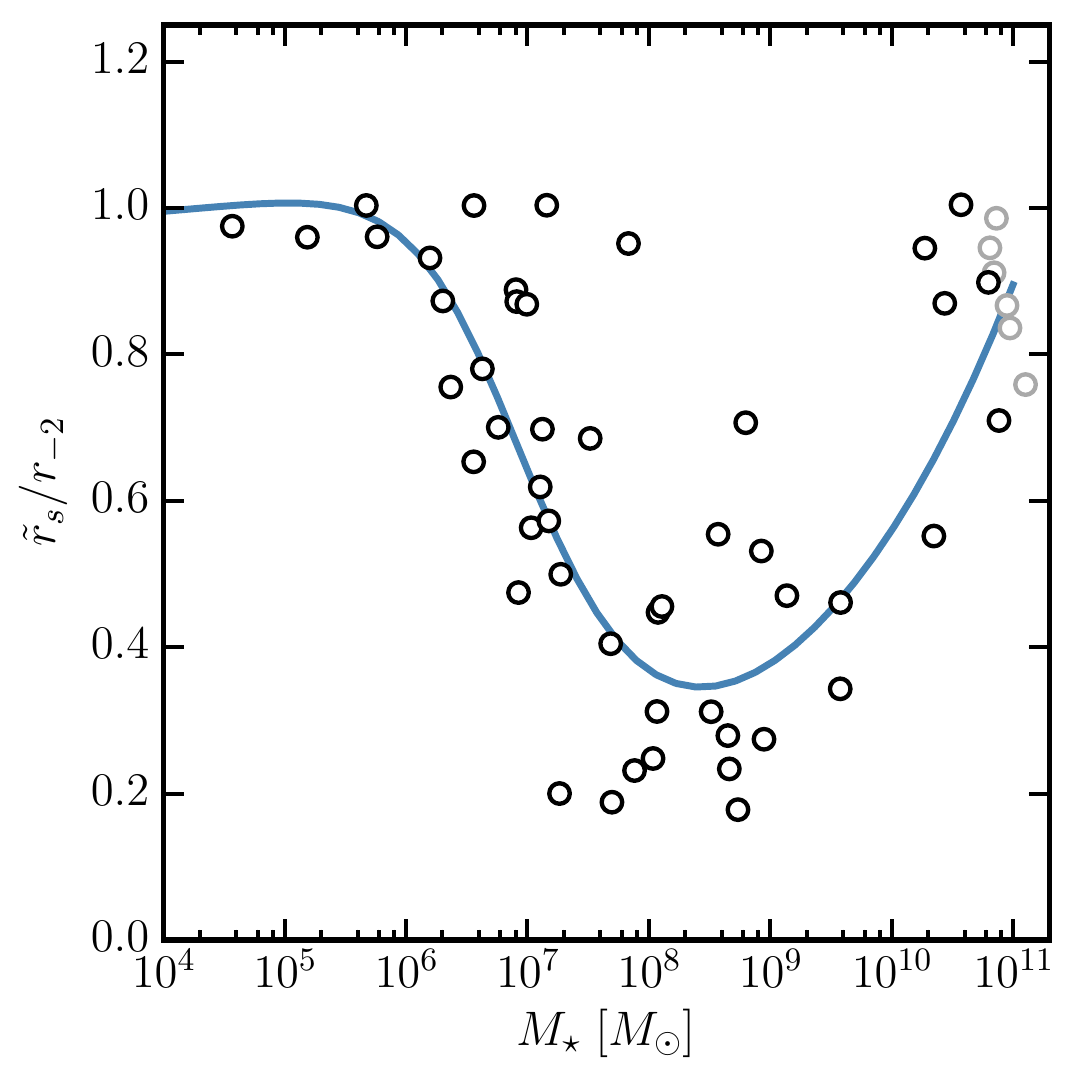}
    \includegraphics[width=\columnwidth]{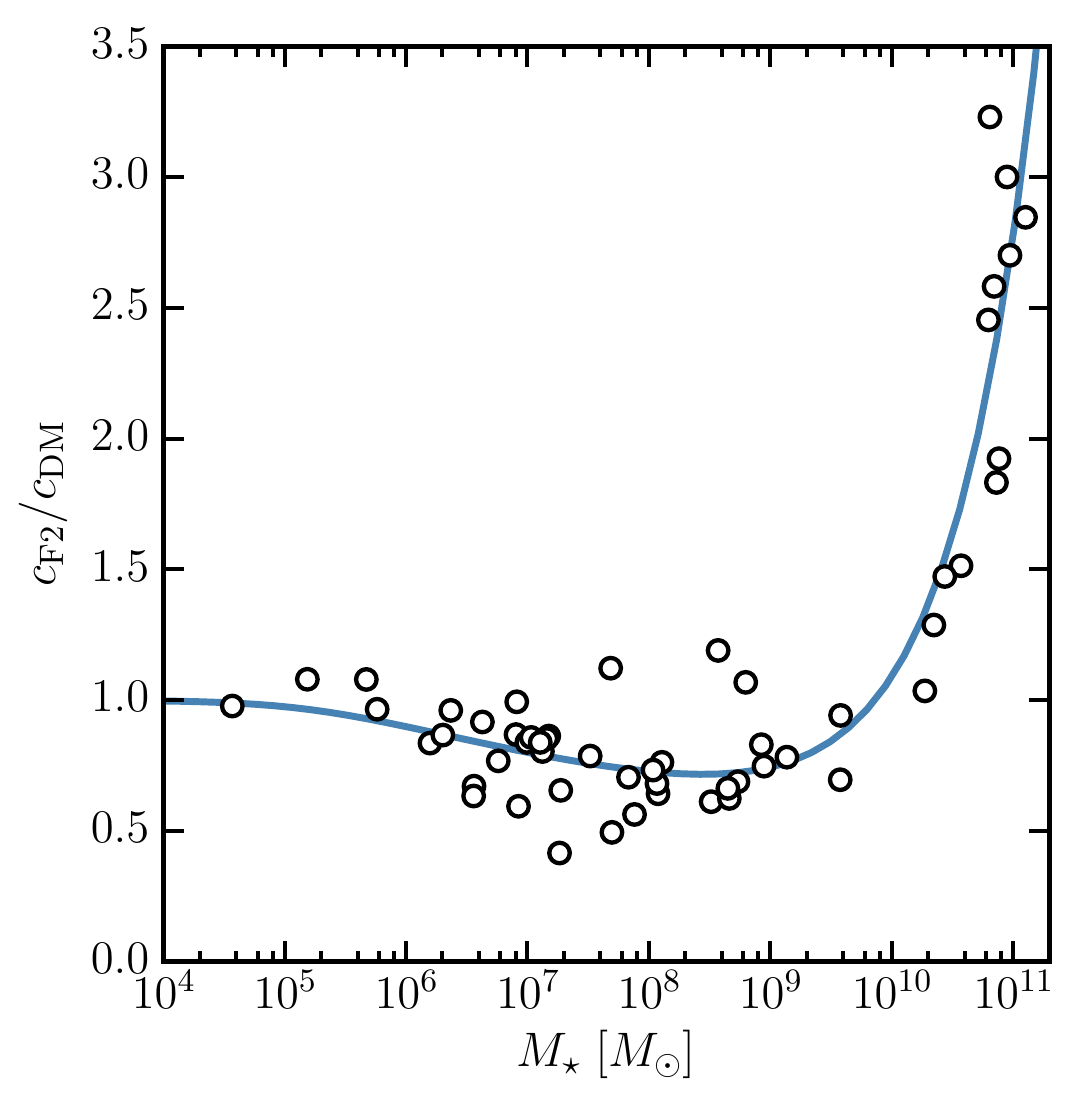}
    \caption{
        Halo concentration as a function of $M_{\star}$. 
        {\bf \em Left}:
        The ratio of the scale parameter $\tilde{r}_{s}$ to $r_{-2}$ of the galaxy dark matter profile shows similar trends seen in earlier Fig.~\ref{fig:9}. The largest difference again correlates with the largest cores at $M_{\star} = 10^{8-9}\ M_{\odot}$. The solid blue curve depicts the fit of the points using the best-fit parameters given in Table~\ref{tab:rsr2.fit} when using Eq.~\protect\eqref{eq:rs.fit} for $x=M_{\star}$.
        {\bf \em Right}:
        The comparison between concentrations as a function of stellar mass. The reduction of concentration is apparent for a large span of $M_{\star}$. The solid blue curve depicts the fit of points using best-fit parameters from Table~\ref{tab:conc.fit} when using  Eq.~\protect\eqref{eq:conc.fit} for $x=M_{\star}$.
    }
    \label{fig:A1}
\end{figure*}

The analysis presented in Section~\ref{sec:cEinasto} focused on properties recovered by the core-Einasto profile and then characterizing these trends with the $M_{\star}/M_{\rm halo}$ of the simulated FIRE-2 halos. Here, we perform our analysis now on the stellar mass of the galaxies, $M_{\star}$, as this can provide deeper insight to observations of real galaxies comparable to the galaxies analyzed in this article.  

The left plot in Fig.~\ref{fig:A1} depicts the relation of $\tilde{r}_{s}$ to $r_{-2}$ of the galaxies' dark matter profile as a function of $M_{\star}$. We find quite a bit of difference between this implied relationship and the relationship seen previously in Fig.~\ref{fig:8}. Primarily, the values of $\tilde{r}_{s}/r_{-2}$ are more spread out for the ranges of $M_{\star}$ considered here. This is better seen with fitting the data with Eq.~\eqref{eq:rs.fit}. Best fit results are given in Table~\ref{tab:rsr2.fit} and are shown as the blue curve in the left plot. The right plot of Fig.~\ref{fig:A1} shows the ratio between the concentrations of the halos for the galaxies and the DMO analogs. We consider the same definition of the concentration discussed previously in Fig.~\ref{fig:9}. The depletion in concentration spans from $M_{\star} \simeq 10^{6} - 10^{9}\ M_{\odot}$, the most prominent being at  $M_{\star} \simeq 10^{7-8}\ M_{\odot}$. The points are fitted with Eq.~\eqref{eq:conc.fit} with the best fit results, given in Table~\ref{tab:rcore.fit}, are shown as the blue curve in the right plot.

\section{Analytical Properties of core-Einasto Halos}
\label{sec:analytic.cEinasto}
Here we derive formulae in the form concerning the spatial properties of dark matter halos described by Eq.~\eqref{eq:cored.einasto.density}. In the limit of $r_{c} \ \rightarrow 0$, profiles should transform back to a cusped form, i.e., $\rho_{\rm cEin} \rightarrow \rho_{\rm Ein}$.

\subsection{Cumulative mass distribution}
For a spherical averaged volume, the cumulative mass is 
\begin{align}
    M(<r) 
    &= 4\pi \tilde{\rho}_{s} \int_{0}^{r} dr' r'^{2} \exp \Bigg\{-\frac{2}{\alpha_{\epsilon}} \left[\left( \frac{r'+r_{c}}{\tilde{r}_{s}} \right)^{\alpha_{\epsilon}} - 1\right]\Bigg\}
    \, .
\end{align}
Let us set $s = 2(r+r_{c})^{\alpha_{\epsilon}}/\alpha_{\epsilon} \tilde{r}_{s}^{\alpha_{\epsilon}}$, such that algebraically massaging gives us $r = s^{1/\alpha_{\epsilon}} \left(\alpha_{\epsilon}/2\right)^{1/\alpha_{\epsilon}} \tilde{r}_{s} - r_{c}$. When substituting this into the cumulative mass expression, we have the expanded form of
\begin{align}
    M(<r) 
    =
    \frac{4\pi \tilde{\rho}_{s} e^{2/\alpha_{\epsilon}}}{\alpha_{\epsilon}}
    &
    \Bigg\{
    \tilde{r}_{s}^{3}\left(\frac{\alpha_{\epsilon}}{2}\right)^{3/\alpha_{\epsilon}} \int_{s(0)}^{s(r)} ds\ s^{3/\alpha_{\epsilon} - 1} e^{-s} 
    \\ \nonumber
    & +\
    r_{c}^{2}\tilde{r}_{s} \left(\frac{\alpha_{\epsilon}}{2}\right)^{1/\alpha_{\epsilon}}\int_{s(0)}^{s(r)} ds\ s^{1/\alpha_{\epsilon} - 1} e^{-s}
    \\ \nonumber
    & -\
    2r_{c}\tilde{r}_{s}^{2}\left(\frac{\alpha_{\epsilon}}{2}\right)^{2/\alpha_{\epsilon}} \int_{s(0)}^{s(r)} ds\ s^{2/\alpha_{\epsilon} - 1} e^{-s}
    \Bigg\}
    \, .
\end{align}
We can define the integral parametrization as
\begin{align}
    \tilde{\gamma}_{\beta}[x_{1},x_{2}]
    :=
    \left(\frac{\alpha_{\epsilon} \tilde{r}_{s}^{\alpha_{\epsilon}}}{2}\right)^{\beta} \gamma_{\beta} [x_{1},x_{2}]
    \, ,
\end{align}
which is a characterization variant of the lower incomplete gamma function:
\begin{align}
    \gamma_{\beta}[x_{1},x_{2}]
    =
    \int_{x_{1}}^{x_{2}} ds\ s^{\beta - 1}e^{-s}
    \, .
\end{align}
This allows us to write the expression for the integrated mass in a more compact form
\begin{align}
    M(<r) 
    =
    \frac{4\pi \tilde{\rho}_{s} e^{2/\alpha_{\epsilon}}}{\alpha_{\epsilon}}
    &
    \Bigg\{
    \tilde{\gamma}_{3/\alpha_{\epsilon}}\left[s(0)\ ,s(r)\right]\
    +\
    r_{c}^{2}\tilde{\gamma}_{1/\alpha_{\epsilon}}\left[s(0)\ ,s(r)\right]
    \nonumber \\
    &\quad -\
    2r_{c}\tilde{\gamma}_{2/\alpha_{\epsilon}}\left[s(0)\ ,s(r)\right]
    \Bigg\}
    \, .
    \label{eq:cored.einasto.integrated.mass}
\end{align}
In the limit of $r_{c}\rightarrow 0$, we return back to the analytic form of the cumulative mass for the Einasto profile
\begin{align}
    M_{\rm Ein}(<r) 
    &=
    \lim_{r_{c}\rightarrow 0} M_{\rm cEin}(<r) 
    \nonumber \\
    &=
    \frac{4\pi \tilde{\rho}_{s} e^{2/\alpha_{\epsilon}}\tilde{r}_{s}^{3}}{\alpha_{\epsilon}}
    \left(\frac{\alpha_{\epsilon}}{2}\right)^{3/\alpha_{\epsilon}} 
    \gamma_{3/\alpha_{\epsilon}}\left[0\ ,s(r)\right]
    \label{eq:einasto.integrated.mass}
    \, ,
\end{align}
where we then retrieve the lower incomplete gamma function in this limit
\begin{align}
    \gamma_{\beta}\left[0,x\right]
    &=
    \int_{0}^{x} ds\ s^{\beta - 1} e^{-s} 
    \, .
\end{align}

\subsection{Gravitational potential}
The gravitational potential of a spherically symmetric mass distribution,  $\rho(r)$, can be found through the expression \citep{binney2011galactic},
\begin{align}
    \Psi(r)
    &=
    4\pi G 
    \left[
    \int_{0}^{r}dr'\ r'^{2} \rho(r') 
    + \int_{r}^{\infty} dr'\ r' \rho(r')
    \right]
    \, .
\end{align}
It follows for the cored-Einasto, 
\begin{align}
    \Psi(r)
    &=
    \frac{4\pi G \tilde{\rho}_{s} e^{2/\alpha_{\epsilon}}}{\alpha_{\epsilon}}
    \Bigg\{
    \frac{1}{r}
    \Big(
    \tilde{\gamma}_{3/\alpha_{\epsilon}}\left[s(0)\ ,s(r)\right]
    \\
    &\quad +
    r_{c}^{2}\tilde{\gamma}_{1/\alpha_{\epsilon}}\left[s(0)\ ,s(r)\right]
    -
    2r_{c}\tilde{\gamma}_{2/\alpha_{\epsilon}}\left[s(0)\ ,s(r) \right]
    \Big)
    \nonumber \\
    &\quad +
    \tilde{\Gamma}_{2/\alpha_{\epsilon}}\left[s(r)\right]
    -
    r_{c} \tilde{\Gamma}_{1/\alpha_{\epsilon}}\left[s(r)\right]
    \Bigg\}
    \, ,
\end{align}
where we have defined 
\begin{align}
    \tilde{\Gamma}_{\beta}[s(r)]
    =
    \left(\frac{\alpha_{\epsilon} \tilde{r}_{s}^{\alpha_{\epsilon}}}{2}\right)^{\beta} \Gamma_{\beta}\left[s(r)\right]
    \, ,
\end{align}
such that
\begin{align}
    \Gamma_{\beta}[x]
    =
    \int_{x}^{\infty} ds\ s^{\beta-1}s^{-s}
\end{align}
is the upper incomplete Gamma function.

\subsection{Energy of induced core formation}
The transformation from a cusp inner region to a core is presumed to be from highly energetic stellar feedback. After the dark matter cusp is removed we would infer that the halo settles in a new equilibrium state. Dark matter in dynamical equilibrium will then satisfy the virial theorem, i.e. $E=\mathcal{W}/2$. Here, $\mathcal{W}$ is the magnitude of the gravitational potential energy associated with the mass distribution:
\begin{align}
    \mathcal{W}
    &= 
    -\int_{0}^{r_{\rm vir}} dr'
    \frac{G M(<r')}{r'} 4\pi r'^{2} \rho(r') 
    \, .
\end{align}
For the core-Einasto, the gravitational energy is
\begin{align}
    \mathcal{W}_{\rm cEin}
    &=
    -\Bigg(\frac{16\pi^{2} G^{2} \tilde{\rho}_{s}^{2} e^{4/\alpha_{\epsilon}}}{\alpha_{\epsilon}}
    \Bigg)
    \int_{0}^{r_{\rm vir}} dr'
    e^{-s(r')}
    \times
    \\
    &
    \Bigg\{ 
    \tilde{\gamma}_{3/\alpha_{\epsilon}}\left[s(0)\ ,s(r')\right]  
    +\ 
    r_{c}^{2} \tilde{\gamma}_{1/\alpha_{\epsilon}}\left[s(0)\ ,s(r')\right]
    \nonumber \\
    &\qquad -
    2 r_{c} \tilde{\gamma}_{2/\alpha_{\epsilon}}\left[s(0)\ ,s(r')\right] \Bigg\}
    \nonumber
    \, ,
\end{align}
while for the cusp nature, the Einasto profile has
\begin{align}
    \mathcal{W}_{\rm Ein}
    &=
    -\Bigg(\frac{16\pi^{2} G^{2} \rho_{-2}^{2} e^{4/\alpha_{\epsilon}}}{\alpha_{\epsilon}}\Bigg)\
     \int_{0}^{r_{\rm vir}} dr'
    \exp\Bigg[
    \frac{2}{\alpha_{\epsilon}}\left(\frac{r'}{r_{-2}}\right)^{\alpha_{\epsilon}}\Bigg] 
    \nonumber \\
    &\qquad
    \times \left(\frac{2r_{-2}^{2}}{\alpha_{\epsilon}}\right)^{3/\alpha_{\epsilon}} \gamma_{3/\alpha_{\epsilon}}\left[0\ ,\frac{2}{\alpha_{\epsilon}} \left(\frac{r'}{r_{-2}}\right)^{\alpha_{\epsilon}}\right]  
    \, .
\end{align}
Analytically, we can then quantify a conservative limit for the lower bound of energy needed to transform the inner density via the virial theorem, i.e.,
\begin{align}
    \Delta E 
    &=
    \frac{\Delta \mathcal{W}}{2} = \frac{\mathcal{W}_{\rm cEin} - \mathcal{W}_{\rm Ein}}{2}
    \, .
\end{align}

\section{A Profile For Baryonic Contracted Halos}
\label{sec:cEinasto.BC}
\begin{figure*}
    \centering
    \includegraphics[width=\textwidth]{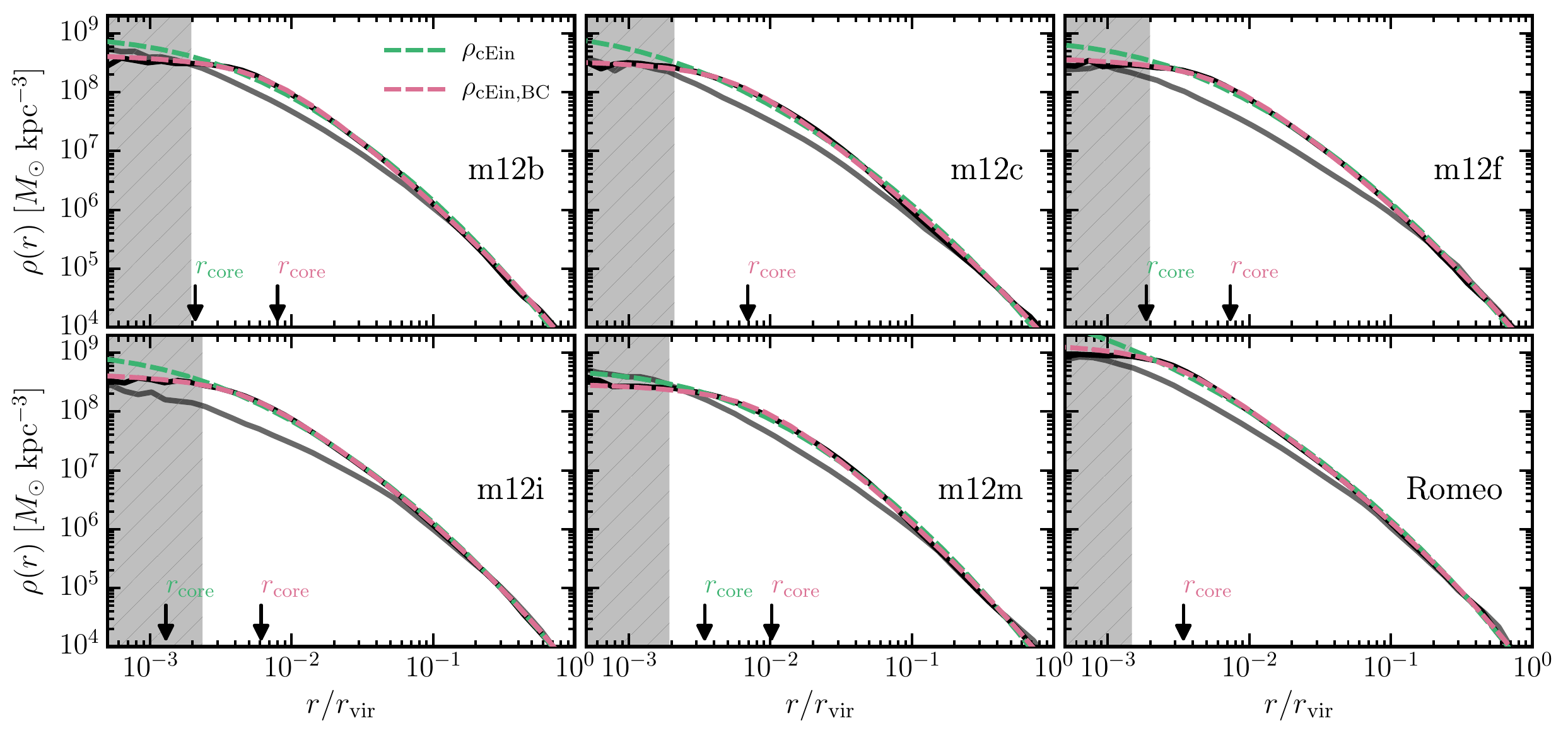}
    \caption{ ---
        {\bf \emph{Refined profiles for cored MW-mass halos}}. 
        As in Fig.~\ref{fig:4}, galaxies are shown as solid black curves while their DMO analogs are solid grey curves. The original $\rho_{\rm cEin}$ fits are plotted as the green dashed curves while $\rho_{\rm cEin,BC}$ fits are plotted as the pink dashed curves. The location of the resulting core radius of each galaxy from both fits is indicate by an arrow with the corresponding color. We see that a radially dependent density component in $\rho_{\rm cEin,BC}$ greatly improves the fits while also accurately predicting the core radius. 
    }
    \label{fig:C1}
\end{figure*}

A major focus of this work is that Eq.~\eqref{eq:cored.einasto.density}, $\rho_{\rm cEin}$, characterizes dark matter profiles with dark matter cores. While a majority of the dwarf galaxies in our sample are well described by $\rho_{\rm cEin}$, a majority of our MW-mass halos (not including m12w, m12z, Louise, and Thelma) are not well fitted by this profile given the inaccurate results of $r_{c}$. This seems to happen for MW-mass halos that have small cores garnished with baryonic contraction to their dark matter distribution in the innermost regions. This motivates us to come up with a profile that accommodates both of these features in galaxies that are this massive.

We would guess that the amplitude of a baryonic-contracted halo has the density amplitude be radially dependent:
\begin{align}
    \tilde{\rho}_{s,\rm BC}(r) = \tilde{\rho}_{s} 
    \Big[ 1 + X \cdot \tanh \Big(\frac{r_{c}}{r} \Big) \Big]
    \, ,
\end{align}
which contributes to the profile at small radii. Here, $X$ is some free variable in the fit that is added to compensate for unusual amplitudes in several of the MW-mass halos. This is written in a way such that at
$r_{c} = 0$, we only have have $\tilde{\rho}_{s,\rm BC} = \tilde{\rho}_{s} = \rho_{-2}$, and at 
$r=0$, we have $\tilde{\rho}_{s, \rm BC} = \tilde{\rho_{s}}(1 + X)$. It would then
\begin{align}
    \rho_{\rm cEin,\rm BC}(r)
    &=
    \tilde{\rho}_{s,\rm BC}(r)
    \times
    \exp \Bigg\{
    - \frac{2}{\alpha_{\epsilon}} \left[ \left(\frac{r + r _{c}}{\tilde{r}_{s}}\right)^{\alpha_{\epsilon}} - 1 \right]
    \Bigg\}
    \, .
    \label{eq:cored.einasto.density.BC}
\end{align}
Additionally, this allows us to parameterize the central core density similar to Eq.~\eqref{eq:cored.einasto.central.density}:
\begin{align}
    \rho_{0,\rm BC} := \rho_{\rm cEin,BC}(0) 
    &= \Big[ 1+X \Big] \rho_{0}
    \, .
\end{align}


\begin{table}
    \centering
    \setlength{\tabcolsep}{8.25pt}
    \captionsetup{justification=centering}
    \caption{Best-fit parameters for Milky Way-mass halos. 
    }
    \begin{tabular}{SSSSSSS} 
    \toprule
    \toprule
    {Halo} &
    {$\tilde{\rho}_{s}$} & 
    {$\tilde{r}_{s}$} & 
    {$X$} & 
    {$r_{c}$} & 
    {$Q_{\rm min}$}
    \\
    {Name} &
    {$[M_{\odot}\ \rm kpc^{-3}]$} & 
    {$[\rm kpc]$} & 
    {} & 
    {$[\rm kpc]$} & 
    \\ 
    \midrule
    {m12b} &
    {$1.1 \times 10^{6}$} & 
    {21.2} & 
    {5.44} & 
    {$^\text{\cmark}$1.77} &
    {0.0236}
    \\
    {m12c} &
    {$4.4 \times 10^{5}$} & 
    {31.5} & 
    {5.50} & 
    {$^\text{\cmark}$1.52} & 
    {0.0349}
    \\
    {m12f} &
    {$1.3 \times 10^{6}$} & 
    {21.3} & 
    {3.73} & 
    {$^\text{\cmark}$1.73} & 
    {0.0230}
    \\
    {m12i} &
    {$2.0 \times 10^{6}$} & 
    {16.5} & 
    {2.37} & 
    {$^\text{\cmark}$1.28} & 
    {0.0085}
    \\
    {m12m} &
    {$1.0 \times 10^{6}$} & 
    {22.1} & 
    {5.50} & {$^\text{\cmark}$2.31} & 
    {0.0306}
    \\
    {Romeo} &
    {$2.8\times 10^{6}$} & 
    {15.1} & 
    {3.54} & {$^\text{\cmark}$0.76} & 
    {0.0168}
    \\
    {Juliet} &
    {$1.6\times 10^{6}$} & 
    {17.2} & 
    {3.93} & {$^\text{\cmark}$0.70} & 
    {0.0187}
    \\
    \bottomrule
    \bottomrule
    \vspace{-1ex}
    \\
    \multicolumn{5}{l}{\textbf{Note.} Use Eq. \protect\eqref{eq:cored.einasto.density.BC} with $\alpha_{\epsilon} = 0.16$.}\\
    \end{tabular}
\label{tab:mw.cores.params}
\end{table}


Fig.~\ref{fig:C1} plots the results for fitting $\rho_{\rm cEin,BC}$ (dashed pink curve) to several of the FIRE-2 MW-mass halos (solid black curve). Also plotted is the DMO analog as the gray curve. The value of $r_{c}$ predicted by $\rho_{\rm cEin,BC}$ is highlighted in the same color and pointed to with its $r_{\rm vir}$ normalization. We list our values for these fits in Table \ref{tab:mw.cores.params}. We can see that for MW-mass halos with both baryonic contraction and a physical core, $\rho_{\rm cEin,BC}$, while not particularly succinct, is the most ideal function we can use to probe $r_{c}$. However, the exact behaviour and physical interpretation of $\tilde{r}_{s}$ is left, now, somewhat ambiguous compared to how it was expected to behave previously in Section~\ref{sec:cEinasto}. The same MW-mass halos that have had their core radii previously predicted with $\rho_{\rm cEin}$ are also plotted in Fig.~\ref{fig:C1} as the green dashed curve. The predicted core radius from this profile is pointed to and highlighted in green. From direct comparison between the analytical fits, we see significant improvements. We have included $r_{\rm c}$ values here in the main text as cyan points in Figs.~\ref{fig:7}~and~\ref{fig:8}.

\section{Comparison with other dark matter profiles}
\label{sec:cNFW}
Here we compare several other dark matter profiles in the literature and compare their fits to our FIRE-2 simulation sample:

\begin{figure*}
    \centering
    \includegraphics[width=\columnwidth]{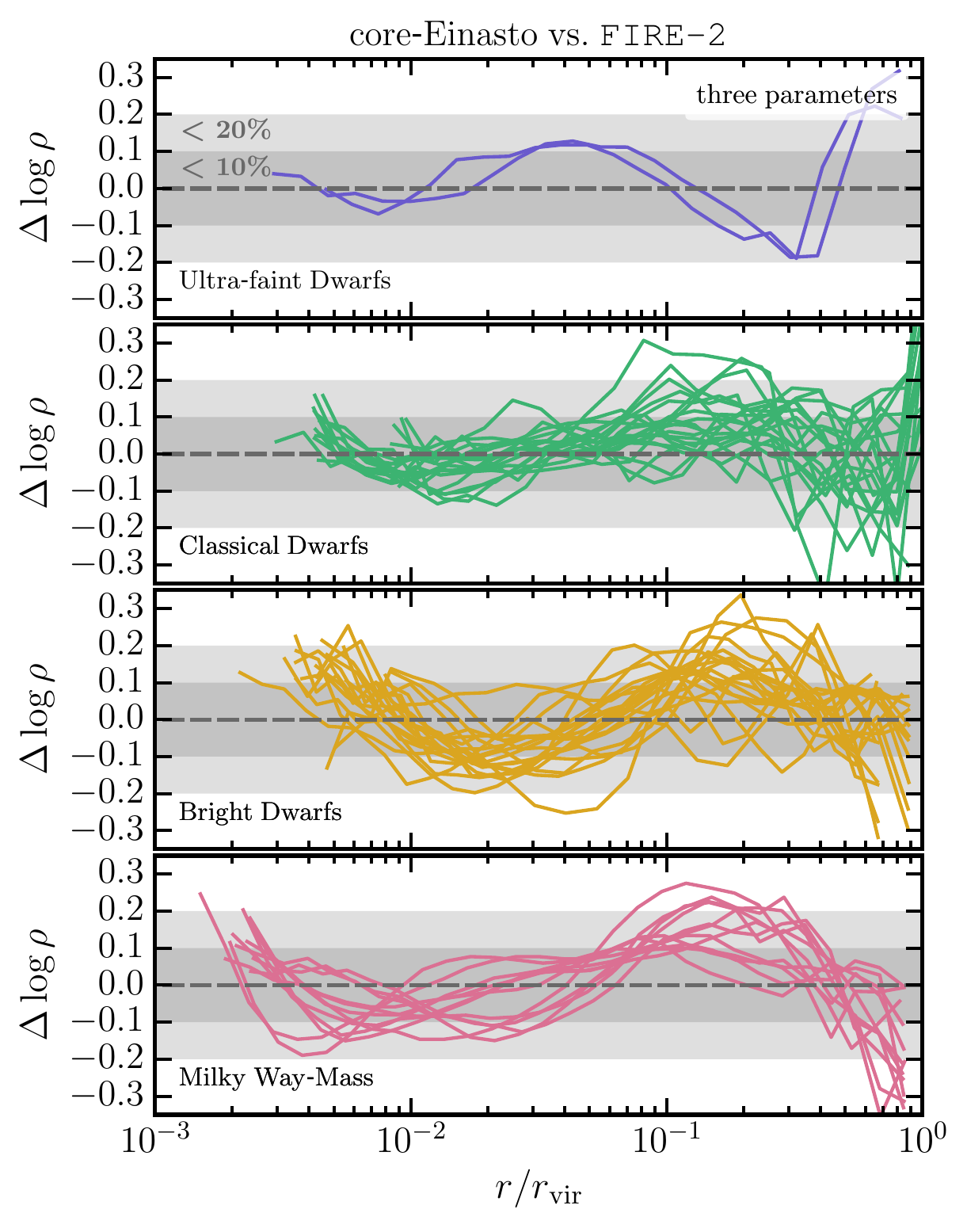}
    \includegraphics[width=\columnwidth]{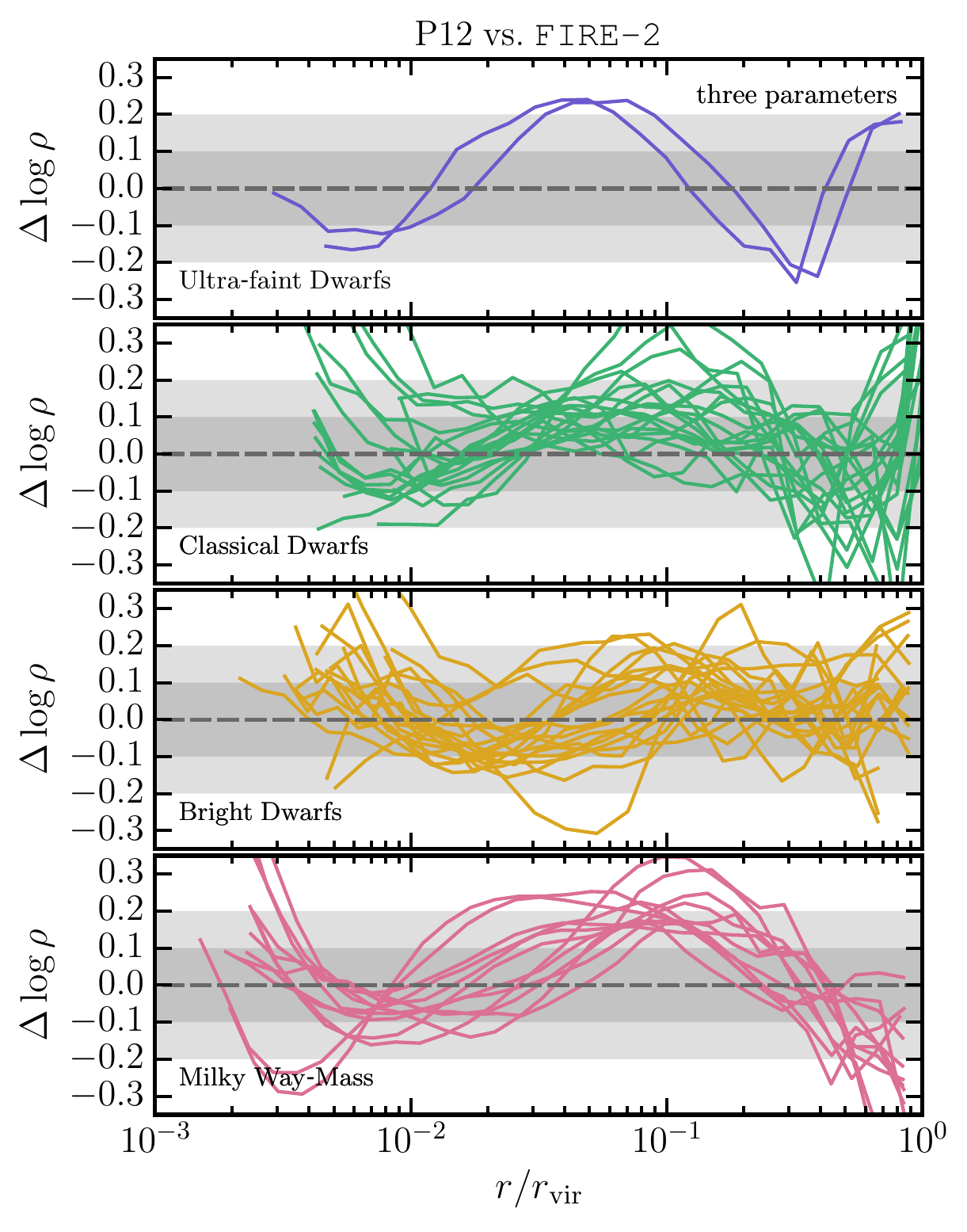}
    \includegraphics[width=\columnwidth]{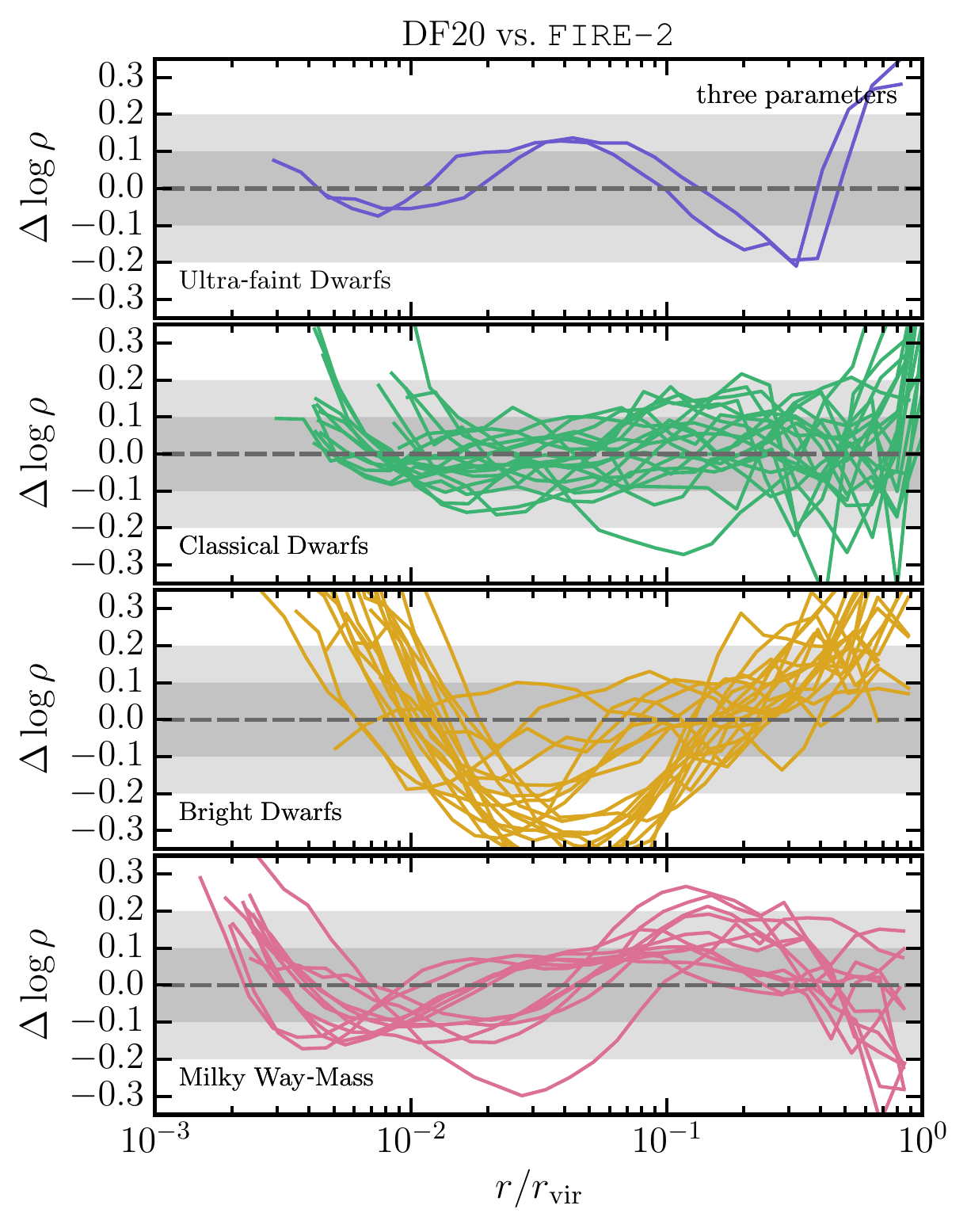}
    \includegraphics[width=\columnwidth]{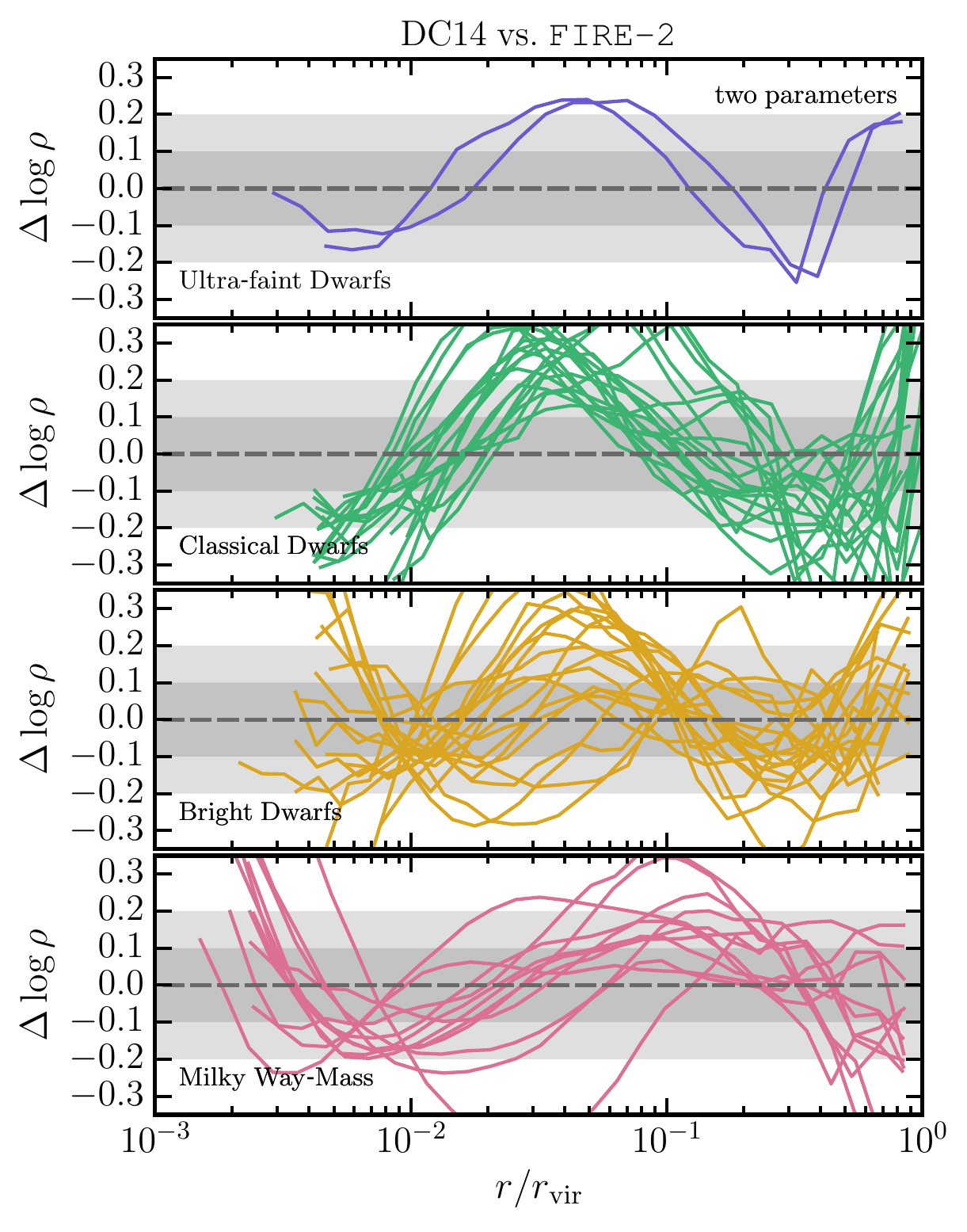}
    \caption{Fit residuals, as in  Fig.~\ref{fig:5}:  residuals for core-Einasto (top left), P12 (top right), DF20 (bottom left), and DC14 (bottom right) profiles for our halo sample.  The number of free parameters in each fit is indicated in the upper right of each panel.  See \ref{sec:cNFW} for the definition of P12, DF20, and DC14.
    }
    \label{fig:D1}
\end{figure*}

\begin{figure}
    \centering
    \includegraphics[width=\columnwidth]{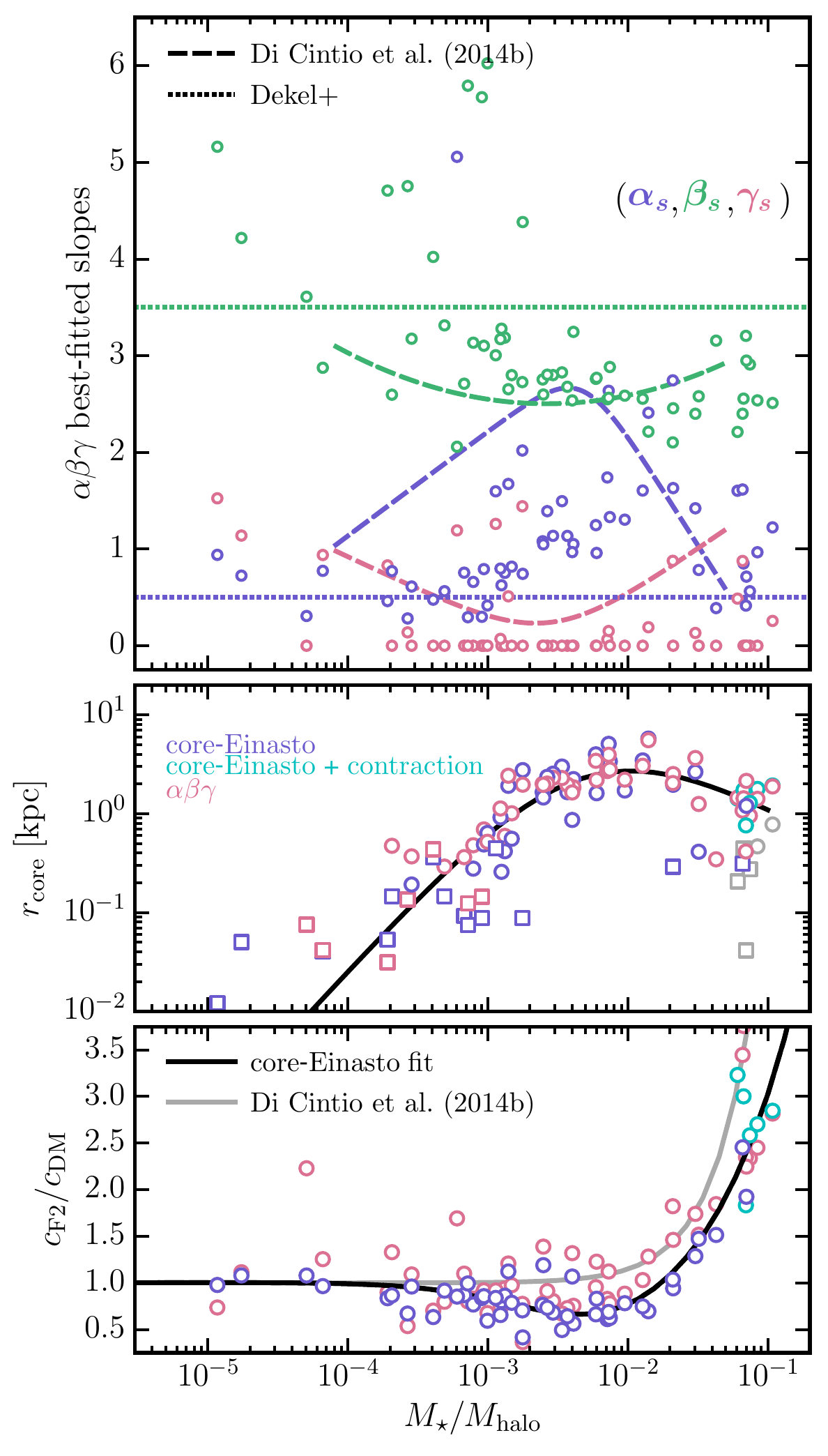}
    \caption{ Properties from the $\alpha\beta\gamma$-profile fits. {\bf \em Top panel}: The best fitting results for the inner slope ($\gamma_{s}$; pink), the outer slope ($\beta_{s}$; green), and the transitioning slope ($\alpha_{s}$; blue).
    {\bf \em Middle panel}:
    The core radius as parameterized from the characteristic radius, $r_{-1}$, from the best-fit $\alpha\beta\gamma$-profiles via Eq.~\eqref{eq:abc.rcore}. We assumed that $r_{-1}$ acts as a probe of $r_{c}$ and find agreeable results from $r_{c}$ derived by fitting Eq.~\eqref{eq:cored.einasto.density} in the main text.
    {\bf \em Bottom panel}:
    The ratio between the concentration parameter for the FIRE-2 halos and their DMO analogs using Eq.~\eqref{eq:abc.r-2}. Shown are the core-Einasto (blue) and core-Einasto with contraction (cyan) values presented previously in Fig.~\ref{fig:10}. Both the FIRE-2 halos and DMO analogs are fitted with the $\alpha\beta\gamma$-profile to obtain the concentration parameters shown in pink. The concentration ratio is mostly in agreement with slight differences at the edge of the bright dwarf regime ($M_{\star}/M_{\rm halo}\approx 10^{-2}$).
    }
    \label{fig:D2}
\end{figure}

\begin{itemize}
    \item {\bf \em P12} \cite[][]{penarrubia2012coupling}:
    One commonly adopted dark matter profile that is an extension of the two-parameter NFW profile that accommodates a physical core radius:
    \begin{align}
        \rho_{\bf \rm P12}(r) 
        = \frac{\rho_{0}r_{0}^{3}}{(r_{c} + r)(r_{0}+r)^{2}}
        \label{eq:profile.P12}
        \, ,
    \end{align}
    where $\rho_{0}$ is the characteristic scale density and $r_{0}$ is some scale radius. The form of Eq.~\eqref{eq:profile.P12} transforms back to a NFW profile in the limit of $r_{c} \rightarrow 0$. 
    The form of Eq.~\eqref{eq:profile.P12} is a three-parameter profile with free variables $\rho_{0}$, $r_{0}$, and $r_{c}$. Eq.~\eqref{eq:profile.P12} is fitted with the FIRE-2 halos by utilizing the fitting routine discussed in Section~\ref{sec:cEinasto} and best-fit parameters are obtained by minimizing the figure-of-merit, Eq.~\eqref{eq:merit.function}.
    \vspace{2ex}
    
    \item {\boldmath$\alpha\beta\gamma$} \cite[][]{zhao1996models}: A generic five parameter profile dubbed the ``$\alpha\beta\gamma$-profile'':
    \begin{align}
        \rho_{\alpha\beta\gamma}(r)
        &=
        \frac{\rho_{s}}{ (r/r_{s})^{\gamma_{s}} \left[ 1 + (r/r_{s})^{\alpha_{s}}\right]^{(\beta_{s}-\gamma_{s})/\alpha_{s}}}
        \label{eq:profile.abc}
        \, ,
    \end{align}
    where $r_{s}$ is the scale radius and $\rho_{s}$ is the scale density. The inner and outer regions are parameterized, respectively, by the logarithmic slopes, $-\gamma_{s}$ and $-\beta_{s}$, while $\alpha_{s}$ controls the rate of transition from the inner and outer region. 
    The form of Eq.~\eqref{eq:profile.abc} has five free-parameters $\rho_{s}$, $r_{s}$, $\alpha_{s}$, $\beta_{s}$, and $\gamma_{s}$.
    Eq.~\eqref{eq:profile.abc} is fitted with the FIRE-2 halos by utilizing the routine discussed in Section~\ref{sec:cEinasto} and best-fit parameters are obtained by minimizing the figure-of-merit, Eq.~\eqref{eq:merit.function}.
    \vspace{2ex}
    
    \item {\bf \em DC14} \cite[][]{di2014mass}: The DC14 model takes the generalized form of Eq.~\eqref{eq:profile.abc} and imposes dependence of the slope parameters as a function of $M_{\star}/M_{\rm halo}$:
    \begin{align}
    \alpha_{s}(X) &= 2.94 - \log_{10}\Big[ \big(10^{X+2.33}\big)^{-1.08} + \big(10^{X+2.33}\big)^{2.29} \Big]
    \\
    \beta_{s}(X) &= 4.23 + 1.34X + 0.26X^{2}
    \\
    \gamma_{s}(X) &= -0.06 + \log_{10}\Big[ \big(10^{X+2.56}\big)^{-0.68} + \big(10^{X+2.56}\big) \Big]
    \, ,
    \end{align}
    where $X:=\log_{10}(M_{\star}/M_{\rm halo}$ and is valid in the range of $-4.1 < X < -1.3$. Outside this mass range resorts to a NFW profile, i.e., $(\alpha_{s},\beta_{s},\gamma_{s})=(1,3,1)$. This now leaves Eq.~\eqref{eq:profile.abc} with two free-parameters: $r_{s}$ and $\rho_{s}$. The DC14 profile is fitted with the FIRE-2 halos by utilizing the routine discussed in Section~\ref{sec:cEinasto} and best-fit parameters are obtained by minimizing the figure-of-merit, Eq.~\eqref{eq:merit.function}.
    \vspace{2ex}    

    \item {\bf \em DF20} \cite[][]{dekel2017profile,freundlich2020dekel}: The DF20 model (or the ``Dekel+'' profile) takes the generic double power-law density profile, namely Eq.~\eqref{eq:profile.abc}, and has fixed slopes $\alpha_{s} = 0.5$ and $\beta_{s}=3.5$. This reduces the analytical profile to be fitted based on three free-parameters: $\rho_{s}$, $r_{s}$, and $\gamma_{s}$.
    The form of DF20 is fitted with the FIRE-2 halos by utilizing the routine discussed in Section~\ref{sec:cEinasto} and best-fit parameters are obtained by minimizing the figure-of-merit, Eq.~\eqref{eq:merit.function}.
\end{itemize}

We also attempted a similar analysis using the core profile from \cite{read2016cNFW} with their four free-parameters and found the resulting fits incompatible with our simulated profiles.

\subsection{Resulting profile residuals}
\begin{figure}
    \centering
    \includegraphics[width=\columnwidth]{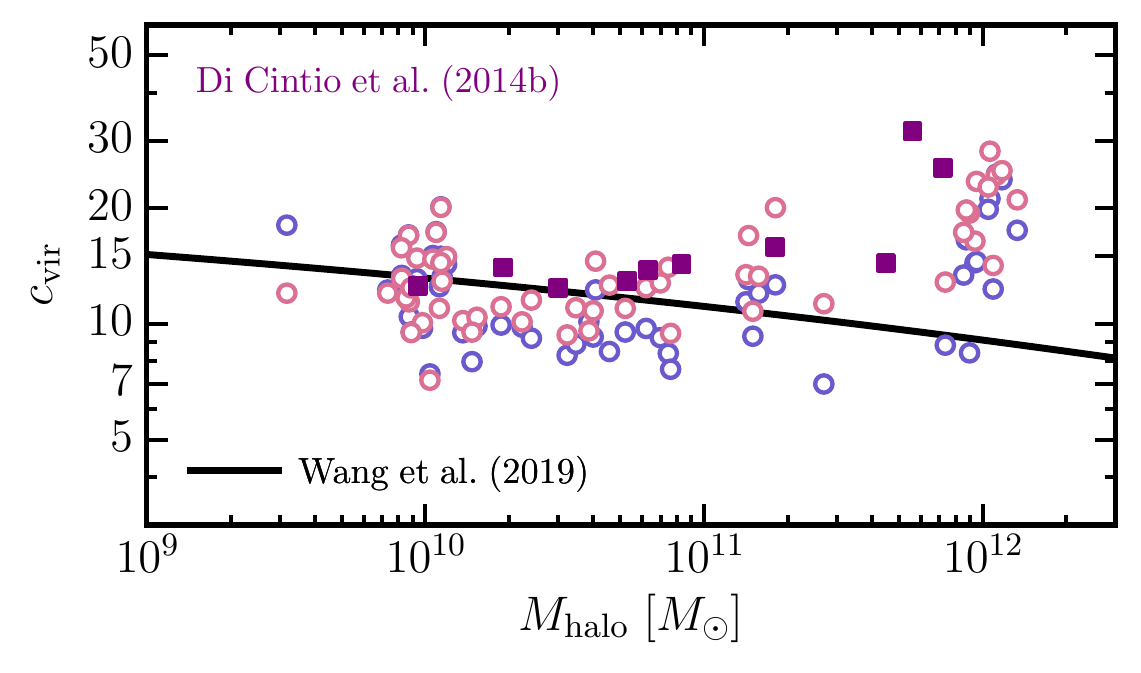}
    \caption{ 
    Concentration as a function of dark matter halo mass. Concentration implied from the core-Einasto profile are given by the blue points while the concentration shown from $\alpha\beta\gamma$-profile fits, i.e., using Eq.~\eqref{eq:abc.r-2}, is given in pink. The solid black curve is the concentration relation from \protect\cite{wang2019zoom}. The purple squares are the results taken from \protect\cite{di2014mass}.
    }
    \label{fig:D3}
\end{figure}

Fig.~\ref{fig:D1} compares the residuals of the FIRE-2 dark matter halos when fitted with the core-Einasto (top left, same fits as presented in the main text), P12 (top right), DF20 (bottom left), and the $\alpha\beta\gamma$ model with DC14 parametrization (bottom right). The interesting comparison to be made is between core-Einasto and P12 since both profiles have three-parameters to be determined, with one being the core radius of the dark matter halo. We see the the core-Einasto does better at fitting the FIRE-2 dark matter halos than the P12 shape. Although, while P12 does not do as well as fitting to our FIRE-2 halos, the form P12 has the advantage of being more analytically friendly when quantifying characteristics of the dark matter halo. For the DF20 model, we find that it is comparable with the core-Einasto fits for the classifications of the ultra-faints, classical dwarfs, and most MW halos. However, DF20 fails to capture the shape of our bright galaxies, i.e., the halos with the largest feedback-induced cores.

Results for the two-parameter DC14 model are shown to be poor fits with the FIRE-2 halos for all of our mass range. This likely has to to do with differences in the dark matter distribution found at fixed stellar mass fractions compared to the simulations explored in DC14.

\subsection{Extended analysis with \texorpdfstring{\boldmath$\alpha\beta\gamma$}{abc}}
Unsurprisingly, the five-parameter $\alpha\beta\gamma$ model provides a superior fit for a majority of our simulated halos compared to our three-parameter core-Einasto profile.  Specifically we find that this profile can do better than $10\%$ for almost all of our galaxies. While the resulting  $\alpha\beta\gamma$-profile fits model the dark matter distribution well, the physical interpretation of the resulting best-fit parameters is less clear. A majority of the fits favor a inner-slope of $\gamma_{s} = 0$ for several cusped profiles, which at times imposes too large of a scale radius to be determined (m10d, m10i, and m10j to name a few). Regardless of the physical interpretation of the resulting parameter fits, we find excellent accuracy modeling our halos, which still enables us to extract characteristics of the best-fit profiles. In the next section, we perform a brief analysis on the best-fit parameters for the $\alpha\beta\gamma$-profile.

In Fig.~\ref{fig:D2}, we present the results when using the $\alpha\beta\gamma$-profile as function of the stellar mass fraction, $M_{\star}/M_{\rm halo}$. The bets-fit parameter fits the $\alpha\beta\gamma$ are also presented for each of our halos in Fig.~\ref{tab:halo.fits}. 

\subsubsection{Best-fit slopes}
The top panel shows the best fitting results for the inner slope ($\gamma_{s}$; pink), the outer slope ($\beta_{s}$; green), and the transitioning slope ($\alpha_{s}$; blue). Also plotted are the trend of the slopes from \cite{di2013dependence} as dashed curves and the dotted curves are having the fixed slopes, $\alpha_{s}=0.5$ and $\beta_{s} = 3.5$, as suggest by \cite{dekel2017profile} and \cite{freundlich2020dekel}. Noticeably, the inner-slope, $\alpha_{s}$, tends to chooses to be zero as a best-fit parameter for a majority of our galaxies. For $\beta_{s}$ and $\gamma_{s}$, a sufficient amount of scatter is seen from the due to allowed large number free-parameters allowed to be fit. Although, trends as a function of $M_{\star}/M_{\rm halo}$, can be somewhat made out. 

\subsubsection{Core radius parametrization}
Notice that in Fig.~\ref{fig:3}, the log-slope profile tends to rise at around $d\log\rho/d\log{r} \simeq -1$, which happens at the radius $r_{-1}$. To play with the idea that the physical core radius can be parameterized by the $\alpha\beta\gamma$-profile, we see how $r_{-1}$ is able to probe the core radius, i.e., 
\begin{align}
    r_{c} \simeq r_{-1} = \left( \frac{1-\gamma_{s}}{\beta_{s}-1} \right)^{1/\alpha_{s}} r_{s}
    \label{eq:abc.rcore}
    \, .
\end{align}
In the middle panel of Fig~\ref{fig:D2}, we plot the previous core radius results from Fig.~\ref{fig:7} for core-Einasto (blue points), the baryonic contracted core-Einasto (cyan points), and the median fit (black curve) while also included the $\alpha\beta\gamma$-profile results using the assumed relation from Eq.~\eqref{eq:abc.rcore}. We see that the $r_{c}$ parametrization from the $\alpha\beta\gamma$-profile follows the median $r_{c}$ curve from core-Einasto parametrization extremely well, implying both excellent agreement with our fitted $r_{c}$ results from core-Einasto and how $r_{-1}$ characterizes the core radius.

\subsubsection{Halo concentration}
Presented in the bottom panel of Fig~\ref{fig:D2} is the ratio between the concentration parameter of the FIRE-2 halos and their DMO analogs. As noted in \cite{di2014mass}, the radius at which the log-slope of the generic five-parameter profile is equal to $-2$, $r_{-2}$, is mapped from the free-parameters via
\begin{align}
    r_{-2} = \left( \frac{2-\gamma_{s}}{\beta_{s}-2} \right)^{1/\alpha_{s}} r_{s}
    \label{eq:abc.r-2}
    \, .
\end{align}
Shown are the points of the core-Einasto (blue) and core-Einasto with contraction (cyan), which are the same values depicted in Fig.~\ref{fig:10} of the main text, while the pink points are the concentration parameters, which is still defined as $c_{\rm vir}=r_{\rm vir}/r_{-2}$, for the best-fit $\alpha\beta\gamma$-profiles. Note that both the FIRE-2 halos and the DMO halos are fitted with $\alpha\beta\gamma$-profile to quantify the $r_{-2}$ values. We mostly find agreement with either methods of quantifying the halo concentration, though several of the halos in the stellar mass fraction range of $10^{-5} - 10^{-3}$ are strongly scattered. In the classical dwarf regime, we find halos that are less concentrated like we found from the core-Einasto model. We somewhat find agreement in the bright dwarf regime, although a sufficient of scatter is present. Though recently, \cite{freundlich2020dekel} reports a similar result at this stellar mass fraction. The MW halos are mostly consistent with our previous findings in the main text.

A different viewed of the concentration parameter can be made by taking the previous points and plotting as a function of $M_{\rm halo}$ in Fig.~\ref{fig:D3}. The purple squares are the halos presented in \cite{di2014mass}.

\section{Simulation Sample}
\label{sec:sim.sample}
Presented in Tables~\ref{tab:halo.sample} and \ref{tab:halo.fits} are the suite of halos simulated using FIRE-2 with their relevant parameters listed at $z=0$. 

\subsection{Global and simulation properties}
Presented in Table~\ref{tab:halo.sample} are the global properties of the FIRE-2 galaxies at $z=0$ as well as the relevant simulation properties. Columns (1-5) contain global properties of the galaxies while columns (6-10) describe the numerical resolution properties of the simulations. All simulations were ran using with $n_{\rm crit}=1000$, the minimum gas density required for star formation in addition to self-shielding, Jeans instability, and self-gravity.
{\bf \emph{References}} , given in the last column, are labeled as such --- A: {\protect\cite{fitts2017fire}}, B: {\protect\cite{graus2019predicted}}, 
C: {\protect\cite{wheeler2019resolved}}, 
D: {\protect\cite{chan2018origin}}, 
E: {\protect\cite{el2017gas}}, 
F: {\protect\cite{hopkins2018fire}}, 
G: {\protect\cite{garrison2018local}}, 
H: {\protect\cite{samuel2020profile}}, 
I: {\protect\cite{wetzel2016reconciling}}.
The individual columns in Table~\ref{tab:halo.sample} are described as follows:

\begin{itemize}
    \item[](1) 
    $M_{\rm halo}$: The mass of the target halo at $z=0$ defined by \protect\cite{bryan1998statistical}. 
    
    \item[](2) 
    $r_{\rm vir}$: The virial radius in physical units of the target halo.
    
    \item[](3)
    $V_{\rm max}$: The maximum circular velocity curve for the dark matter component of the FIRE-2 dark matter halos, i.e., $V_{\rm max} := \mathrm{max}[V_{\rm circ}]$.
    
    \item[](4) 
    $M_{\star}$: Stellar mass (within $10\%$ of $r_{\rm vir}$) of the central galaxy in the target halo.
    
    \item[](5) 
    $r_{1/2}$: The physical radius that encloses half the value of $M_{\star}$ for the central galaxy.
    
    \item[](6)
    $m_{\rm b}$: The mass of baryon particles of the simulation. 
    
    \item[](7) 
    $m_{\rm dm}$: The mass of dark matter particles of the simulation. 
    
    \item[](8) 
    $\epsilon_{\rm dm}$: The dark matter force softening

    \item[](9) 
    $r_{\rm conv}$: Radius of numerical convergence of the DMO analogs, set by Eq.~\eqref{eq:power.radius} and the most conservative criterion as discussed in \cite{hopkins2018fire} .
\end{itemize}

\subsection{Resulting analytical profile fits}
Presented in Table~\ref{tab:halo.fits} are the best-fit parameters for the core-Einasto and $\alpha\eta\gamma$- profile. Columns (1-5) are the results of fitting the simulated density profiles to the core-Einastro profile with $\alpha_\epsilon=0.16$. Columns (6-9) are the results of fitting the simulated density profiles to the $\alpha\beta\gamma$-profile. The individual columns in Table~\ref{tab:halo.fits} are described as follows:
\begin{itemize}
    \item[](1)
    $\tilde{\rho}_{s}$: The scale density fitted as a free parameter for the core-Einasto profile, Eq. \eqref{eq:cored.einasto.density}.
    
    \item[](2)
    $\tilde{r}_{s}$: The scale radius fitted as a free parameter for the core-Einasto profile, Eq. \eqref{eq:cored.einasto.density}.
    
    \item[](3)
    $r_{c}$: The physical core radius of the dark matter profile fitted as a free parameter for the core-Einasto profile, Eq. \eqref{eq:cored.einasto.density}.
    
    \item[](4) 
    $Q_{\rm min,cEin}$: The quoted goodness-of-fit parameter for the core-Einasto fit, i.e., Eq.~\eqref{eq:merit.function}.
    
    \item[](5) 
    $\rho_{s}$: The scale density fitted as a free parameter for the $\alpha\beta\gamma$-profile.
    
    \item[](7-9)
    $(\alpha,\beta,\gamma)$: The three characteristic slopes fitted as a free parameter for the $\alpha\beta\gamma$-profile.
    
    \item[](10) 
    $Q_{\rm min, \alpha\beta\gamma}$: The quoted goodness-of-fit parameter for the $\alpha\beta\gamma$-profile fit, i.e., Eq.~\eqref{eq:merit.function}.
\end{itemize}

For the quoted core radii (column 3 in Table~\ref{tab:halo.fits}), the symbols are defined as follows: (\cmark) -- Verified location of dark matter core in the simulated profile; (\xmark) -- Improper value of dark matter core in the simulated profile if one is physically present; ($\dagger$) -- Dark matter core radius fitted inside the region of conservative numerical convergence, i.e., $r_{c} < r_{\rm conv}$. The exact meaning of these results are discussed in more properly in Section~\ref{sec:rcore.param}.
    

\begin{table*}
    \setlength{\tabcolsep}{10.25pt}
    \renewcommand{\arraystretch}{0.95}
    \centering
        \caption{ 
        Global parameters of the FIRE-2 halos.
        }
    \label{tab:halo.sample}
    \begin{threeparttable}
    \begin{tabular}{SSSSSS|SSSS|S} 
        \toprule
        \toprule
        {Halo} & 
        {$M_{\rm halo}$} & 
        {$r_{\rm vir}$} & 
        {$V_{\rm max}$} & 
        {$M_{\star}$} & 
        {$r_{1/2}$} & 
        {$m_{\rm b}$} & 
        {$m_{\rm dm}$} & 
        {$\epsilon_{\rm dm}$} & 
        {$r_{\rm conv}$} & 
        {Reference} 
        \\
        {Name} & 
        {$[M_{\odot}]$} & 
        {$[\rm kpc]$} & 
        {$[\rm km\ s^{-1}]$} & 
        {$[M_{\odot}]$} & 
        {$[\rm kpc]$} & 
        {$[M_{\odot}]$} & 
        {$[M_{\odot}]$} & 
        {$[\rm pc]$} & 
        {$[\rm kpc]$} &
        {}
        \\ 
        \midrule
        & 
        {(1)} & 
        {(2)} & 
        {(3)} & 
        {(4)} & 
        {(5)} & 
        {(6)} & 
        {(7)} & 
        {(8)} & 
        {(9)} & 
        {} \\ 
        \midrule
        \multicolumn{9}{l}{{\bf \em Ultra-Faint Dwarfs} (2)}\\
        \midrule
        {m10v$_{250}$} & 
        {$8.9 \times 10^{9}$} & 
        {57.7} & 
        {30} & 
        {$1.5 \times 10^{5}$} & 
        {0.35} & 
        {250} & 
        {1300} &
        {29} &
        {0.166} &
        {C}
        \\
        {m10v$_{250}$B} & 
        {$3.2 \times 10^{9}$} & 
        {40.9} &  
        {24} &  
        {$ 3.7 \times 10^{4}$} & 
        {0.42} & 
        {250} &
        {1300} &
        {29} &
        {0.153} & 
        {C}
        \\
        \midrule
        \multicolumn{9}{l}{{\bf \em Classical Dwarfs} (20)} 
        \\
        \midrule
        {m10b} & 
        {$9.3 \times 10^{9}$} & 
        {54.8} & 
        {31} & 
        {$4.7 \times 10^{5}$} &
        {0.34} & 
        {500} & 
        {2500} &
        {50} &
        {0.218} & 
        {A}
        \\
        {m10c} & 
        {$8.8\times 10^{9}$} & 
        {54.1} & 
        {31} & 
        {$5.8\times 10^{5}$} & 
        {0.35} & 
        {500} & 
        {2500} &
        {50} &
        {0.227} &
        {A}
        \\
        {m10d} & 
        {$8.2 \times 10^{9}$} & 
        {50.7} & 
        {32} & 
        {$1.6 \times 10^{6}$} & 
        {0.53} & 
        {500} & 
        {2500} &
        {50} &
        {0.209} & 
        {A}
        \\
        {m10e} & 
        {$9.8 \times 10^{9}$} & 
        {53.8} & 
        {31} & 
        {$2.0 \times 10^{6}$} & 
        {0.62} & 
        {500} & 
        {2500} &
        {50} &
        {0.216} &
        {A}
        \\
        {m10f} & 
        {$8.7 \times 10^{9}$} & 
        {51.5} & 
        {35} & 
        {$4.7 \times 10^{6}$} & 
        {0.75} & 
        {500} & 
        {2500} &
        {50} &
        {0.202} &
        {A}
        \\
        {m10g} & 
        {$7.3 \times 10^{9}$} & 
        {48.6} & 
        {32} & 
        {$5.7 \times 10^{6}$} & 
        {0.95} & 
        {500} & 
        {2500} &
        {50} &
        {0.215} &
        {A}
        \\
        {m10h} & 
        {$1.2 \times 10^{10}$} &
        {57.2} & 
        {37} & 
        {$8.1 \times 10^{6}$} & 
        {0.83} & 
        {500} & 
        {2500} &
        {50} &
        {0.207} &
        {A}
        \\
        {m10i} & 
        {$1.1 \times 10^{10}$} &
        {56.3} & 
        {40} & 
        {$8.2 \times 10^{6}$} & 
        {0.57} & 
        {500} & 
        {2500} &
        {50} &
        {0.195} &
        {A}
        \\
        {m10j} & 
        {$1.1 \times 10^{10}$} &
        {55.4} & 
        {37} & 
        {$9.9 \times 10^{6}$} & 
        {0.70} & 
        {500} & 
        {2500} &
        {50} &
        {0.194} &
        {A}
        \\
        {m10k} & 
        {$1.1 \times 10^{10}$} &
        {56.4} & 
        {38} & 
        {$1.1 \times 10^{7}$} & 
        {1.14} & 
        {500} & 
        {2500} &
        {50} &
        {0.207} &
        {A}
        \\
        {m10l} & 
        {$1.1 \times 10^{10}$} &
        {56.1} & 
        {37} & 
        {$1.3\times 10^{7}$} & 
        {0.78} & 
        {500} & 
        {2500} &
        {50} &
        {0.202} &
        {A}
        \\
        {m10m} & 
        {$1.1 \times 10^{10}$} &
        {56.1} & 
        {38} & 
        {$1.5 \times 10^{7}$} & 
        {0.96} & 
        {500} & 
        {2500} &
        {50} &
        {0.208} &
        {A}
        \\
        {m10q$_{250}$} & 
        {$8.2 \times 10^{9}$} & 
        {56.2} & 
        {33} & 
        {$2.3 \times 10^{6}$} & 
        {0.81} & 
        {250} & 
        {1300} &
        {29} &
        {0.150}  &
        {C}
        \\
        {m10xc$_{\rm A}$} & 
        {$8.5 \times 10^{9}$} & 
        {53.1} & 
        {35} & 
        {$8.5 \times 10^{6}$} & 
        {1.80} & 
        {4000} &
        {20000} &
        {100} &
        {0.455} &
        {B}
        \\
        {m10xd$_{\rm A}$} & 
        {$2.4 \times 10^{10}$} &
        {75.5} & 
        {38} & 
        {$1.4 \times 10^{7}$} &
        {1.90} & 
        {4000} & 
        {20000} &
        {100} &
        {0.476} & 
        {B}
        \\   
        {m10xe$_{\rm A}$} & 
        {$1.4 \times 10^{10}$} &
        {62.5} & 
        {35} & 
        {$3.6 \times 10^{6}$} &
        {1.27} & 
        {4000} & 
        {20000} &
        {100} &
        {0.529} &
        {B}
        \\
        {m10xe$_{\rm B}$} & 
        {$1.1 \times 10^{10}$} &
        {58.6} & 
        {38} & 
        {$1.3 \times 10^{7}$} & 
        {1.90} & 
        {4000} & 
        {20000} &
        {100} &
        {0.488} &
        {B}
        \\ 
        {m10xe$_{\rm C}$} & 
        {$1.0 \times 10^{10}$} &
        {57.0} & 
        {34} & 
        {$1.8 \times 10^{7}$} & 
        {3.00} & 
        {4000} & 
        {20000} &
        {100} &
        {0.474} &
        {B}
        \\
        {m10xe$_{\rm D}$} & 
        {$8.9 \times 10^{9}$} & 
        {53.9} & 
        {34} & 
        {$3.6 \times 10^{6}$} & 
        {1.47} & 
        {4000} & 
        {20000} &
        {100} &
        {0.482} &
        {B}
        \\
        {m10xg$_{\rm A}$} & 
        {$1.5 \times 10^{10}$} &
        {64.4} & 
        {40} & 
        {$1.9\times 10^{7}$} & 
        {2.20} & 
        {4000} & 
        {20000} &
        {100} &
        {0.465} &
        {B}
        \\
        \midrule
        \multicolumn{9}{l}{{\bf \em Bright Dwarfs} (20)}\\
        \midrule
        {m10xa} & 
        {$1.9 \times10^{10}$} & 
        {69.4} & 
        {45} & 
        {$7.6 \times 10^{7}$} & 
        {3.18} & 
        {4000} & 
        {20000} &
        {100} &
        {0.453} &
        {B}
        \\
        {m10xb} & 
        {$2.2 \times 10^{10}$} &
        {73.5} & 
        {42} & 
        {$3.3 \times 10^{7}$} & 
        {2.39} & 
        {4000} & 
        {20000} &
        {100} &
        {0.480}  &
        {B}
        \\
        {m10xc} & 
        {$3.2 \times 10^{10}$} &
        {82.9} & 
        {48} & 
        {$1.2 \times 10^{8}$} & 
        {3.26} & 
        {4000} & 
        {20000} &
        {100} &
        {0.451}  &
        {B}
        \\
        {m10xd} & 
        {$3.9 \times 10^{10}$} &
        {88.5} & 
        {53} & 
        {$6.8 \times 10^{7}$} & 
        {4.04} & 
        {4000} & 
        {20000} &
         {100} &
        {0.437} &
        {B}
        \\
        {m10xe} & 
        {$4.5 \times 10^{10}$} &
        {93.6} & 
        {56} & 
        {$3.3 \times 10^{8}$} & 
        {4.17} & 
        {4000} & 
        {20000} &
        {100} &
        {0.448}  &
        {B}
        \\
        {m10xf} & 
        {$5.2 \times 10^{10}$} &
        {97.7} & 
        {58} & 
        {$1.3 \times 10^{8}$} & 
        {3.33} & 
        {4000} & 
        {20000} &
        {100} &
        {0.453} &
        {B}
        \\
        {m10xg} & 
        {$6.2 \times 10^{10}$} &
        {103} & 
        {65} & 
        {$4.6 \times 10^{8}$} & 
        {3.98} & 
        {4000} & 
        {20000} &
        {100} &
        {0.443}  &
        {B}
        \\
        {m10xh} & 
        {$7.4 \times 10^{10}$} & 
        {110} & 
        {68} & 
        {$5.4 \times 10^{8}$} & 
        {6.04} & 
        {4000} & 
        {20000} &
        {100} &
        {0.434}  &
        {B}
        \\
        {m10xh$_{\rm A}$} & 
        {$1.5 \times 10^{10}$} &
        {63.9} & 
        {38} & 
        {$5.0 \times 10^{7}$} & 
        {3.14} & 
        {4000} & 
        {20000} &
        {100} &
        {0.464} &
        {B}
        \\
        {m10xi} & 
        {$7.6 \times 10^{10}$} &
        {111} & 
        {64} & 
        {$4.5 \times 10^{8}$} & 
        {5.16} & 
        {4000} & 
        {20000} &
        {100} &
        {0.441} &
        {B}
        \\
        {m10z} & 
        {$3.5 \times 10^{10}$} & 
        {90.5} & 
        {49} & 
        {$4.9 \times 10^{7}$} & 
        {3.20} & 
        {2100} & 
        {10000} &
        {43} &
        {0.370} & 
        {D}
        \\
        {m11a} & 
        {$4.0 \times 10^{10}$} &
        {95.0} & 
        {52} & 
        {$1.2 \times 10^{8}$} & 
        {2.63} & 
        {2100} & 
        {10000} &
        {43} &
        {0.314}  & 
        {D}
        \\
        {m11b} & 
        {$4.1 \times 10^{10}$} &
        {95.6} & 
        {59} & 
        {$1.1 \times 10^{8}$} & 
        {2.39} & 
        {2100} & 
        {10000} &
        {43} &
        {0.314}  & 
        {D}
        \\
        {m11c} & 
        {$1.4 \times 10^{11}$} &
        {145} & 
        {80} & 
        {$8.5 \times 10^{8}$} & 
        {2.78} & 
        {2100} & 
        {10000} &
        {43} &
        {0.673}  & 
        {F}
        \\
        {m11d} & 
        {$2.7 \times 10^{11}$} &
        {179} & 
        {88} & 
        {$3.8 \times 10^{9}$} & 
        {6.01} & 
        {7100} & 
        {35000} &
        {40} &
        {0.502}  & 
        {E}
        \\
        {m11e} & 
        {$1.4 \times 10^{11}$} &
        {146} & 
        {83} & 
        {$1.4 \times 10^{9}$} & 
        {3.36} & 
        {7100} & 
        {35000} &
        {40} &
        {0.481}  & 
        {E}
        \\
        {m11h} & 
        {$1.8 \times 10^{11}$} &
        {157} & 
        {90} & 
        {$3.8 \times 10^{9}$} & 
        {3.92} & 
        {7100} & 
        {35000} &
        {40} &
        {0.503} & 
        {E}
        \\
        {m11i} & 
        {$7.0 \times 10^{10}$} &
        {114} & 
        {62} & 
        {$8.9 \times 10^{8}$} & 
        {3.35} & 
        {7100} & 
        {35000} &
        {40} &
        {0.548} & 
        {E}
        \\
        {m11q} & 
        {$1.6 \times 10^{11}$} &
        {153} & 
        {80} & 
        {$6.3 \times 10^{8}$} & 
        {2.35} & 
        {7100} & 
        {35000} &
        {40} &
        {0.523} &
        {D} 
        \\
        {m11q$_{880}$}  & 
        {$1.5 \times 10^{11}$} &
        {114} & 
        {80} & 
        {$3.7 \times 10^{8}$} & 
        {2.83} & 
        {880} & 
        {4400} &
        {20} &
        {0.225} & 
        {E}
        \\
        \midrule
        \multicolumn{9}{l}{{\bf \em Milky Way-Mass} (12)}\\
        \midrule
        {m12b} & 
        {$1.1 \times 10^{12}$} & 
        {224} &
        {183} & 
        {$9.4 \times 10^{10}$} & 
        {2.66} & 
        {7100} & 
        {35000} &
        {40} &
        {0.437} & 
        {G}
        \\
        {m12c} & 
        {$1.1 \times 10^{12}$} & 
        {219} & 
        {157} & 
        {$6.5 \times 10^{10}$} & 
        {3.37} & 
        {7100} & 
        {35000} &
        {40} &
        {0.461} &
        {G}
        \\
        {m12f} & 
        {$1.3 \times 10^{12}$} & 
        {237} & 
        {184} & 
        {$8.9 \times 10^{10}$} & 
        {3.60} & 
        {7100} & 
        {35000} &
        {40} &
        {0.471} & 
        {F}
        \\
        {m12i} & 
        {$9.4 \times 10^{11}$} & 
        {210} & 
        {162} & 
        {$7.0 \times 10^{10}$} & 
        {2.80} & 
        {7100} & 
        {35000} &
        {40} &
        {0.496} & 
        {I}
        \\
        {m12m} & 
        {$1.2 \times 10^{12}$} & 
        {227} & 
        {187} & 
        {$1.3 \times 10^{11}$} & 
        {4.88} & 
        {7100} & 
        {35000} &
        {40} &
        {0.439} & 
        {F}
        \\
        {m12r} & 
        {$9.0 \times 10^{11}$} & 
        {211} & 
        {136} & 
        {$1.9 \times 10^{10}$} & 
        {4.37} & 
        {7100} & 
        {35000} &
        {40} &
        {0.476} & 
        {H}
        \\
        {m12w} & 
        {$9.5 \times 10^{11}$} & 
        {215} & 
        {157} & 
        {$5.5 \times 10^{10}$} & 
        {3.04} & 
        {7100} & 
        {35000} &
        {40} &
        {0.507} & 
        {H}
        \\
        {m12z} & 
        {$7.3 \times 10^{11}$} & 
        {195} & 
        {130} & 
        {$2.2 \times 10^{10}$} & 
        {4.71} & 
        {4200} & 
        {22000} &
        {33} &
        {0.383} &
        {G}
        \\
        {Thelma} & 
        {$1.1 \times 10^{12}$} & 
        {220} & 
        {178} & 
        {$7.7 \times 10^{10}$} & 
        {4.36} & 
        {4000} & 
        {20000} &
        {32} &
        {0.366} &
        {G}
        \\
        {Louise} & 
        {$8.5 \times 10^{11}$} & 
        {203} & 
        {159} & 
        {$2.7 \times 10^{10}$} & 
        {3.27} & 
        {4000} & 
        {20000} &
        {32} &
        {0.359} &
        {G}
        \\
        {Romeo} & 
        {$1.0 \times 10^{12}$} & 
        {222} & 
        {188} & 
        {$7.3 \times 10^{10}$} & 
        {4.18} & 
        {3500} & 
        {20000} &
        {31} &
        {0.329} &
        {G}
        \\
        {Juliet} & 
        {$8.7 \times 10^{11}$} & 
        {209} & 
        {164} & 
        {$3.7 \times 10^{10}$} & 
        {2.14} & 
        {3500} & 
        {20000} &
        {31} &
        {0.339} &
        {G}
        \\
        \bottomrule
        \bottomrule
    \end{tabular}
    \end{threeparttable}
\end{table*}



\begin{table*}
    \setlength{\tabcolsep}{11.25pt}
    \renewcommand{\arraystretch}{0.95}
    \centering
        \caption{ 
        Resulting profile fits.
        }
    \label{tab:halo.fits}
    \begin{threeparttable}
    \begin{tabular}{SSSSS|SSSSSS} 
        \toprule
        \toprule
        {Halo} & 
        {$\tilde{\rho}_{s}$} & 
        {$\tilde{r}_{s}$} & 
        {$r_{c}$} & 
        {$Q_{\rm min}$ } & 
        {$\rho_{s}$} & 
        {$r_{s}$} & 
        {$\alpha_{s}$} &
        {$\beta_{s}$} & 
        {$\gamma_{s}$} & 
        {$Q_{\rm min}$} 
        \\
        {Name} & 
        {$[M_{\odot}\,\rm kpc^{-3}]$} & 
        {$[\rm kpc]$} & 
        {$[\rm kpc]$} & 
        {cEin} & 
        {$[M_{\odot}\ \rm kpc^{-3}]$} & 
        {$[\rm kpc]$} & 
        {} &
        {} &
        {} &
        {$\alpha\beta\gamma$}
        \\ 
        \midrule
        {}    & 
        {(1)} & 
        {(2)} & 
        {(3)} & 
        {(4)} & 
        {(5)} & 
        {(6)} & 
        {(7)} & 
        {(8)} & 
        {(9)} & 
        {(10)} \\
        \midrule
        \multicolumn{8}{l}{{\bf \em Ultra-Faint Dwarfs} (2)}\\
        \midrule
        {m10v$_{250}$} & 
        {$5.5 \times 10^{5}$} & 
        {5.82} & 
        {$^{\dagger}$0.05} & 
        {0.0480} & 
        {$3.9 \times 10^{5}$} &
        {25.0} &
        {0.73} & 
        {4.22} &
        {1.14} &
        {0.0462}
        \\
        {m10v$_{250}$B} & 
        {$2.7 \times 10^{6}$} & 
        {2.21} & 
        {$^{\dagger}$0.01} & 
        {0.0453} & 
        {$9.32 \times 10^{4}$} &
        {25.0} &
        {0.94} & 
        {5.16} &
        {1.53} &
        {0.0328}
        \\
        \midrule
        \multicolumn{9}{l}{{\bf \em Classical Dwarfs} (20)} 
        \\
        \midrule
        {m10b} & 
        {$1.4 \times 10^{6}$} & 
        {4.02} & 
        {0.00} & 
        {0.0414} & 
        {$2.4 \times 10^{10}$} &
        {1.70} &
        {0.31} &
        {3.61} &
        {0.00} &
        {0.0313}
        \\
        {m10c} & 
        {$9.4 \times 10^{5}$} & 
        {4.75} & 
        {$^{\dagger}$0.04} & 
        {0.0398} & 
        {$1.0 \times 10^{7}$} &
        {3.44} &
        {0.78} & 
        {2.88} & 
        {0.93} &
        {0.0384}
        \\
        {m10d} & 
        {$1.9 \times 10^{6}$} & 
        {3.53} & 
        {$^{\dagger}$0.05} & 
        {0.0317} & 
        {$6.2 \times 10^{6}$} &
        {25.0} &
        {0.46} & 
        {4.71} & 
        {0.83} &
        {0.0328}
        \\
        {m10e} & 
        {$1.0 \times 10^{6}$} & 
        {4.82} & 
        {$^{\dagger}$0.15} & 
        {0.0336} & 
        {$1.8 \times 10^{8}$} &
        {0.87} &
        {0.77} &
        {2.60} & 
        {0.00} &
        { 0.0232}
        \\
        {m10f} & 
        {$5.6 \times 10^{6}$} & 
        {2.35} & 
        {$^{\dagger}$0.15} & 
        {0.0575} & 
        {$8.8 \times 10^{8}$} &
        {1.29} &
        {0.56} &
        {3.31} & 
        {0.00} &
        {0.0596}
        \\
        {m10g} & 
        {$3.6 \times 10^{6}$} & 
        {2.78} & 
        {$^\text{\cmark}$0.28} & 
        {0.0429} & 
        {$2.3 \times 10^{8}$} &
        {1.51} &
        {0.66} & 
        {3.13} & 
        {0.00} &
        {0.0448}
        \\
        {m10h} & 
        {$2.7 \times 10^{6}$} & 
        {3.55} & 
        {$^{\dagger}$0.10} & 
        {0.0418} & 
        {$4.7 \times 10^{8}$} &
        {0.74} &
        {0.76} & 
        {2.71} &
        {0.00} &
        {0.0283}
        \\
        {m10i} & 
        {$6.1 \times 10^{6}$} & 
        {2.43} & 
        {$^{\dagger}$0.07} & 
        {0.0465} & 
        {$1.7\times 10^{10}$} &
        {25.0} &
        {0.30} & 
        {5.79} &
        {0.00} &
        {0.0483}
        \\
        {m10j} & 
        {$4.6 \times 10^{6}$} & 
        {2.77} & 
        {$^{\dagger}$0.10} & 
        {0.0260} & 
        {$1.2 \times 10^{10}$} &
        {25.0} &
        {0.30} & 
        {5.67} &
        {0.00} &
        {0.0296}
        \\
        {m10k} & 
        {$7.8\times 10^{6}$} & 
        {2.39} & 
        {$^\text{\cmark}$0.49} & 
        {0.0360} & 
        {$1.4 \times 10^{10}$} &
        {1.77} &
        {0.79} & 
        {3.10} &
        {0.00} &
        {0.0363}
        \\
        {m10l} & 
        {$5.8 \times 10^{6}$} & 
        {2.54} & 
        {$^\text{\cmark}$0.26} & 
        {0.0360} & 
        {$4.0 \times 10^{8}$} &
        {1.63} &
        {0.63} & 
        {3.28} &
        {0.00} &
        {0.0365}
        \\
        {m10m} & 
        {$1.0 \times 10^{7}$} & 
        {2.14} & 
        {$^\text{\cmark}$0.42} & 
        {0.0421} & 
        {$2.1 \times 10^{8}$} &
        {1.68} &
        {0.75} & 
        {3.19} &
        {0.00} &
        {0.0465}
        \\
        {m10q$_{250}$} & 
        {$4.0 \times 10^{6}$} & 
        {2.64} & 
        {$^\text{\cmark}$0.19} & 
        {0.0262} & 
        {$4.4 \times 10^{8}$} &
        {1.32} &
        {0.61} & 
        {3.18} &
        {0.00} &
        {0.0224}
        \\
        {m10xc$_{\rm A}$} & 
        {$1.1 \times 10^{7}$} & 
        {1.99} & 
        {$^\text{\cmark}$0.64} & 
        {0.0262} & 
        {$4.4 \times 10^{8}$} &
        {25.0} &
        {0.42} & 
        {6.02} &
        {0.00} &
        {0.0259}
        \\
        {m10xd$_{\rm A}$} & 
        {$5.3 \times 10^{5}$} & 
        {8.25} & 
        {0.00} & 
        {0.0734} & 
        {$1.3 \times 10^{7}$} &
        {1.56} &
        {5.06} & 
        {2.06} &
        {1.19} &
        {0.0186}
        \\   
        {m10xe$_{\rm A}$} & 
        {$6.2 \times 10^{5}$} & 
        {6.60} & 
        {0.00} & 
        {0.0413} & 
        {$2.7 \times 10^{9}$} &
        {25.0} &
        {0.28} & 
        {4.75} &
        {0.14} &
        {0.0422}
        \\
        {m10xe$_{\rm B}$} & 
        {$4.6 \times 10^{6}$} & 
        {2.90} & 
        {$^{\dagger}$0.45} & 
        {0.0278} & 
        {$1.6 \times 10^{6}$} &
        {6.46} &
        {1.60} & 
        {3.01} &
        {1.26} &
        {0.0186}
        \\ 
        {m10xe$_{\rm C}$} & 
        {$4.1 \times 10^{7}$} & 
        {1.54} & {$^\text{\cmark}$2.80} & 
        {0.0196} & 
        {$1.5 \times 10^{7}$} &
        {10.1} &
        {0.74} & 
        {4.38} &
        {0.00} &
        {0.0187}
        \\
        {m10xe$_{\rm D}$} & 
        {$4.1 \times 10^{6}$} & 
        {2.80} & 
        {$^{\dagger}$0.36} & 
        {0.0601} & 
        {$4.6\times 10^{8}$} &
        {4.45} &
        {0.48} & 
        {4.02} &
        {0.00} &
        {0.0618}
        \\
        {m10xg$_{\rm A}$} & 
        {$4.0 \times 10^{6}$} & 
        {3.26} & 
        {$^\text{\cmark}$0.92} & 
        {0.0222} & 
        {$4.3 \times 10^{7}$} &
        {3.27} &
        {0.80} & 
        {3.17} &
        {0.07} &
        {0.0194}
        \\
        \midrule
        \multicolumn{9}{l}{{\bf \em Bright Dwarfs} (20)}\\
        \midrule
        {m10xa} & 
        {$5.4 \times 10^{7}$} & 
        {1.62} & 
        {$^\text{\cmark}$2.24} & 
        {0.0240} & 
        {$2.4 \times 10^{7}$} &
        {3.99} &
        {1.05} & 
        {3.25} &
        {0.00} &
        {0.0180}
        \\
        {m10xb} & 
        {$1.9 \times10^{6}$} & 
        {5.13} & 
        {$^\text{\cmark}$0.56} & 
        {0.0248} & 
        {$7.5 \times 10^{7}$} &
        {2.07} &
        {0.81} & 
        {2.80} &
        {0.00} &
        {0.0224}
        \\
        {m10xc} & 
        {$4.3 \times 10^{6}$} & 
        {4.47} & 
        {$^\text{\cmark}$1.65} & 
        {0.0346} & 
        {$2.1 \times 10^{7}$} &
        {3.10} &
        {1.14} & 
        {2.68} &
        {0.00} &
        {0.0276}
        \\
        {m10xd} & 
        {$8.3 \times 10^{5}$} & 
        {8.30} & 
        {$^{\dagger}$0.09} & 
        {0.0325} & 
        {$8.1 \times 10^{5}$} &
        {10.9} &
        {2.02} & 
        {2.73} &
        {1.44} &
        {0.0210}
        \\
        {m10xe} & 
        {$1.4 \times 10^{7}$} & 
        {3.32} & 
        {$^\text{\cmark}$2.77} & 
        {0.0586} & 
        {$1.2 \times 10^{10}$} &
        {3.61} &
        {1.74} &
        {2.55} & 
        {0.06} &
        {0.0206}
        \\
        {m10xf} & 
        {$5.7 \times 10^{6}$} & 
        {4.67} & 
        {$^\text{\cmark}$1.65} & 
        {0.0334} & 
        {$3.1 \times 10^{9}$} &
        {3.37} &
        {1.08} & 
        {2.75} &
        {0.00} &
        {0.0268}
        \\
        {m10xg} & 
        {$5.1 \times 10^{7}$} & 
        {2.48} & 
        {$^\text{\cmark}$3.38} & 
        {0.0453} & 
        {$2.0 \times 10^{7}$} &
        {4.50} &
        {1.33} & 
        {2.88} &
        {0.00} &
        {0.0304}
        \\
        {m10xh} & 
        {$8.7 \times 10^{7}$} & 
        {2.33} & 
        {$^\text{\cmark}$5.09} & 
        {0.0740} & 
        {$7.8 \times 10^{6}$} &
        {4.98} &
        {2.64} & 
        {2.57} &
        {0.15} &
        {0.0174}
        \\
        {m10xh$_{\rm A}$} & 
        {$6.2 \times 10^{7}$} & 
        {1.51} & 
        {$^\text{\cmark}$3.00} & 
        {0.0433} & 
        {$9.8 \times 10^{6}$} &
        {3.43} &
        {1.49} & 
        {2.82} &
        {0.00} &
        {0.0205}
        \\
        {m10xi} & 
        {$1.5 \times 10^{7}$} & 
        {4.05} & 
        {$^\text{\cmark}$3.99} & 
        {0.0389} & 
        {$1.2 \times 10^{7}$} &
        {5.40} &
        {1.25} & 
        {2.76} &
        {0.00} &
        {0.0297}
        \\
        {m10z} & 
        {$5.6 \times 10^{6}$} & 
        {4.13} & 
        {$^\text{\cmark}$1.91} & 
        {0.0315} & 
        {$5.0 \times 10^{6}$} &
        {5.01} &
        {1.67} & 
        {2.65} &
        {0.51} &
        {0.0206}
        \\
        {m11a} & 
        {$ 1.4 \times 10^{7}$} & 
        {3.20} & 
        {$^\text{\cmark}$2.54} & 
        {0.0286} & 
        {$1.9 \times 10^{7}$} &
        {3.91} &
        {1.14} & 
        {2.80} &
        {0.00} &
        {0.0200}
        \\
        {m11b} & 
        {$6.2 \times 10^{7}$} & 
        {1.93} & 
        {$^\text{\cmark}$2.36} & 
        {0.0426} & 
        {$2.9 \times 10^{7}$} &
        {3.08} &
        {1.39} & 
        {2.80} &
        {0.00} &
        {0.0100}
        \\
        {m11c} & 
        {$4.6 \times 10^{6}$} & 
        {6.73} & 
        {$^\text{\cmark}$1.61} & 
        {0.0271} & 
        {$5.1 \times 10^{7}$} &
        {3.98} &
        {0.96} & 
        {2.77} &
        {0.00} &
        {0.0254}
        \\
        {m11d} & 
        {$5.1 \times 10^{6}$} & 
        {8.81} & 
        {$^\text{\cmark}$5.75} & 
        {0.0594} & 
        {$5.5 \times 10^{6}$} &
        {6.56} &
        {2.41} & 
        {2.21} &
        {0.19} &
        {0.0195}
        \\
        {m11e} & 
        {$8.9 \times 10^{6}$} & 
        {5.26} & 
        {$^\text{\cmark}$1.72} & 
        {0.0546} & 
        {$4.6 \times 10^{7}$} &
        {3.15} &
        {1.30} & 
        {2.58} &
        {0.00} &
        {0.0399}
        \\
        {m11h} & 
        {$9.3 \times 10^{6}$} & 
        {5.73} & 
        {$^\text{\cmark}$1.96} & 
        {0.0562} & 
        {$3.9 \times 10^{7}$} &
        {3.17} &
        {1.63} & 
        {2.46} &
        {0.00} &
        {0.0169}
        \\
        {m11i} & 
        {$2.0 \times 10^{7}$} & 
        {3.40} & 
        {$^\text{\cmark}$3.46} & 
        {0.0495} & 
        {$1.3 \times 10^{7}$} &
        {4.02} &
        {1.60} & 
        {2.56} &
        {0.00} &
        {0.0244}
        \\
        {m11q} & 
        {$2.1 \times 10^{6}$} & 
        {8.97} & 
        {$^\text{\cmark}$0.86} & 
        {0.0463} & 
        {$8.3 \times 10^{7}$} &
        {2.56} &
        {0.97} & 
        {2.54} &
        {0.00} &
        {0.0465}
        \\
        {m11q$_{880}$}  & 
        {$4.5 \times 10^{6}$} & 
        {6.81} & 
        {$^\text{\cmark}$1.46} & 
        {0.0336} & 
        {$6.0 \times 10^{7}$} &
        {3.07} &
        {1.05} & 
        {2.60} &
        {0.00} &
        {0.0265}
        \\
        \midrule
        \multicolumn{9}{l}{{\bf \em Milky Way-Mass} (12)}\\
        \midrule
        {m12b} & 
        {$6.5 \times 10^{6}$} & 
        {11.18} & 
        {$^\text{\xmark}$0.47} & 
        {0.0528} & 
        {$5.8 \times 10^{8}$} &
        {2.21} &
        {0.97} & 
        {2.54} &
        {0.00} &
        {0.0236}
        \\
        {m12c} & 
        {$2.1 \times 10^{6}$} & 
        {17.14} & 
        {$^\text{\xmark}$0.21} & 
        {0.0690} & 
        {$1.3 \times 10^{9}$} &
        {2.48} &
        {1.60} & 
        {2.21} &
        {0.48} &
        {0.0124}
        \\
        {m12f} & 
        {$4.3 \times 10^{6}$} & 
        {13.76} & 
        {$^\text{\xmark}$0.44} & 
        {0.0450} & 
        {$6.1 \times 10^{8}$} &
        {2.41} &
        {0.85} & 
        {2.56} &
        {0.00} &
        {0.0224}
        \\
        {m12i} & 
        {$3.3\times 10^{6}$} & 
        {13.60} & 
        {$^\text{\xmark}$0.27} & 
        {0.0255} & 
        {$1.6 \times 10^{9}$} &
        {3.00} &
        {0.56} & 
        {2.91} &
        {0.00} &
        {0.0201}
        \\
        {m12m} & 
        {$7.3 \times 10^{6}$} & 
        {10.96} & 
        {$^\text{\xmark}$0.78} & 
        {0.0539} & 
        {$1.8 \times 10^{8}$} &
        {3.37} &
        {1.22} & 
        {2.51} &
        {0.25} &
        {0.0128}
        \\
        {m12r} & 
        {$8.1 \times 10^{5}$} & 
        {23.74} & 
        {$^{\dagger}$0.29} & 
        {0.0611} & 
        {$2.2 \times 10^{7}$} &
        {4.56} &
        {2.75} & 
        {2.10} &
        {0.88} &
        {0.0184}
        \\
        {m12w} & 
        {$3.2 \times 10^{6}$} & 
        {13.31} & 
        {$^{\dagger}$0.31} & 
        {0.0451} & 
        {$4.0 \times 10^{7}$} &
        {4.84} &
        {1.61} & 
        {2.40} &
        {0.87} &
        {0.0222}
        \\
        {m12z} & 
        {$3.7 \times 10^{6}$} & 
        {12.19} & 
        {$^\text{\cmark}$2.64} & 
        {0.0432} & 
        {$2.1\times 10^{7}$} &
        {5.12} &
        {1.42} & 
        {2.40} &
        {0.13} &
        {0.007}
        \\
        {Thelma} & 
        {$5.0 \times 10^{6}$} & 
        {12.66} & 
        {$^\text{\cmark}$1.20} & 
        {0.0212} &
        {$3.0 \times 10^{8}$} &
        {5.46} &
        {0.71} & 
        {2.95} &
        {0.00} &
        {0.026}
        \\
        {Louise} & 
        {$3.4 \times 10^{6}$} & 
        {13.15} & 
        {$^\text{\cmark}$0.41} & 
        {0.0371} & 
        {$6.0\times 10^{8}$} &
        {2.25} &
        {0.78} & 
        {2.58} &
        {0.00} &
        {0.0222}
        \\
        {Romeo} & 
        {$5.6 \times 10^{6}$} & 
        {11.59} & 
        {$^\text{\xmark}$0.04} & 
        {0.0400} & 
        {$1.6\times 10^{10}$} &
        {2.78} &
        {0.42} & 
        {3.21} &
        {0.00} &
        {0.0332}
        \\
        {Juliet} & 
        {$3.6 \times 10^{6}$} & 
        {12.70} & 
        {$^\text{\xmark}$0.00} & 
        {0.0433} & 
        {$1.8 \times 10^{8}$} &
        {2.48} &
        {0.39} & 
        {3.16} &
        {0.00} &
        {0.0338}
        \\
        \bottomrule
        \bottomrule
    \end{tabular}
    \end{threeparttable}
\end{table*}


\bsp
\label{lastpage}
\end{document}